\documentclass[12pt,prd]{article}
\usepackage{jheppub}
\usepackage{rotating}
\usepackage{array}
\usepackage{amsmath}
\usepackage[normalem]{ulem}
\usepackage{slashed}
\usepackage{booktabs}
\usepackage[pdftex,table]{xcolor}
\usepackage{units}
\usepackage{xfrac}
\usepackage{mathtools}
\usepackage{empheq}
\usepackage[]{units}
\usepackage{multirow}
\usepackage{amssymb}
\usepackage{url}
\usepackage{comment}

\usepackage{tikz}

\def\beq{\begin{equation}}
\def\eeq{\end{equation}}
\newcommand{\bea}{\begin{eqnarray}\begin{aligned}}
\newcommand{\eea}{\end{aligned}\end{eqnarray}}
\def\bitem{\begin{itemize}}
\def\eitem{\end{itemize}}

\definecolor{darkpurple}{rgb}{0.5, 0.2, 0.8}
\definecolor{darkblue}{rgb}{0.0, 0.0, 0.8}
\definecolor{darkgreen}{rgb}{0.0, 0.4, 0.0}
\definecolor{darkred}{rgb}{0.5, 0.0, 0.0}
\usepackage{hyperref}
\hypersetup{
    linktocpage,
     colorlinks,
     citecolor=darkgreen,
     linkcolor= darkgreen,
     urlcolor=darkgreen
}

\newcommand{\ym}{ \widehat{m} }

\newcommand{\Am}{A_{\text{mass}}}

\abstract{
The ABCD method is one of the most widely used data-driven background estimation techniques in high energy physics. Cuts on two statistically-independent classifiers separate signal and background into four regions, so that background in the signal region can be estimated simply using the other three control regions. Typically, the independent classifiers are chosen ``by hand" to be intuitive and physically motivated variables. Here, we explore the possibility of automating the design of one or both of these classifiers using machine learning. We show how to use state-of-the-art decorrelation methods to construct powerful yet independent discriminators. Along the way, we uncover a previously unappreciated aspect of the ABCD method: its accuracy hinges on having low signal contamination in control regions not just overall, but {\it relative} to the signal fraction in the signal region. We demonstrate the method with three examples: a simple model consisting of three-dimensional Gaussians; boosted hadronic top jet tagging; and a recasted search for paired dijet resonances. In all cases, automating the ABCD method with machine learning significantly improves performance in terms of ABCD closure, background rejection and signal contamination. 
}

\keywords{}

\begin{document}
\title{ABCDisCo: Automating the ABCD Method with  Machine Learning}

\author[1]{Gregor Kasieczka,} 
\author[2]{Benjamin Nachman,}
\author[3]{Matthew D. Schwartz,}
\author[2,4,5]{and David Shih,}
\affiliation[1]{\normalsize Institut f\"ur Experimentalphysik, Universit\"at Hamburg,\\Luruper Chaussee 149, D-22761 Hamburg, Germany}
\affiliation[2]{\normalsize Physics Division, Lawrence Berkeley National Laboratory, Berkeley, CA 94720, USA}
\affiliation[3]{\normalsize Department of Physics, Harvard University, Cambridge, MA 02138}
\affiliation[4]{\normalsize NHETC, Department of Physics and Astronomy, Rutgers University, Piscataway, NJ 08854, USA}
\affiliation[5]{\normalsize Berkeley Center for Theoretical Physics, University of California, Berkeley, CA 94720, USA}

\emailAdd{gregor.kasieczka@uni-hamburg.de}
\emailAdd{bpnachman@lbl.gov}
\emailAdd{shih@physics.rutgers.edu}
\emailAdd{schwartz@g.harvard.edu}

\maketitle
 
\newpage
\section{Introduction}
\label{sec:intro}

A key component of high energy physics data analysis, whether for Standard Model (SM) measurements or searches beyond the SM, is background estimation.
While powerful simulations and first-principles calculations exist and are constantly improving, they still remain inadequate for the task of precisely estimating backgrounds in many situations. For example, events with a large number of hadronic jets have high-multiplicity SM backgrounds whose cross sections are difficult to estimate.  Therefore methods for {\it data-driven} background estimation remain a crucial part of the experimental toolkit. 
The idea behind all data-driven background estimation strategies is to extrapolate or interpolate from some control regions which are background dominated into a signal region of interest. 

\begin{figure}[t]
    \centering
    \hspace{-10mm}
      \resizebox{16cm}{!}{
\begin{tikzpicture}
\node at (0,0) {    
\includegraphics[trim=5cm 2cm 5cm 3cm,clip=true,width=28cm]{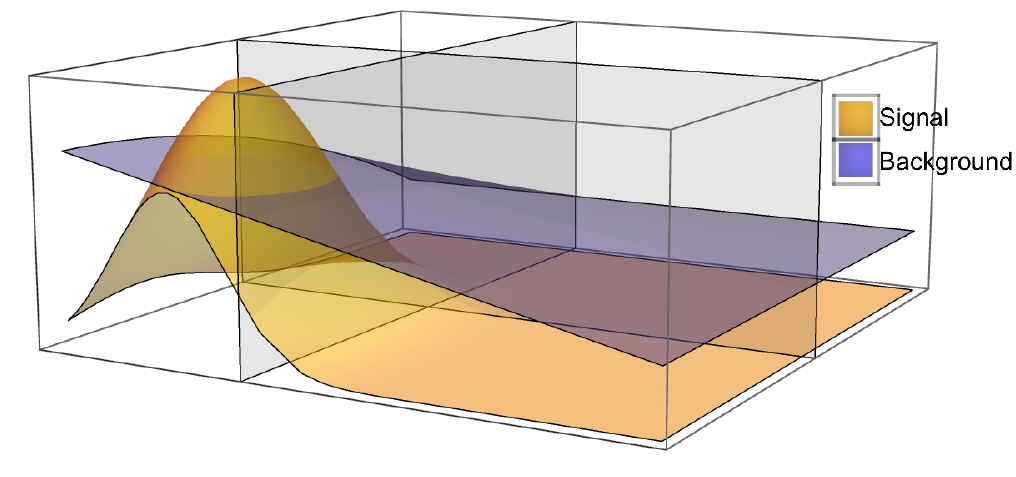}    };
\begin{scope}[shift={(0,0)}]
    \node[scale=3] at (-6,-5) {$f$};
    \draw[<-,line width=3] (-5,-5.1) to (-3,-5.4);
    \node[scale=3] at (7,-5) {$g$};
    \draw[<-,line width=3] (7.7,-4.5) to (9,-3.5);
    \node[scale=4,darkred] at (-9,5) {$A$};
    \node[scale=4,darkred] at (-3.5,6) {$B$};
    \node[scale=4,darkred] at (1,4) {$C$};
    \node[scale=4,darkred] at (4,5.5) {$D$};
    \end{scope}
\node at (22,0) {    
{\includegraphics[trim=0cm 0cm 0cm 0cm,clip=true,width=14cm]{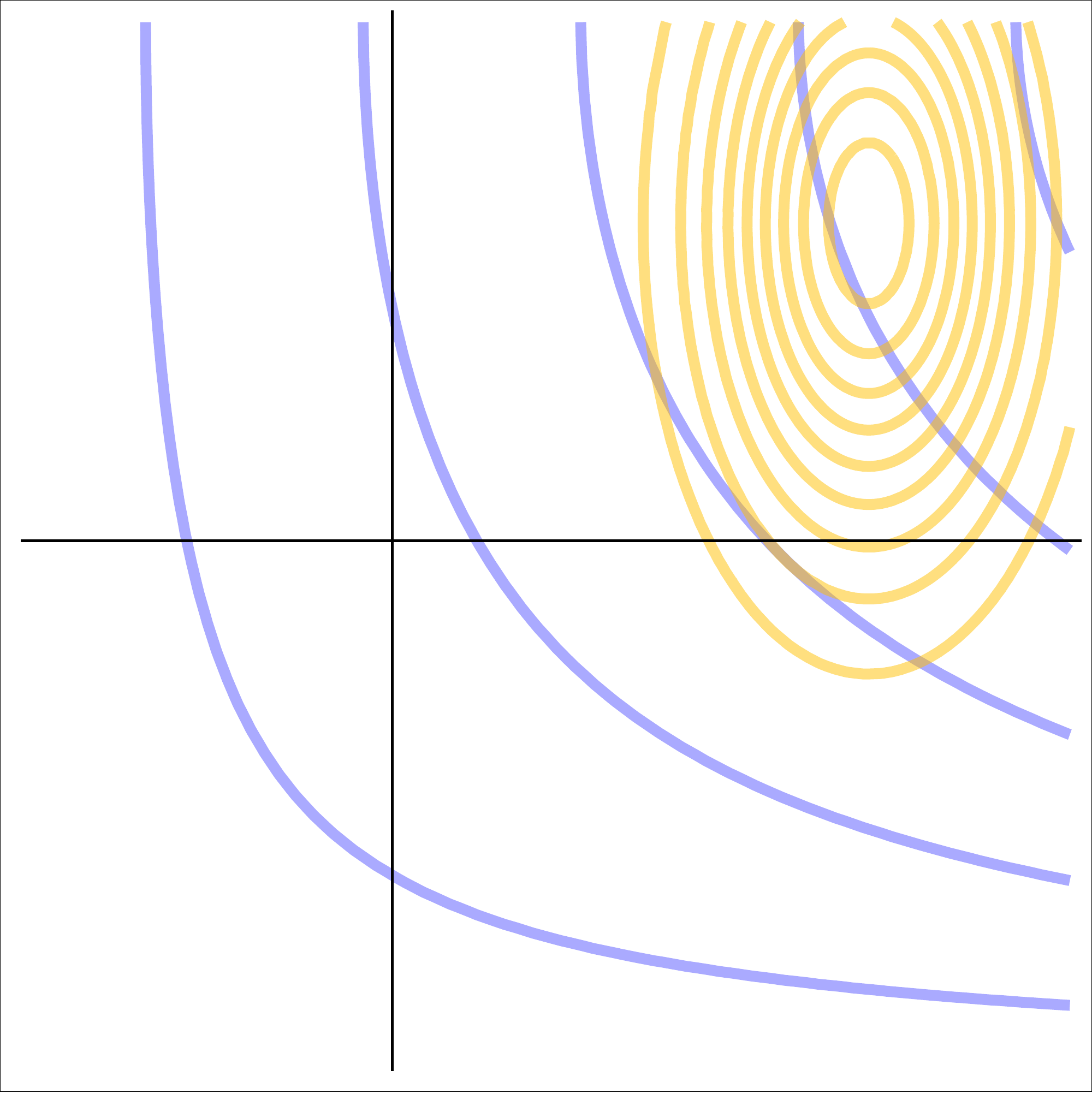}    }};
\begin{scope}[shift={(22,0)}]
    \node[scale=3] at (-4.2,-8) {$f$};
    \draw[->,line width=3] (-3.5,-8) to (-1.5,-8);
    \node[scale=3] at (-7.7,-2.2) {$g$};
    \draw[->,line width=3] (-7.7,-1.5) to (-7.7,0.5);
    \node[scale=4,darkred] at (-4,-3) {$D$};
    \node[scale=4,darkred] at (-4,3) {$C$};
    \node[scale=4,darkred] at (4,-3) {$B$};
    \node[scale=4,darkred] at (4,3) {$A$};
    \end{scope}
\end{tikzpicture}}
    \caption{The ABCD method is used to estimate the background in region $A$ as $N_A = \dfrac{N_B N_C}{N_D}$. It requires the signal to be relatively localized in region $A$ and the observables to be independent on background. The shaded planes (left) or lines (right) denote thresholds which isolate the signal in region $A$.}
    \label{fig:abcd3d}
\end{figure}
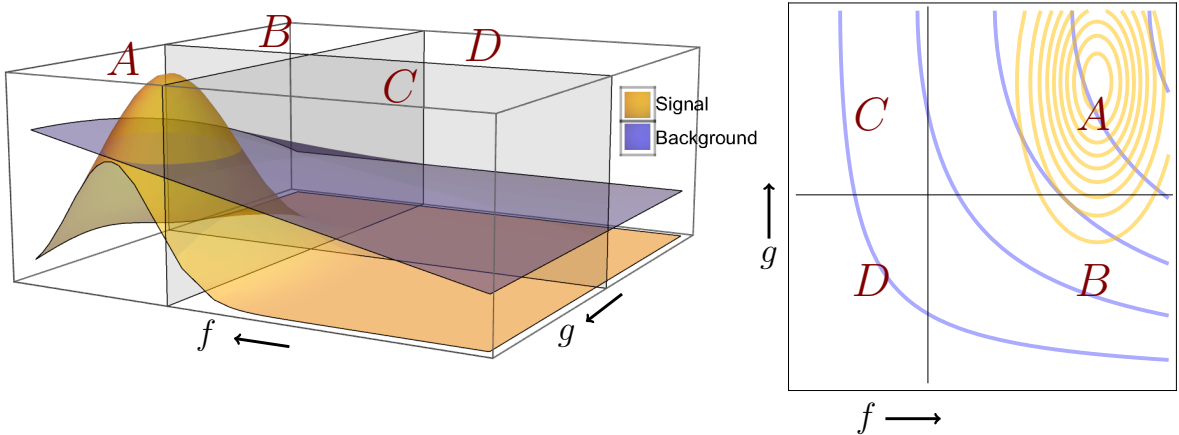

One classic (see e.g. Ref.~\cite{Abe:1990sd}) data-driven background method which is used in a multitude~\cite{atlasexoticstwiki,atlassusytwiki,cmsexoticstwiki,cmssusytwiki,cmsb2gtwiki} of physics analyses at the Large Hadron Collider (LHC) and elsewhere is the {\it ABCD method}. The idea of the ABCD method is to pick two observables $f$ and $g$ (for example, the invariant mass of a dijet system and the rapidity of that system) which are approximately statistically independent for the background, and which are effective discriminators of signal versus background. Simple thresholds on these observables partition events into four regions. Three of these regions, called $B$, $C$ and $D$, are background dominated. The fourth, $A$, is the signal region. If the observables are independent then the background in the signal region can be predicted from the other three regions via:
\beq
\label{eq:abcdmain}
N_A = \frac{N_B N_C}{ N_D},
\eeq
where $N_i$ is the number of events in region $i$.  This setup is depicted schematically for signal and background distributions in Fig.~\ref{fig:abcd3d}.

Typically, the observables $f$ and $g$ for the ABCD method are chosen to be simple, physically well-motivated features such as mass, $H_T$, missing $E_T$, etc. Their independence is always ensured manually, e.g.\ by choosing features that are known physically to have little correlation or by trial-and-error.\footnote{There are examples where $f$ or $g$ are chosen automatically, as is the case when one of them is a neural network (see e.g. Ref.~\cite{Aad:2020hzm}).  However, such analyses do not have an automated procedure for ensuring that $f$ and $g$ are independent and the departure from Eq.~(\ref{eq:abcdmain}) can be significant. } In some cases, independence can be guaranteed by using completely orthogonal sources of information, such as 
measurements from different sub-detectors or properties of independently produced particles.  However, more often than not, the features are not 100\% independent and one has to apply a residual correction derived from simulations.  
Ideally, this simulation correction has small uncertainties --- either because the effect itself is small, or 
because the correction is robust.
But such corrections, together with the fact that simple kinematic features are typically not optimal discriminants of signal versus background, generally limit the effectiveness of the ABCD method and the sensitivity of the analysis in question. 
(See~\cite{Choi:2019mip}, however, for a proposal for extending the ABCD method using higher-order information when the features are not independent.) 

In this paper, we will explore the systematic application of deep learning to the ABCD method.  Deep learning has already demonstrated impressive success in finding observables that are effective at discrimination~\cite{Larkoski:2017jix,Guest:2018yhq,Kasieczka:2019dbj,hepmllivingreview,Cogan:2014oua,Almeida:2015jua,deOliveira:2015xxd,ATL-PHYS-PUB-2017-017,Lin:2018cin,Komiske:2018oaa,Barnard:2016qma,Komiske:2016rsd,Kasieczka:2017nvn,Macaluso:2018tck,Nguyen:2018ugw,ATL-PHYS-PUB-2019-028,Andrews:2018nwy,Guest:2016iqz,Louppe:2017ipp,Cheng:2017rdo,Henrion:DLPS2017,Ju:2020xty,Martinez:2018fwc,Moreno:2019bmu,Qasim:2019otl,Chakraborty:2019imr,Chakraborty:2020yfc,1797439,1801423,Komiske:2018cqr,Qu:2019gqs,Datta:2019,Datta:2017rhs,Datta:2017lxt,Komiske:2017aww,Butter:2017cot,Chen:2019uar,Fraser:2018ieu,Datta:2019ndh,Moreno:2019neq,Stoye:DLPS2017,Chien:2018dfn,Kasieczka:2018lwf,1806025,Diefenbacher:2019ezd,Nakai:2020kuu,Sirunyan:2017ezt,bielkov2020identifying,Baldi:2014kfa,10.1088/2632-2153/ab9023,1792136,deOliveira:2018lqd,Paganini:DLPS2017,Hooberman:DLPS2017,Belayneh:2019vyx}
and that are uncorrelated with other observables~\cite{Louppe:2016ylz,Dolen:2016kst,Moult:2017okx,Stevens:2013dya,Shimmin:2017mfk,Bradshaw:2019ipy,ATL-PHYS-PUB-2018-014,DiscoFever,Xia:2018kgd,Englert:2018cfo,Wunsch:2019qbo,Rogozhnikov:2014zea,Sirunyan:2019nfw,clavijo2020adversarial,Aguilar-Saavedra:2017rzt,Sirunyan:2020lcu}. Building on previous success, we will aim to use deep learning to automate the selection of features used in the ABCD method, simultaneously optimizing their discrimination power while ensuring their independence.

The main tool we will use in automating the ABCD method will be a recently proposed method for training decorrelated deep neural networks~\cite{DiscoFever}. This method uses a well-known statistical measure of non-linear dependence known as {\it Distance Correlation} (DisCo)  \cite{szekely2007, szekely2009, SzeKely:2013:DCT:2486206.2486394,szekely2014}. DisCo is a function of two random variables (or samples thereof) and is zero if and only if the variables are statistically independent, otherwise it is positive. Therefore it can be added as a regularization term in the loss function of a neural network to encourage the neural network output to be decorrelated against any other feature. In~\cite{DiscoFever} it was shown that DisCo decorrelation achieves state-of-the-art decorrelation performance while being easier and more stable to train than approaches based on adversarial methods. Therefore it is ideally suited to automating the ABCD method. 

We will propose two new ideas for automating the ABCD method, which we will call {\it Single DisCo} and {\it Double DisCo}, respectively. In Single DisCo, we will train a single neural network classifier on signal and background and use DisCo regularization to force it to be independent in the background of a second, fixed feature (such as invariant mass). In Double DisCo, we will train {\it two} neural network classifiers and use DisCo regularization to force them to be independent of one another.  %

We will study three examples to illustrate the effectiveness of these methods. The first example is a simple model where signal and background are drawn from three-dimensional Gaussian distributions. Here the aim is to understand many of the features of Single and Double DisCo in a fully controlled environment. The second example is boosted hadronic top tagging, where often sideband interpolation in mass is employed. For the ABCD method we treat a window selection on the mass as a classifier variable.  Thus we use the invariant mass  as the Single DisCo fixed feature, and we then show how Double DisCo can improve on this by combining mass with other information to produce more effective classification. Finally, we examine a search that currently uses the conventional ABCD method: the ATLAS paired dijet resonance search, motivated by RPV squark decays~\cite{Aaboud:2017nmi} (for a similar search by CMS, see~\cite{Sirunyan:2018rlj}).
We show that significant performance gains
are possible using Single and Double DisCo. 

In the course of our study of the ABCD method, we will uncover a hitherto unappreciated limitation of the method, which we call {\it normalized signal contamination}. Usually, practitioners are concerned with the overall signal-to-background ratio in the control regions; if this is small then they are usually satisfied. We point out that in fact another relevant quantity for the significance calculation is the signal-to-background ratio in the control regions {\it relative} or {\it normalized} to the signal-to-background ratio in the signal region. In other words, the requirement of signal contamination is actually
\beq
{N_{i,s}\over N_{i,b}}\ll {N_{A,s}\over N_{A,b}}
\eeq
{\it in addition} to ${N_{i,s}\over N_{i,b}}\ll 1$ (where $N_{i,s}$ and $N_{i,b}$ are the numbers of signal and background events in region $i=A,B,C,D$). In many analyses (e.g.~\cite{Aaboud:2017nmi}), the signal fraction in the signal region is quite small, meaning that even a small amount of signal contamination in the control regions can bias the $p$-values reported by the search. We will show that Single and Double DisCo not only improve the discrimination power and background closure of the ABCD method but can also significantly reduce the level of signal contamination at the same time. 

\clearpage
This paper is organized as follows.  Section~\ref{sec:ABCDrequirements} reviews the ABCD method and Sec.~\ref{sec:automating} describes how the method can be automated using deep learning.  Numerical results for examples described above are presented in Sec.~\ref{sec:results}. The paper ends with conclusions and outlook in Sec.~\ref{sec:conclusions}.

\section{The ABCD method}
\label{sec:ABCDrequirements}

The ABCD method starts with two features $f$ and $g$. Imposing thresholds $f_c$ and $g_c$ divides the feature space into four rectangular regions, $A$, $B$, $C$ and $D$ with corresponding event counts:
\bea
\label{eq:regiondefinitions}
N_{A,\ell} &= N_{\ell}\,\Pr (f \geq f_c \,\,\,\text{and}\,\,\, g \geq g_c|\ell)\cr
N_{B,\ell} &= N_{\ell}\,\Pr (f \geq f_c \,\,\,\text{and}\,\,\, g<g_c|\ell)\cr
N_{C,\ell} &= N_{\ell}\,\Pr (f<f_c \,\,\,\text{and}\,\,\, g \geq g_c|\ell)\cr
N_{D,\ell} &= N_{\ell}\,\Pr (f<f_c \,\,\,\text{and}\,\,\, g<g_c|\ell),
\eea
where $N_\ell=N_{A,\ell} +N_{B,\ell} +N_{C,\ell}+ N_{D,\ell} $ is the total number of events of type $\ell$ and $\ell\in\{\text{signal ($s$)},\text{background ($b$)},\text{all ($a$)}\}$ and $\Pr(\cdot)$ is the probability.  The regions $B,C$ and $D$ can be used to predict $N_A$:
\beq\label{eq:ABCDpred}
 N_{A,b}^\text{predicted}\equiv {N_{B,a} N_{C,a}\over N_{D,a}}\,.
\eeq
For the ABCD method to be valid, we would need $N_{A,b} =N_{A,b}^\text{predicted}$.

There are two requirements for $N_{A,b}^\text{predicted}$ to be accurate.  First, the Bernoulli random variables $f<f_c$ and $g<g_c$ must be independent for the background in order to guarantee that 
\begin{equation}
    N_{A,b} =  {N_{B,b} N_{C,b}\over N_{D,b}}\,.
    \label{Njustb}
\end{equation}
To see this, note that~\eqref{Njustb} is equivalent to
\beq\label{eq:independence1}
N_b\times N_{A,b} =(N_{A,b}+N_{B,b})\times(N_{A,b}+N_{C,b})\,.
\eeq
Then, substituting in Eq.~\eqref{eq:regiondefinitions} to Eq.~\eqref{eq:independence1} yields
\beq
\Pr (f \geq f_c \,\,\,\text{and}\,\,\, g \geq g_c|b) = \Pr (f \geq f_c|b)\times \Pr (g \geq g_c|b),
\eeq
which is a definition of independence.  While it is sufficient to have one set of thresholds, having a range over which independence holds adds robustness to the estimation procedure.  If the ABCD method holds for all values of $f_c$ and $g_c$, then $f$ and $g$ themselves must be independent.  
Note that this condition is stronger than requiring zero {\it linear} correlation. Two random variables can have zero linear correlation yet be nonlinearly dependent.  In general, such a case would invalidate Eq.~\eqref{Njustb}. 

The second requirement for the ABCD method involves the signal and the background:
\begin{equation}
    \frac{N_{B,a} N_{C,a}}{ N_{D,a}}=\frac {N_{B,b} N_{C,b}}{ N_{D,b}}\,.
\end{equation}
In particular, if the signal contamination in regions $B,C$ and $D$ is large, then Eq.~\eqref{eq:ABCDpred} will not hold. But what does large mean in this context? 
Typically, large signal contamination is taken to be an overall statement, i.e.
\beq\label{eq:deltasmall}
\delta_i \equiv {N_{i,s}\over N_{i,b}} \ll 1\,.
\eeq
for regions $i=B,C,D$. 
However, we will now show that in addition to this criterion, another relevant quantity is {\it normalized signal contamination}
\beq\label{eq:rdef}
r \equiv \delta_A^{-1}(\delta_B+\delta_C-\delta_D) = \left({N_{A,s}\over N_{A,b}}\right)^{-1}\left(
{N_{B,s}\over N_{B,b}}+{N_{C,s}\over N_{C,b}}-{N_{D,s}\over N_{D,b}}
\right)\,,
\eeq
and for the ABCD method to be valid, it must satisfy 
\beq\label{eq:rsmall}
|r|\ll 1\,.
\eeq
Note that this is often a much stronger requirement than (\ref{eq:deltasmall}).  It is not enough that the signal fractions in each control region are small -- they must be {\it small compared to the signal fraction in the signal region}. In many searches (e.g. the RPV stop search in Sec.~\ref{sec:RPV}), signal to background can be quite small in the signal region, meaning that this can be a significant (and underappreciated) 
constraint on the ABCD method. 

To see why (\ref{eq:rsmall}) is required, suppose that the ABCD method closes exactly, so that
Eq.~\eqref{Njustb} holds,  but there is some signal contamination in all four regions.  Then,
\bea
\label{eq:expansion}
N_{A,b}^{\text{predicted}}
&=N_{B,b}\,(1+\delta_B)\times\frac{N_{C,b}\,(1+\delta_C)}{N_{D,b}\,(1+\delta_D)}\\
&=N_{B,b}\times \frac{N_{C,b}}{N_{D,b}} \left[ 1+\delta_B+\delta_C-\delta_D +\mathcal{O}(\delta^2)\right],\\
&=N_{A,b} \left[ 1+\delta_B+\delta_C-\delta_D +\mathcal{O}(\delta^2)\right],
\eea
This will be compared with the number of events in region $A$, $N_{A,a}=N_{A,b}(1+\delta_A)$, to decide whether there is an excess or not. In order to detect the signal in $A$, one needs 
Eq.~\eqref{eq:rsmall} to be satisfied.  Note that we are still assuming that $\delta_{B,C,D}\ll 1$ in order for the subleading terms in Eq.~(\ref{eq:expansion}) to be negligible.

Another point is that generally $\delta_D$ can be neglected compared to $\delta_B$ and $\delta_C$ (as it is diagonally opposite and should therefore be doubly-suppressed).
So we expect $r>0$ and an {\it overestimate} of the background in the signal region.  This will make it much harder to discover new physics. 

Finally, let us make the connection between the normalized signal contamination and classifier performance.  For the fixed thresholds $f_c$ and $g_c$, the signal ($\epsilon_s$) and background ($\epsilon_b$) efficiencies 
for each individual classifier 
can be computed as:
\bea
\epsilon_{f,b}&=\frac{N_{A,b}+N_{B,b}}{N_b} \stackrel{\text{independence}}{=} \frac{N_{A,b}}{N_{A,b}+N_{C,b}}\\
    \epsilon_{g,b}&=\frac{N_{A,b}+N_{C,b}}{N_b} \stackrel{\text{independence}}{=} \frac{N_{A,b}}{N_{A,b}+N_{B,b}}  \\
    \epsilon_{f,s}&=\frac{N_{A,s}+N_{B,s}}{N_s}\\
    \epsilon_{g,s}&=\frac{N_{A,s}+N_{C,s}}{N_s} 
\eea
With these definitions and neglecting $N_{D,s}$, Eq.~\eqref{eq:rdef} can be re-written as

\beq\label{eq:frewrite}
r=\frac{(1-\epsilon_{f,s})}{(1-\epsilon_{f,s}+\epsilon_{g,s})}
\frac{\epsilon_{f,b}}{(1-\epsilon_{f,b})}+
\frac{(1-\epsilon_{g,s})}{(1-\epsilon_{g,s}+\epsilon_{f,s})}
\frac{\epsilon_{g,b}}{(1-\epsilon_{g,b})}\,.
\eeq
The two terms in Eq.~\eqref{eq:frewrite} are nearly the {\it diagnostic odds ratio} 
and importantly are minimized for a given signal efficiency when the background efficiency is as small as possible.  This demonstrates that ``classification performance" and ``signal contamination" are synonymous in this context --- the better a classifier is, the more likely it will be that there is a threshold which ensures a small relative signal contamination. 

\begin{figure}[t]
\centering
\includegraphics[height=0.37\textwidth]{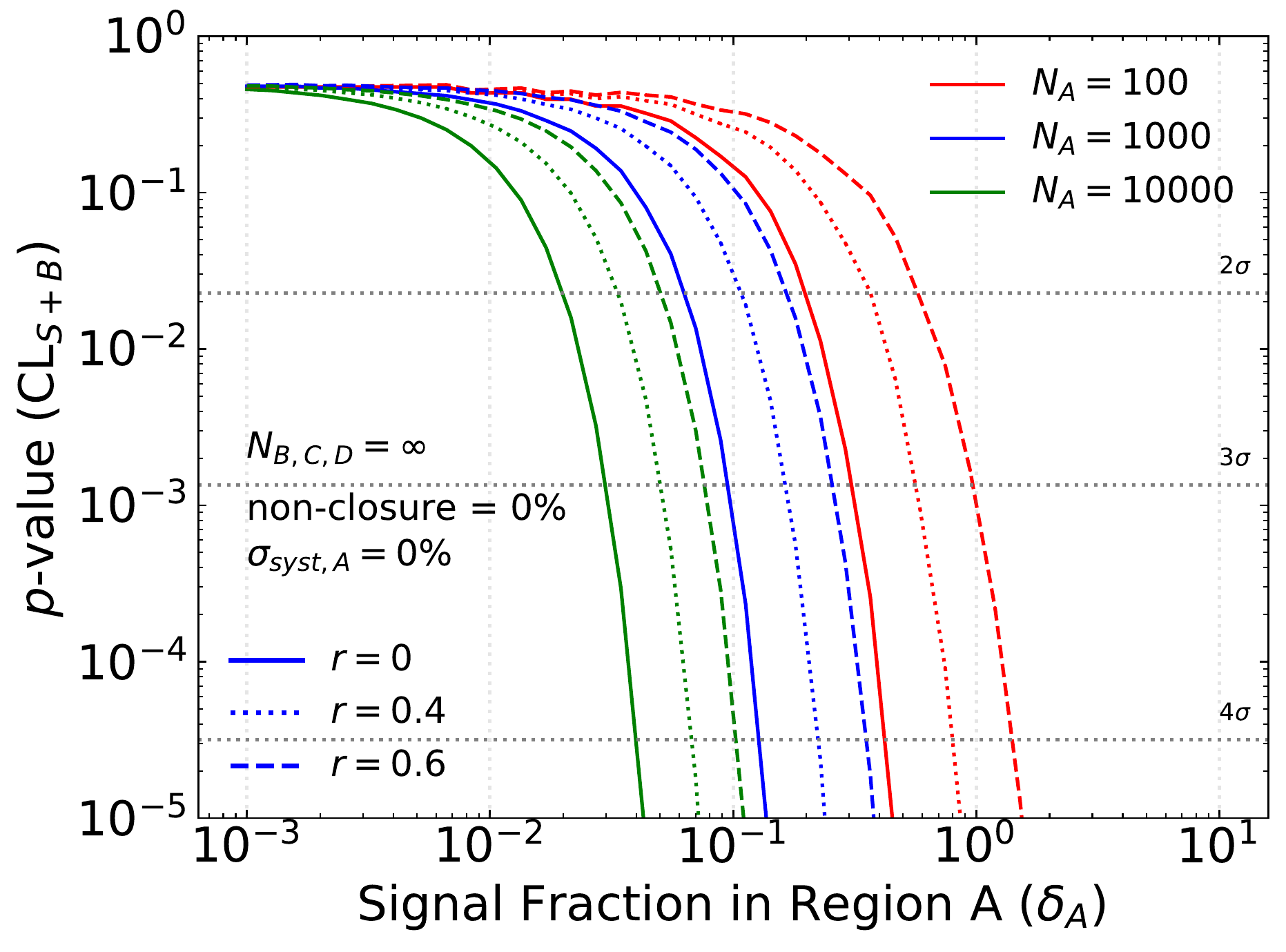}%
\includegraphics[height=0.37\textwidth]{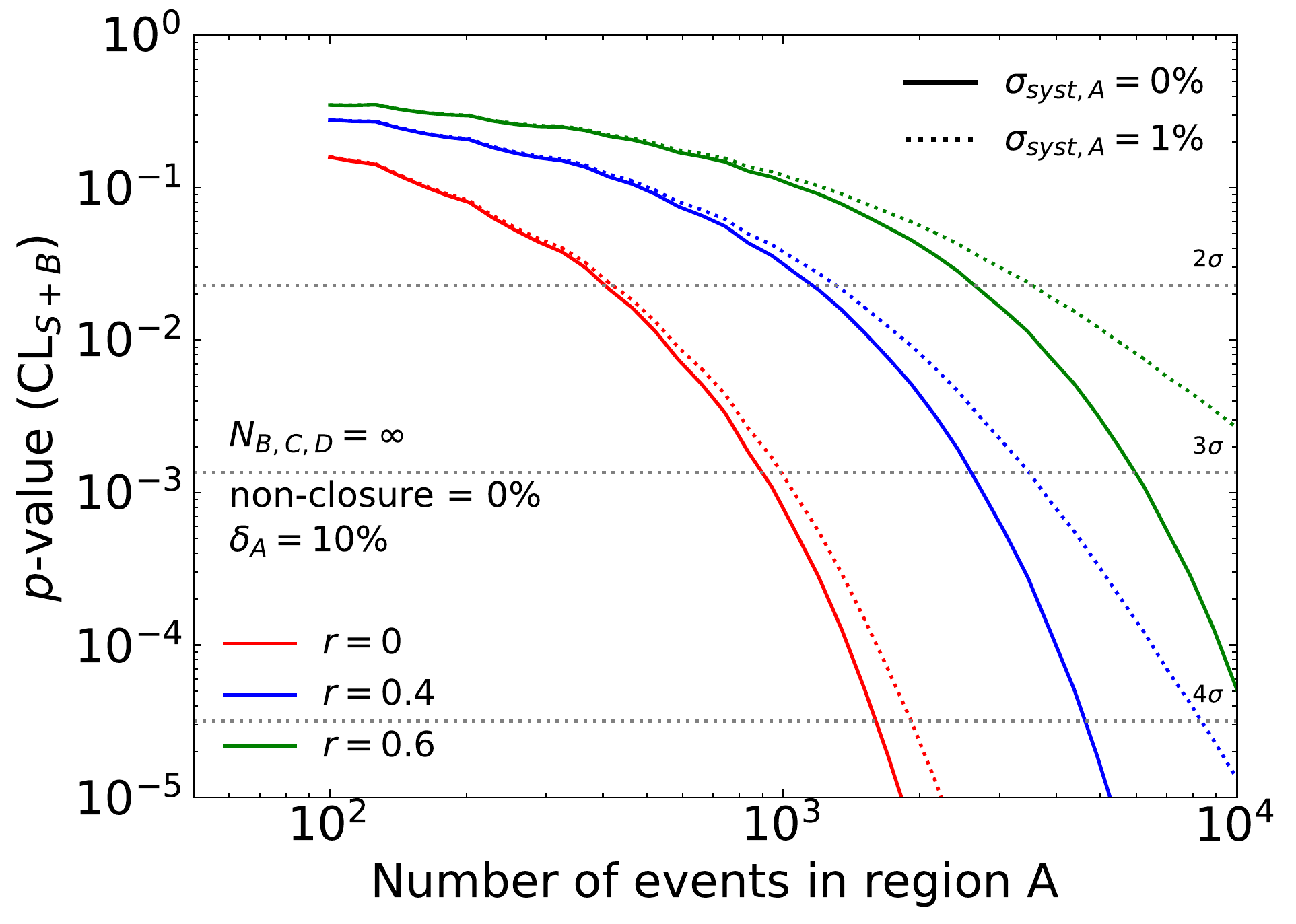}
\caption{The $p$-value (CL${}_\text{S+B}$) for the ABCD method as a function of $\delta_A$ ($N_A$), the signal fraction in region $A$ (the number of background events in region $A$) for the left (right) plot. It is assumed that there is no uncertainty from regions $C$ and $D$.}
\label{fit:abcdindependence}
\end{figure}

To illustrate these points, we show in Fig.~\ref{fit:abcdindependence} the effect of signal contamination on the $p$-value. The left plot shows the interplay between the relative signal contamination $r$ and the number of events $N_A$ in the signal region as a function of $\delta_A$.  For example, if the signal fraction in the signal region is $\delta_A=10\%$ and $N_A=1000$, the true $p$-value is $0.0015$ while the reported value assuming negligible signal contamination would be $0.03$ or $0.1$ with an unaccounted for signal contamination of 4\% and 6\% in region $B$, respectively.  
 
 Correctly accounting for this signal contamination would require having a signal-model-dependent ABCD estimation.  This could be done (see e.g.~\cite{Aaboud:2018gmx}), but would be much more complicated than most applications of the ABCD method.  Adding an uncertainty to account for potential signal contamination is also not ideal - this is shown in the bottom plot of Fig.~\ref{fit:abcdindependence}.  Once again, for $N_A=1000$ and $\delta_A=10\%$, the true $p$-value is $0.0015$ and a signal contamination of 4\% in region $B$ results in a $p$-value of $0.03$.  Adding an uncertainty of 5\% increases this to $0.16$ and an uncertainty of 10\% further increases the $p$-value to $0.29$.  So while this would result in a conservative $p$-value, it means that potential discoveries would be masked.

\section{Automating the ABCD Method}
\label{sec:automating}

Having described the requirements for the ABCD method 
(two strong classifiers that are independent for background), we now turn to the main idea of the paper: automating the ABCD method with machine learning. 

Typically, when the ABCD method is used in experimental analyses, the two features are chosen by hand, based on physical intuition. Usually the features are simple  quantities, such as mass, $H_T$, $p_T$, or missing $E_T$. In the remainder of the paper, we will investigate the benefits of allowing the ABCD features to be more complicated functions of the inputs. These functions will be obtained by training neural networks with suitable loss functions that ensure the ABCD objectives. We will see that machine learning has the potential to greatly improve the performance of the ABCD method. 

The basic idea is that we want to train a classifier $f(X)$ where $X$ are the input features (either low level inputs such as four vectors or images, or high level inputs such as $p_T$, mass, etc.) that is forced to be decorrelated against another classifier $g(X)$. This will achieve the first ABCD requirement of independent features. If the two classifiers are both good discriminants, this will satisfy the second ABCD requirement. 

\clearpage

One can imagine two versions of this idea, both of them new:
\begin{enumerate}

\item The second classifier is a simple, existing high-level variable (e.g.\ mass). In this case the problem is basically identical to the one that has been solved in the literature on decorrelation. We then just have to apply
these approaches to the
the ABCD method. 

\item The second classifier is also a neural network. In this case we need to train two neural networks simultaneously while keeping them decorrelated from one another. This requires us to go beyond the usual literature on decorrelation against a fixed feature.

\end{enumerate}

Regardless of whether $g(X)$ is fixed or learned, decorrelation can be achieved by any of the numerous methods that have been proposed~\cite{Louppe:2016ylz,Dolen:2016kst,Moult:2017okx,Stevens:2013dya,Shimmin:2017mfk,Bradshaw:2019ipy,ATL-PHYS-PUB-2018-014,DiscoFever,Xia:2018kgd,Englert:2018cfo,Wunsch:2019qbo,Rogozhnikov:2014zea,Sirunyan:2019nfw,clavijo2020adversarial,Aguilar-Saavedra:2017rzt,Sirunyan:2020lcu}. In this paper we will use the Distance Correlation (DisCo) method~\cite{DiscoFever}. DisCo decorrelation proceeds through a positive-definite regularization term that penalizes statistical dependence. It achieves state-of-the-art performance while being significantly easier to train than adversarial decorrelation methods which rely on saddle-point extremization.  

For the Single DisCo ABCD method, we take the loss function to be the same as in~\cite{DiscoFever}:
\beq
\label{eq:mdisco}
\mathcal{L}[f(X)]=\mathcal{L}_\text{classifier}[f(X),y] + \lambda \,\text{dCorr}^2_{y=0}[f(X),X_0],
\eeq
where $X$ are the features used for classification, $y\in\{0,1\}$ are the labels, $X_0$ is the feature that one wants to be decorrelated from $f(X)$ ($X_0$ could be part of $X$), and $\mathcal{L}_\text{classifier}$ is the classifier loss such as the commonly used binary cross entropy. The subscript $y=0$ in the second term of Eq.~\eqref{eq:mdisco} ensures that the decorrelation is only applied to the background (class 0).  Furthermore, $\lambda\geq 0$ is a hyperparameter that determines the decorrelation strength. 
The function $\text{dCorr}^2[f,g]$ is the squared distance correlation defined in~\cite{szekely2007, szekely2009, SzeKely:2013:DCT:2486206.2486394,szekely2014} (see App.~\ref{sec:discodef}). It has the property that $0\leq \text{dCorr}[f,g]\leq 1$ and $\text{dCorr}[f,g]=0$ if and only if $f$ and $g$ are independent. For Single DisCo, $g(X)=X_0$.

In practice, $f$ is parameterized as a neural network and Eq.~\eqref{eq:mdisco} is minimized using gradient-based methods.  The distance correlation is computed for batches of data used to stochastically estimate the gradient.  In the limit of small numbers of events, the naive distance covariance computed by replacing expectation values with sample averages is a biased estimator of the true distance correlation.  Analogously to the case of sample variance (in which a factor 
of $\frac{1}{N-1}$ instead of $\frac{1}{N}$ --- where $N$
denotes the minibatch-size ---
is inserted to remove bias), there is an analytic low-$N$ correction to the distance covariance that is unbiased~\cite{szekely2014,szekely2009}.  Numerical results suggest that this correction is useful when $N$ is low, but for sufficiently large training datasets with large enough batches, the correlation has little impact on the results.

For the Double Disco ABCD method, we use the loss function
\beq
\label{eq:DDloss}
\mathcal{L}[f,g] = \mathcal{L}_\text{classifier}[f(X),y] + \mathcal{L}_\text{classifier}[g(X),y] +\lambda \,\text{dCorr}_{y=0}^2[f(X),g(X)],
\eeq
where now $f$ and $g$ are two neural networks that are trained simultaneously. When $\lambda=0$, the loss will be minimized when $f=g$ is the optimal classifier (up to degeneracies).  When $\lambda\rightarrow \infty$, $f$ and $g$ will be forced to be independent even if one or both of them does not classify well at all. In practice, if $\lambda$ is taken too large, the DisCo term will tend to overwhelm the training and poor classification performance will result. Thus there should be an optimal $\lambda$ at some finite value which we can be determined by scanning over $\lambda$. 

\section{Applications}
\label{sec:results}

This section explores the efficacy of Single and Double DisCo in some applications of the ABCD method.

\subsection{Simple Example: Three-Dimensional Gaussian Random Variables}
\label{sec:toy}

We begin with a simple example to build some intuition and validate our methods. Consider a three-dimensional space $(X_0,X_1,X_2)$, where the signal and background are both multivariate Gaussian distributions. We choose
the means  $\vec\mu$ and a covariance matrix ${\bf \Sigma}$ for background and signal as
\beq
\vec\mu_b=(0,0,0),\qquad {\bf \Sigma}_b= \sigma_b^2
\begin{pmatrix}
1 & \rho_b & 0 \cr \rho_b & 1 & 0 \cr 0 & 0 & 1\end{pmatrix}\,,
\qquad \sigma_b=1.5, \quad \rho_b=-0.8\,,
\eeq
and
\beq
\vec\mu_s=(2.5,2.5,2),\qquad {\bf \Sigma}_s= \sigma_s^2\begin{pmatrix} 1 & 0 & 0 \cr 0 & 1 & 0 \cr 0 & 0 & 1\end{pmatrix}\,,\quad \sigma_s = 1.5
\,.\eeq 
So for the background, all three features are centered at the origin and features $X_0$ and $X_1$ are correlated with each other but independent of $X_2$. For the signal, all three features are independent but are centered away from the origin.  The first feature $X_0$ will play the role of the known feature for Single DisCo in Sec.~\ref{sec:automating}.  

All of the neural networks presented in this section use three hidden layers with 128 nodes per layer.  The rectified linear unit (ReLU) activation function is used for the intermediate layers and the output is a sigmoid function.  A hyperparameter of $\lambda=1000$ is used for both Single and Double DisCo to ensure total decorrelation.  The Single Disco training converged after 100 epochs while the Double DisCo training required 200 epochs.  Other networks only needed ten epochs.  The Double DisCo networks were trained using a single neural network with a two-dimensional output.  All models were trained using Tensorflow~\cite{tensorflow} through Keras~\cite{keras} with Adam~\cite{adam} for optimization.  Two million examples were generated with 15\% used for testing.  A batch size of 1\% of the total was used for all networks to ensure an accurate calculation of the DisCo term in the relevant loss functions.

\begin{figure}[t]
\centering
\includegraphics[height=0.37\textwidth]{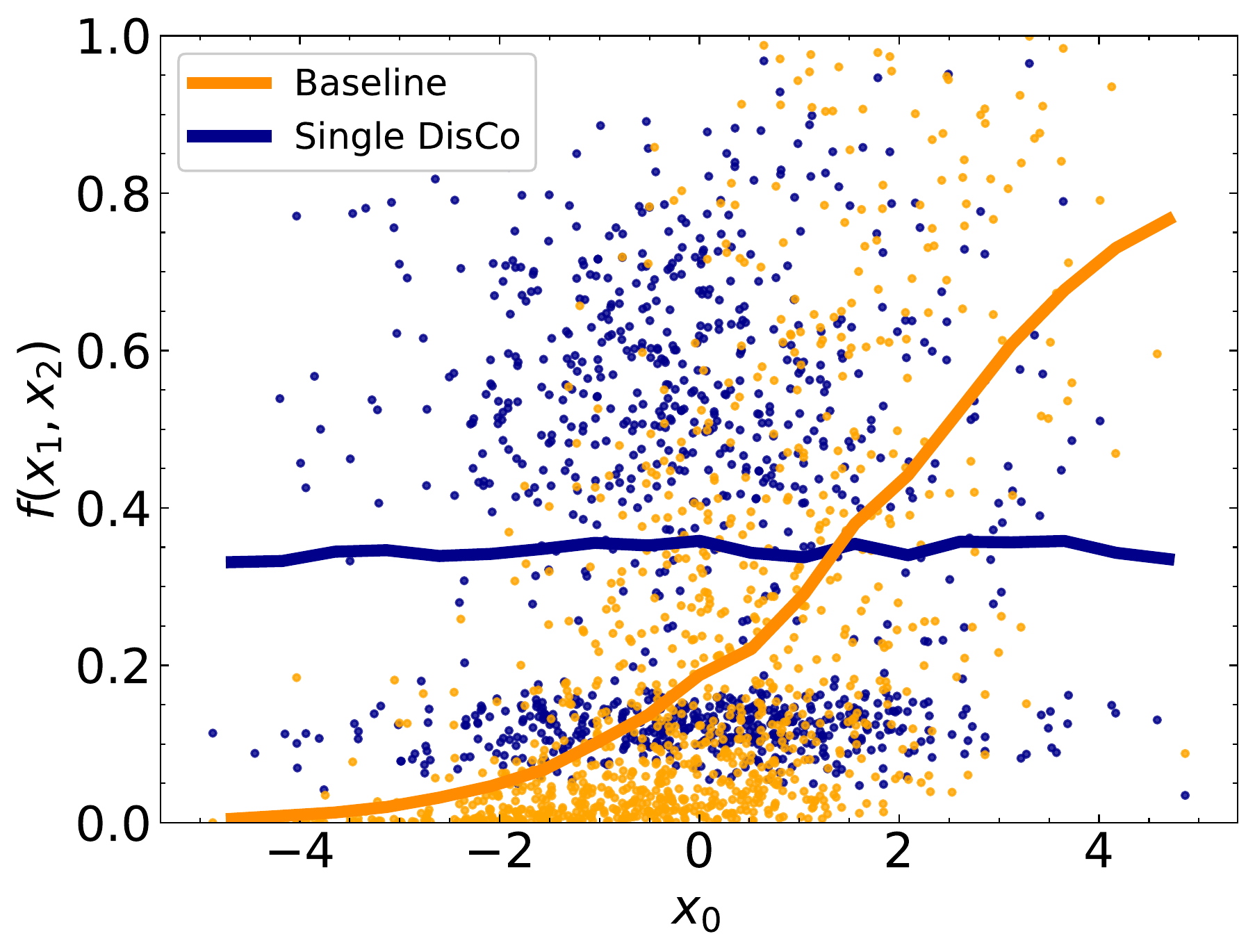}
\includegraphics[height=0.37\textwidth]{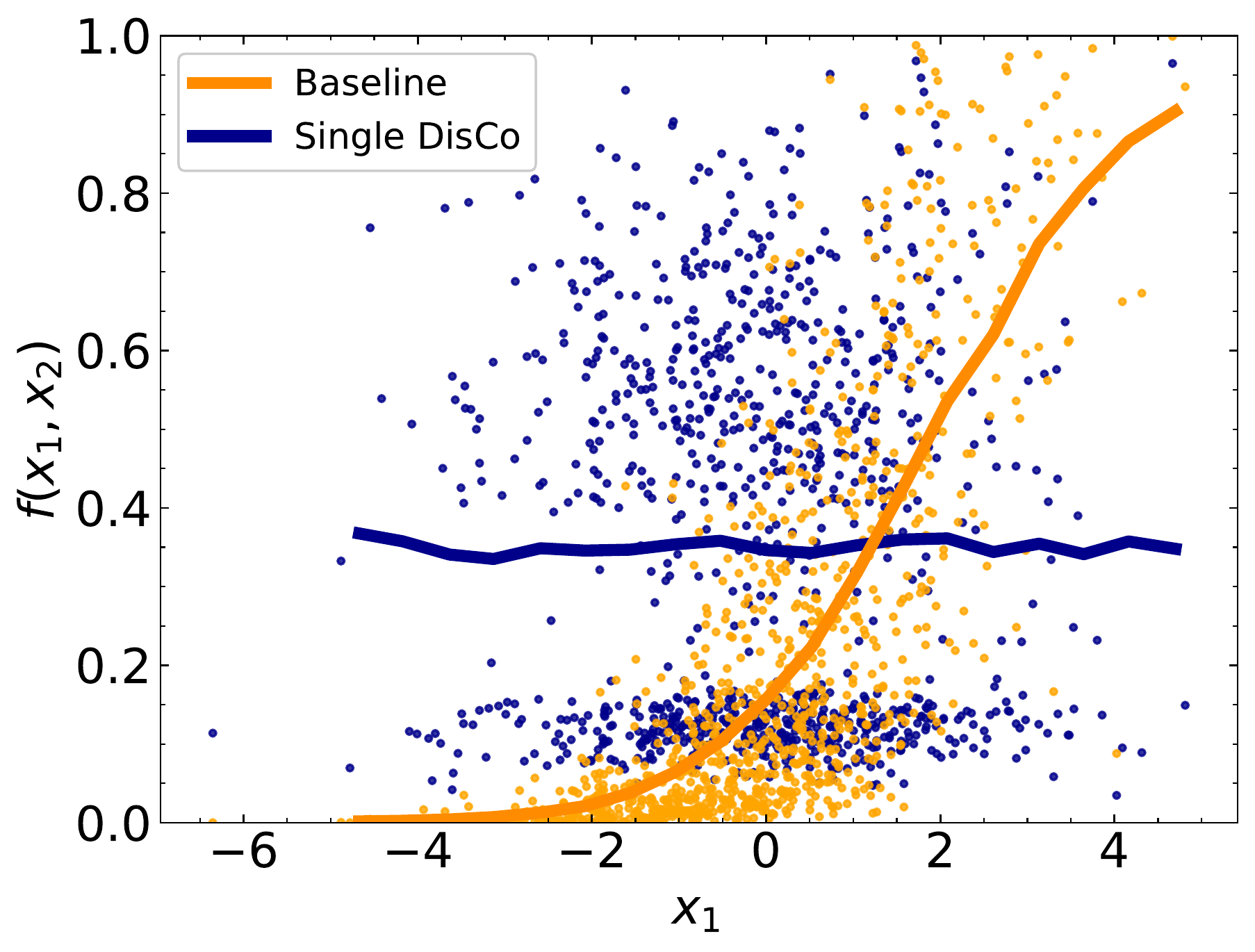} \\\includegraphics[height=0.37\textwidth]{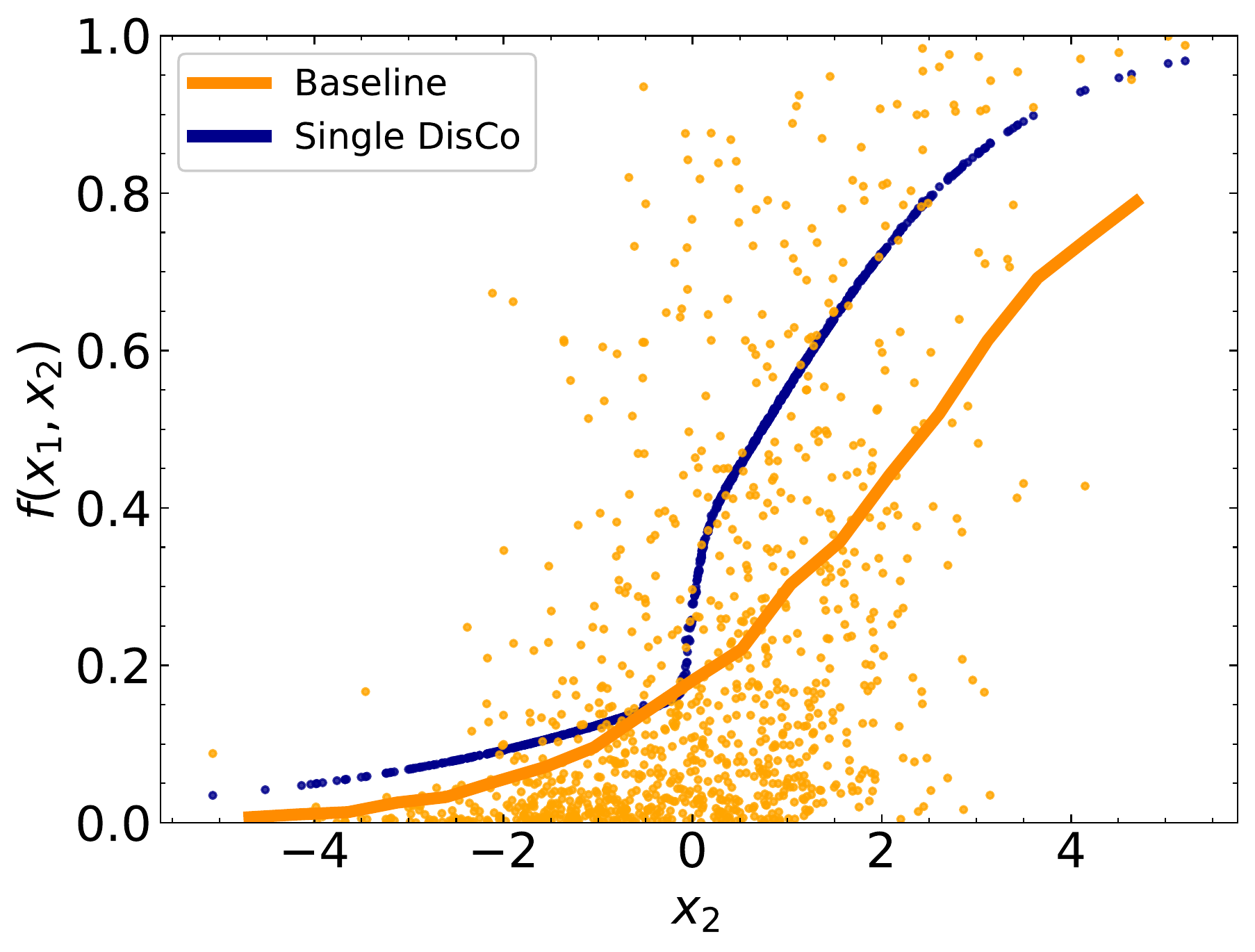}
\caption{Scatter plots showing the relationship (or lack thereof) between the three random variables $X_0$, $X_1$, and $X_2$ and 1) a baseline classifier $f_\text{BL}(X_1,X_2)$ trained on $X_1$ and $X_2$ with no regularization and 2) a classifier $f_\text{SD}(X_1,X_2)$ trained with the Single DisCo loss function that penalizes correlations with $X_0$. Only the background events are shown in these plots. 
The solid lines are the averages of the classifiers over events with the same value of $X_0$, $X_1$ or $X_2$. In the third panel, the scatter of the Single Disco classifier is already a line, so no average is needed.}
\label{results:toysingleDisCo}
\end{figure}

We first consider two classifiers: a baseline classifier $f_\text{BL}(X_1,X_2)$ trained only on $X_1$ and $X_2$ and a Single DisCo classifier $f_\text{SD}(X_1,X_2)$ which includes a penalty for correlations between $f_\text{SD}$ and $X_0$.  The values of these classifiers for events drawn from the distributions are plotted in
Fig.~\ref{results:toysingleDisCo} against the $X_0$, $X_1$, or $X_2$ values of these events. We see that even though $X_0$ was not used in the training of the baseline, the classifier output is still correlated with $X_0$ because of the correlations between $X_0$ and $X_1$.  In contrast to the baseline classifier, the Single DisCo classifier is independent of both $X_0$ and $X_1$ and is simply a function of $X_2$.  Intuitively, it makes sense that a classifier that must be independent of $X_0$ must also be independent of $X_1$.  This is justified rigorously in Appendix~\ref{sec:singleDisCoproof}.

\begin{figure}[h!]
\centering
\includegraphics[width=0.95\textwidth]{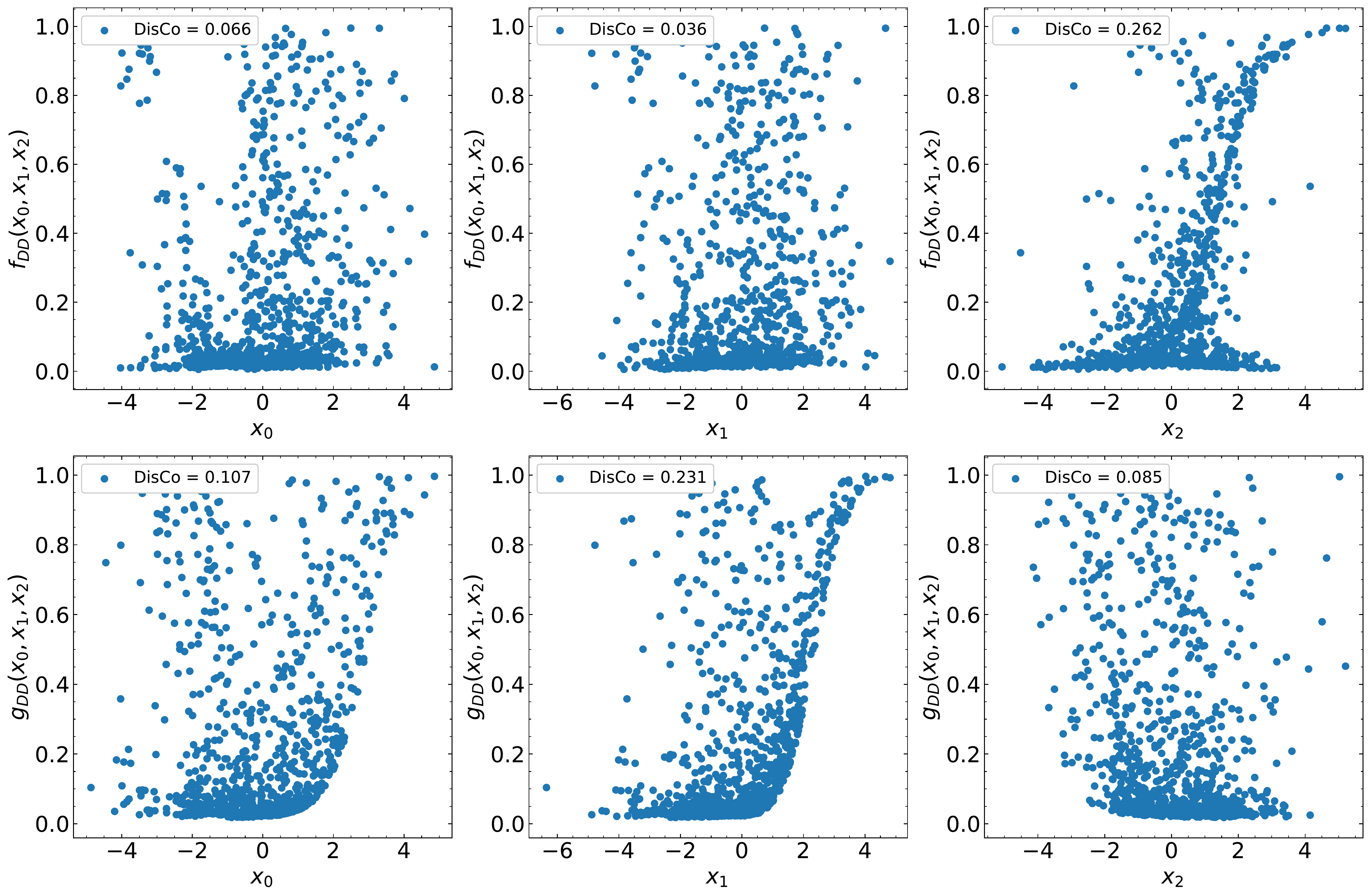}
\includegraphics[width=0.35\textwidth]{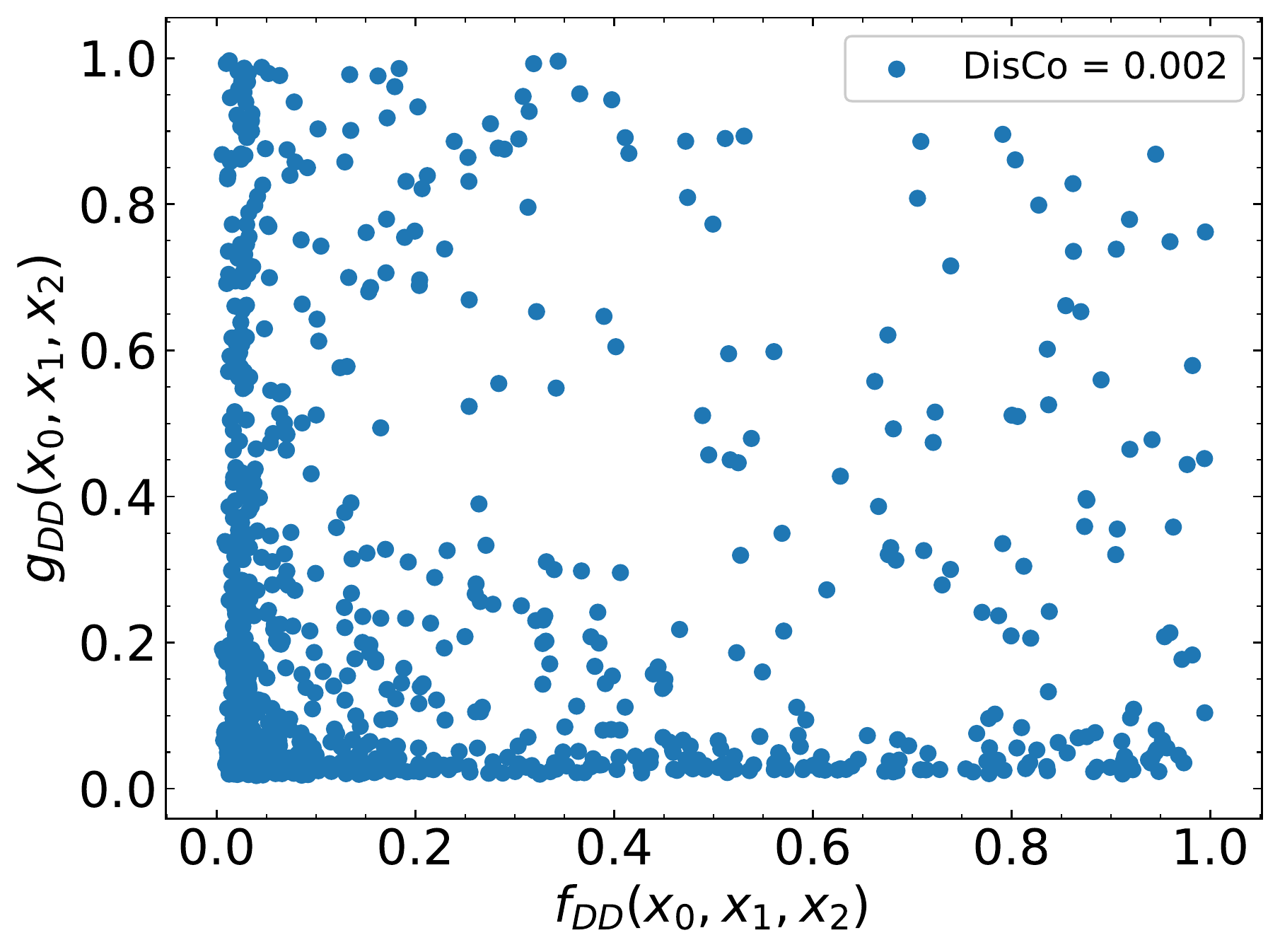}
\caption{Scatter plots showing the relationship between the three random variables $X_0$, $X_1$, $X_2$ and the two Double DisCo neural networks $f_{DD}$ and $g_{DD}$ using only the background.  The distance correlation between the two plotted observables is indicated in the legend.}
\label{results:toyDoubleDisCo}
\end{figure}

For Double DisCo, we train two classifiers $f_\text{DD}(X,Y,Z)$ and $g_\text{DD}(X,Y,Z)$ according to the Double DisCo loss function. The results are illustrated in Fig.~\ref{results:toyDoubleDisCo}. The first classifier depends mostly on $Z$ and the second classifier depends mostly on $X$ and $Y$. However,  the residual dependence on all three observables is not a deficit of the training procedure: even though the three random variables are separable into two independent subsets $(X,Y)$ and $Z$, the two classifiers learned by Double DisCo are non-trivial functions of all three variables. There is a large freedom in choosing the two functions $f_\text{DD}$ and $g_\text{DD}$ with a very small distance correlation and also excellent classification performance. Evidently, Double DisCo prefers to partition the information differently than the naive partitioning in order to achieve better classification performance. 

\begin{figure}[h!]
\centering
\includegraphics[height=0.37\textwidth]{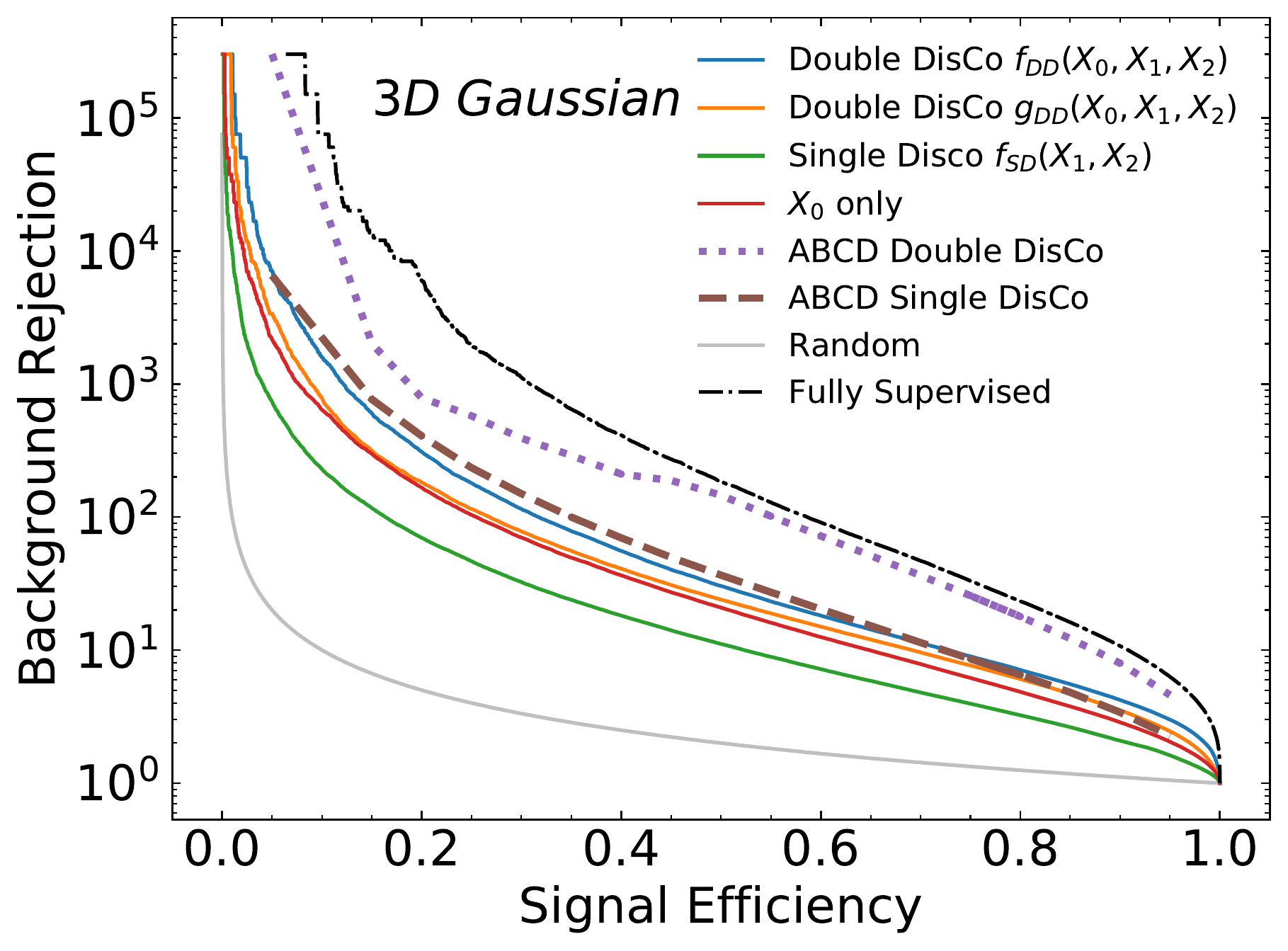}\includegraphics[height=0.37\textwidth]{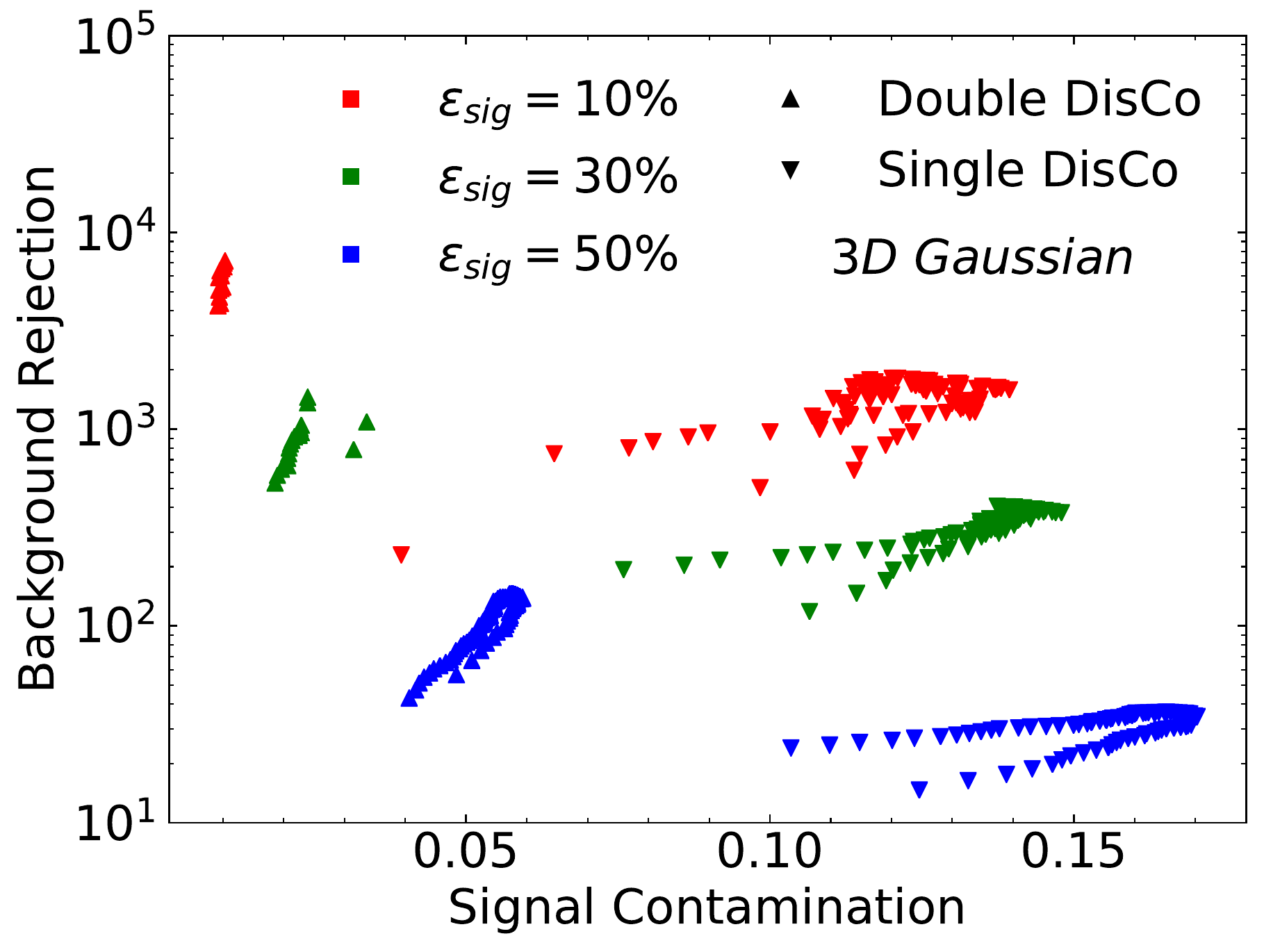} 
\caption{Performance metrics for the Gaussian random variable model. Left: A receiver operating characteristic (ROC) curve.  The lines marked ABCD DisCo are derived by scanning over rectangular thresholds on the two classifiers for points with ABCD closure within 10\%.  In the Single DisCo case, one of the two classifiers is simply a NN trained with only $X_0$ (marked `$X_0$' only in the legend).  Right: a scatter plot between background rejection and the normalized signal contamination for ABCD closure within 20\%. For comparison, the left plot also shows the performance of the two Double DisCo functions separately, the Single DisCo function on its own, as well as a fully supervised classifier using all the available information all at once.}
\label{results:toysingleDisCo_performance}
\end{figure}

Fig.~\ref{results:toysingleDisCo_performance} shows the performance of the Single and Double DisCo classifiers.  The curve for the ABCD method is constructed by scanning 100 values of independent thresholds on the two features, evenly spaced in percentile of one classifier or the other to ensure a fixed signal efficiency. Above 50\% signal efficiency, the ABCD Double DisCo has nearly the same performance as the fully supervised classifier using all of the available information.  The Single DisCo performance is much lower than the Double DisCo performance and is comparable to the best of the two Double DisCo classifiers.  The right plot of Fig.~\ref{results:toysingleDisCo_performance} demonstrates that Double DisCo is not only more effective at rejection background, but it also has a lower signal contamination.

\clearpage

\subsection{Boosted Tops}

Next we turn to a physical example: boosted, hadronically decaying, tops. When top quarks are highly boosted, their hadronic decay products can be collimated into a single large jet and jet substructure methods are often necessarily to distinguish them from QCD jet backgrounds~\cite{Kaplan:2008ie}.
One can estimate these backgrounds using sidebands in the jet mass around %
the top mass $m_t$. 
For the application of Single and Double DisCo, we will first reframe this estimation as an ABCD method and  map mass to a variable where the signal peaks at 1 and the background peaks at a lower value:
\beq
\ym \equiv 1-{|m_{\text{jet}}-m_{t}|\over m_t}\,.
\eeq
For our studies we will use the community top tagging comparison sample~\cite{Kasieczka:2019dbj,Butter:2017cot}.  %
There are 2 million jets total, 1 million each of signal (top jets) and background 
(light quark and gluon QCD jets).  Of these, half are used for training and the other half for validation.

\begin{figure}[t!]
\centering
\includegraphics[height=0.2\textwidth]{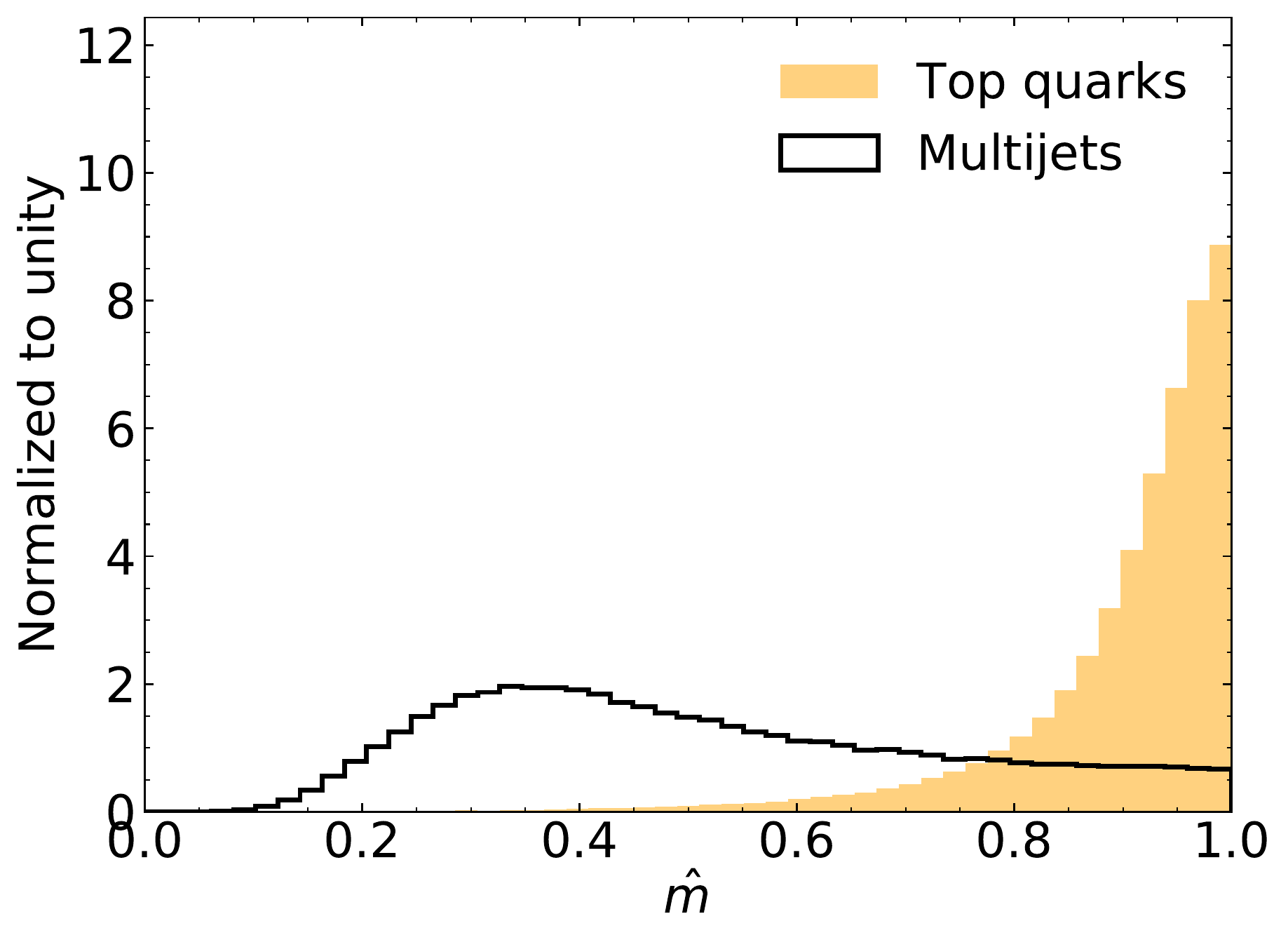}
\includegraphics[height=0.2\textwidth]{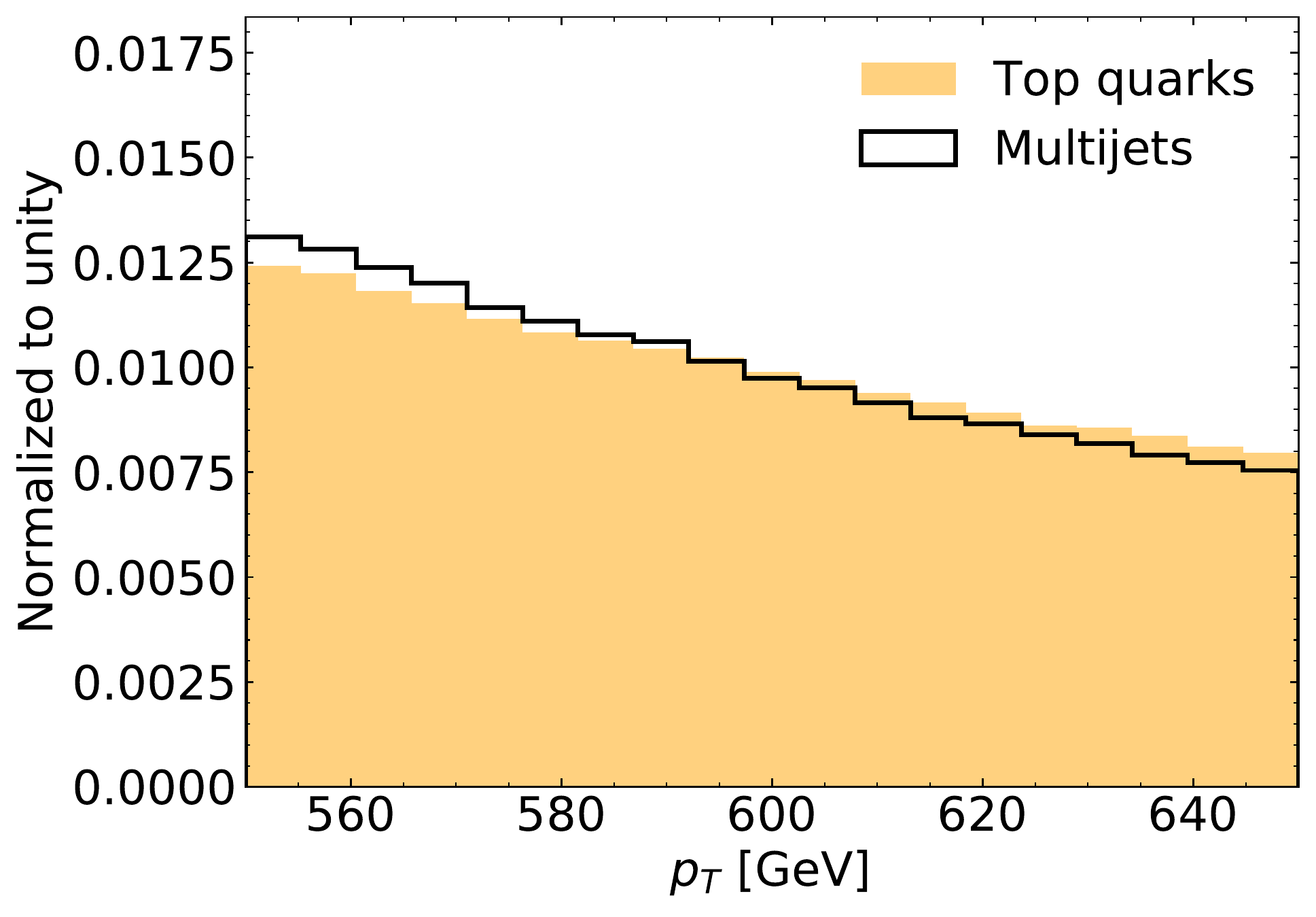}
\includegraphics[height=0.2\textwidth]{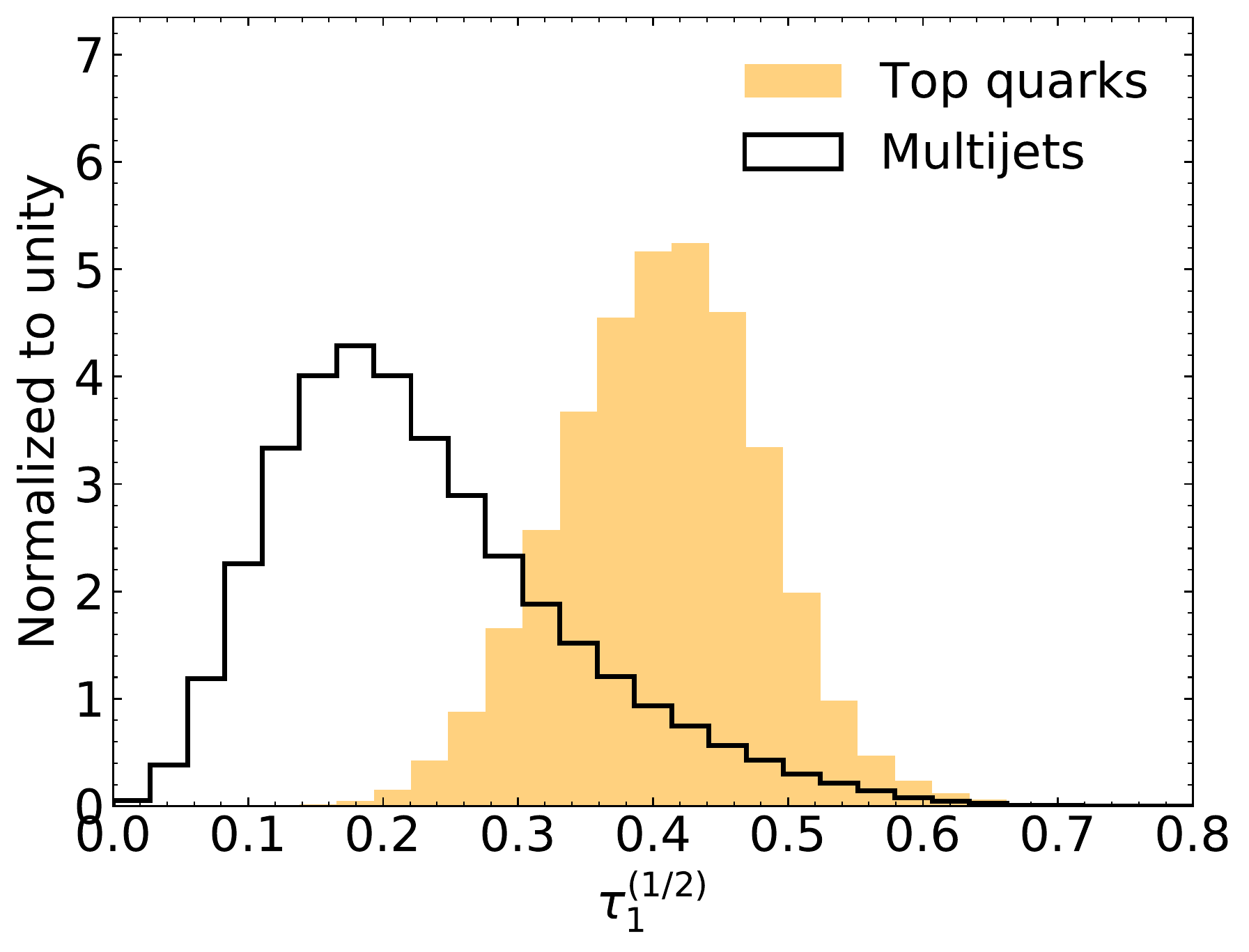}\\
\includegraphics[height=0.2\textwidth]{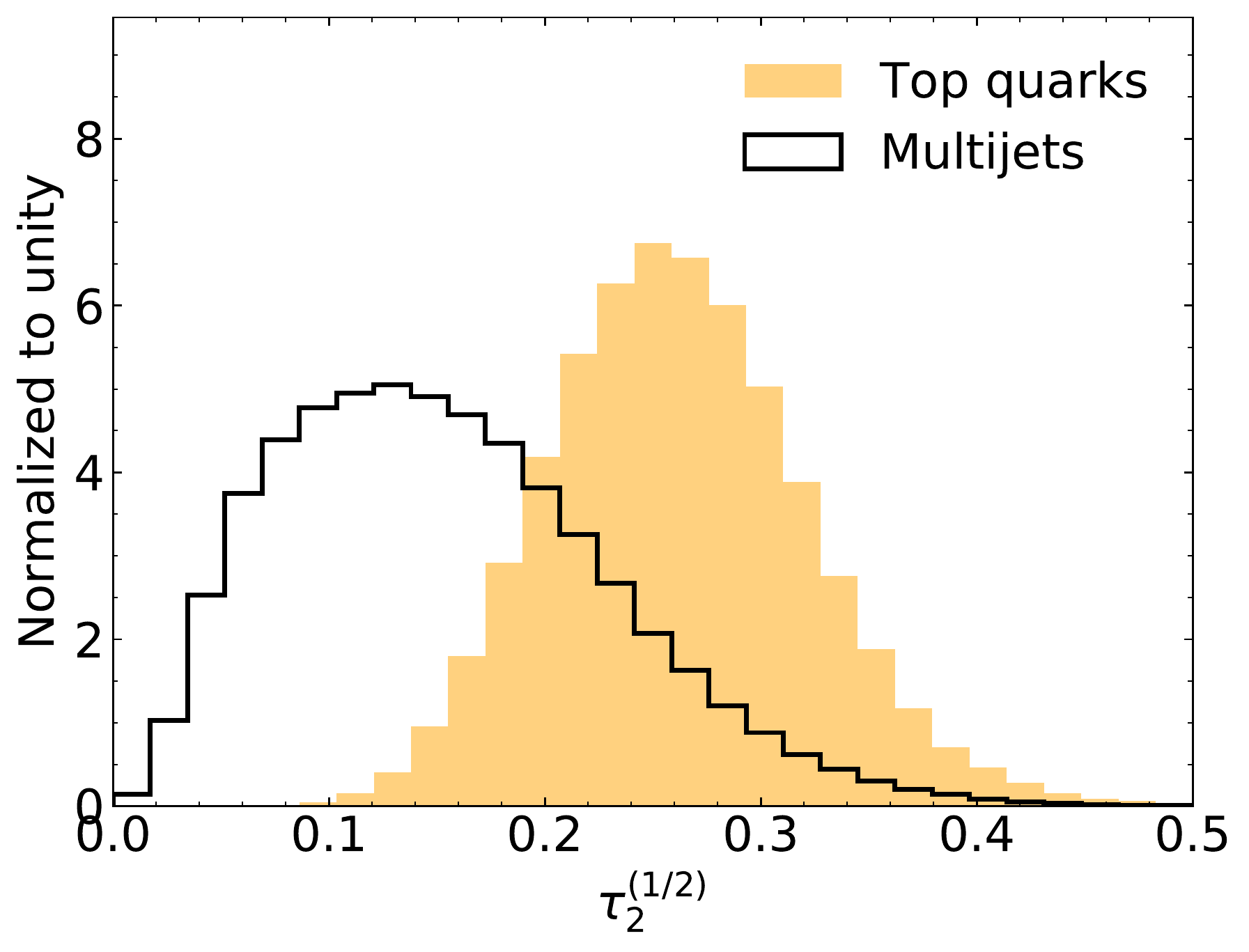}
\includegraphics[height=0.2\textwidth]{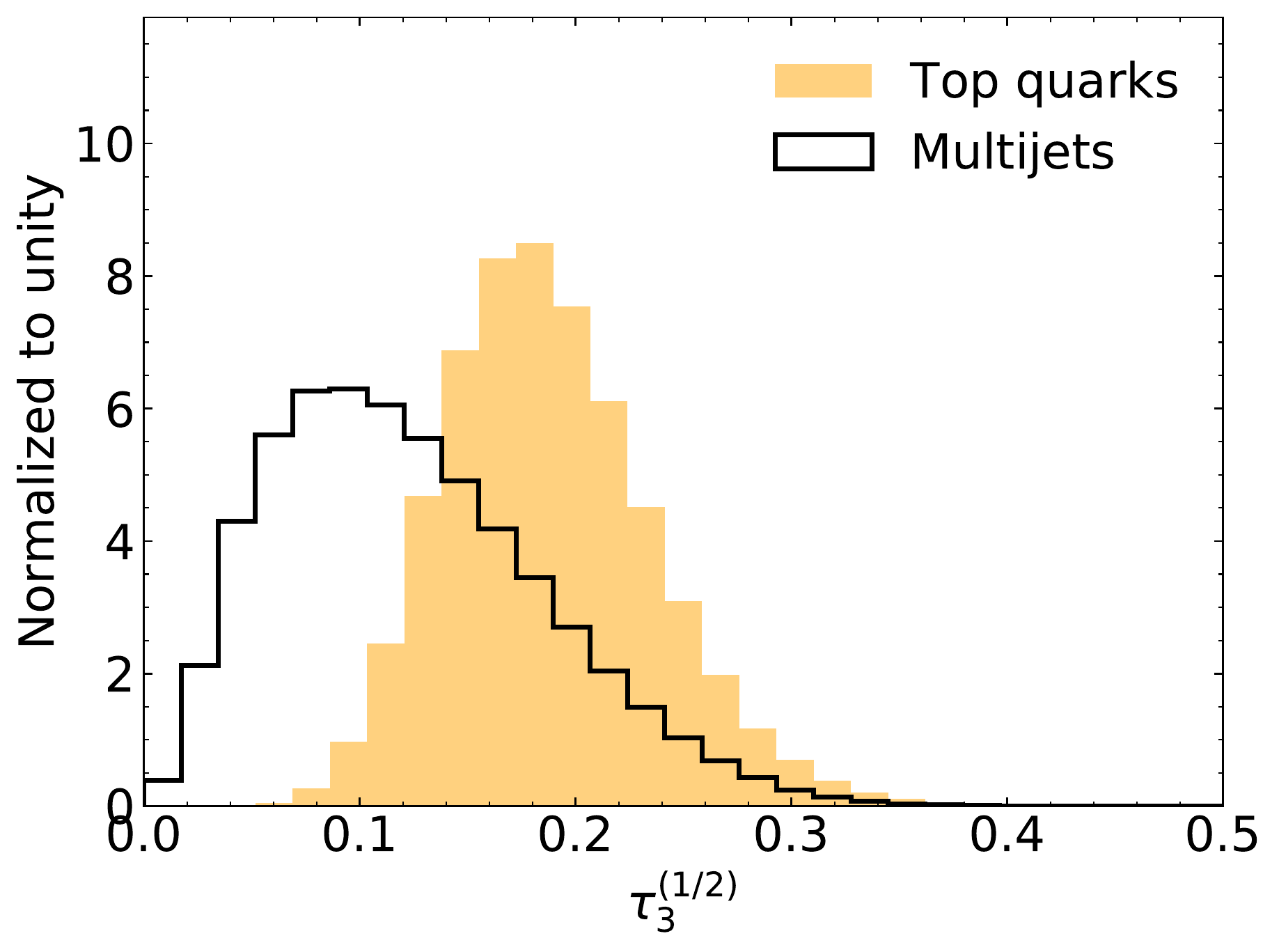}
\includegraphics[height=0.2\textwidth]{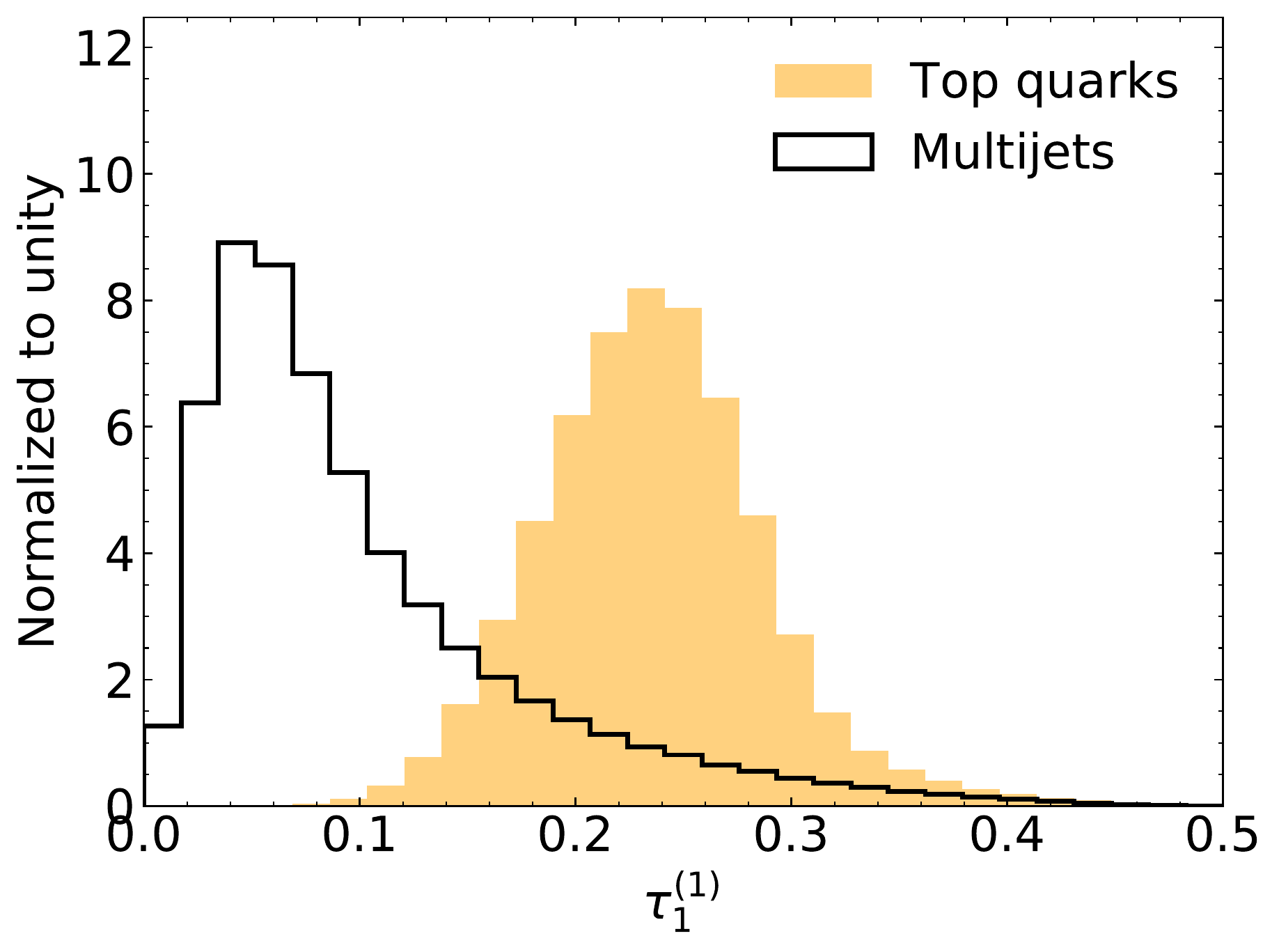}\\
\includegraphics[height=0.2\textwidth]{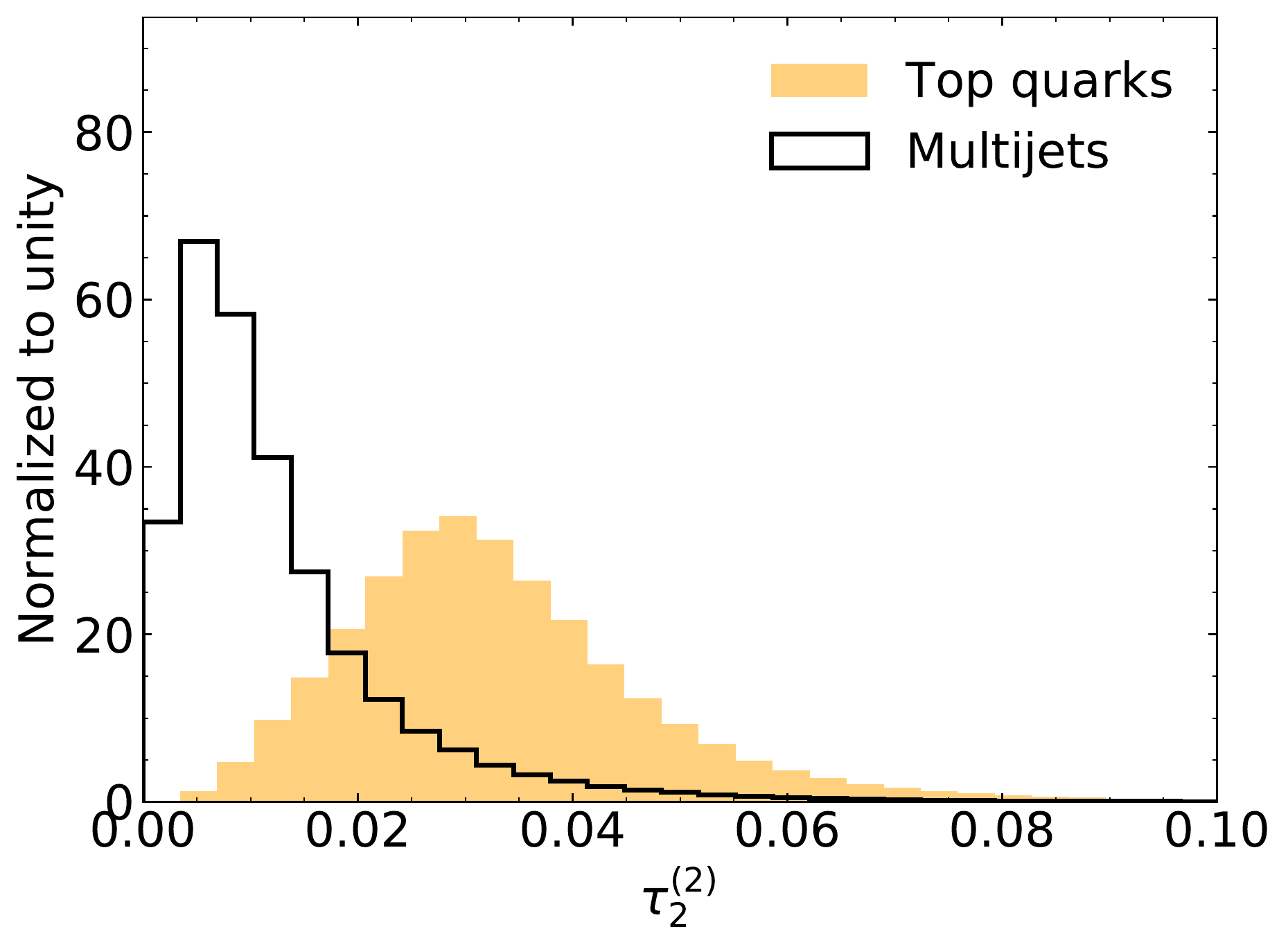}
\includegraphics[height=0.2\textwidth]{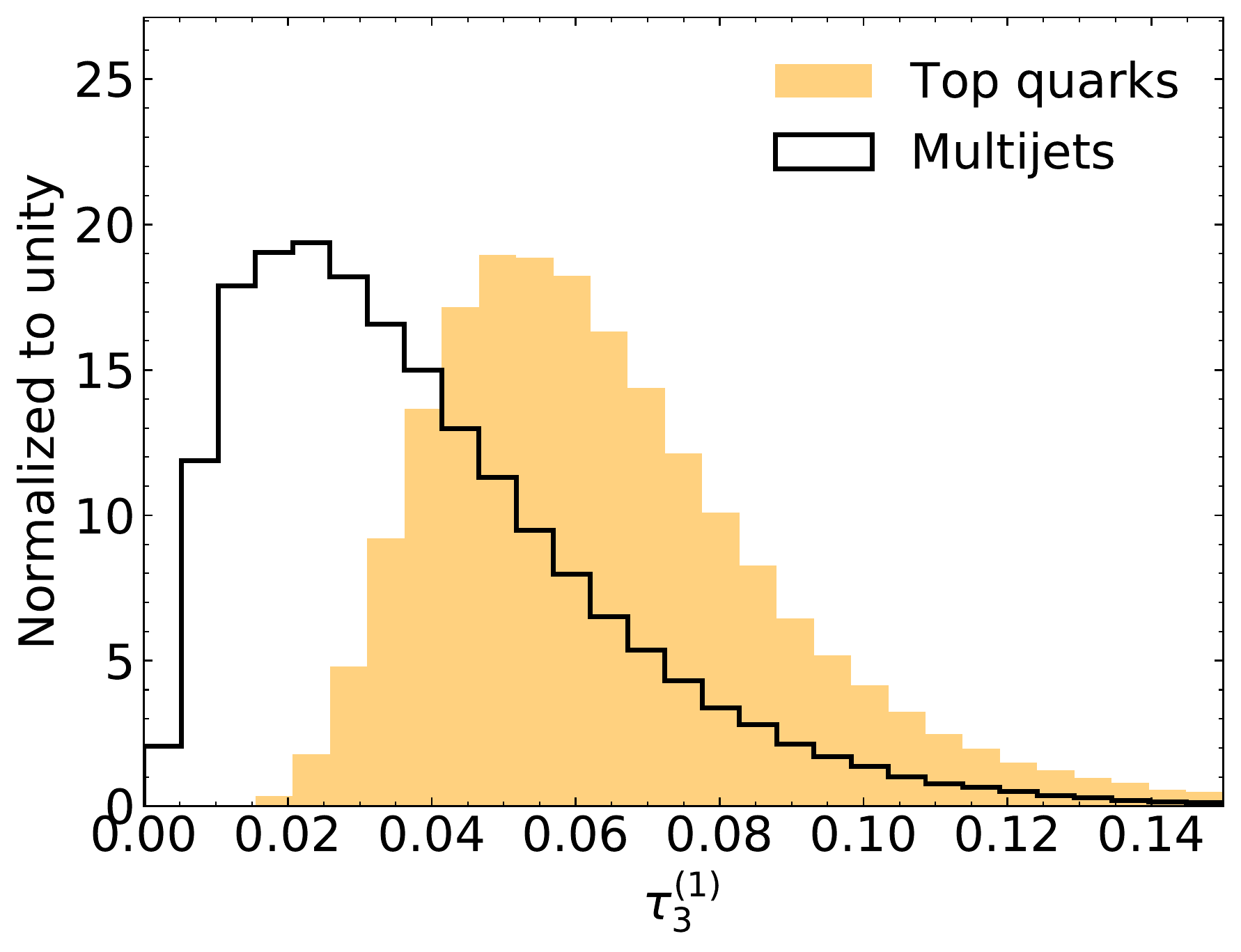}
\includegraphics[height=0.2\textwidth]{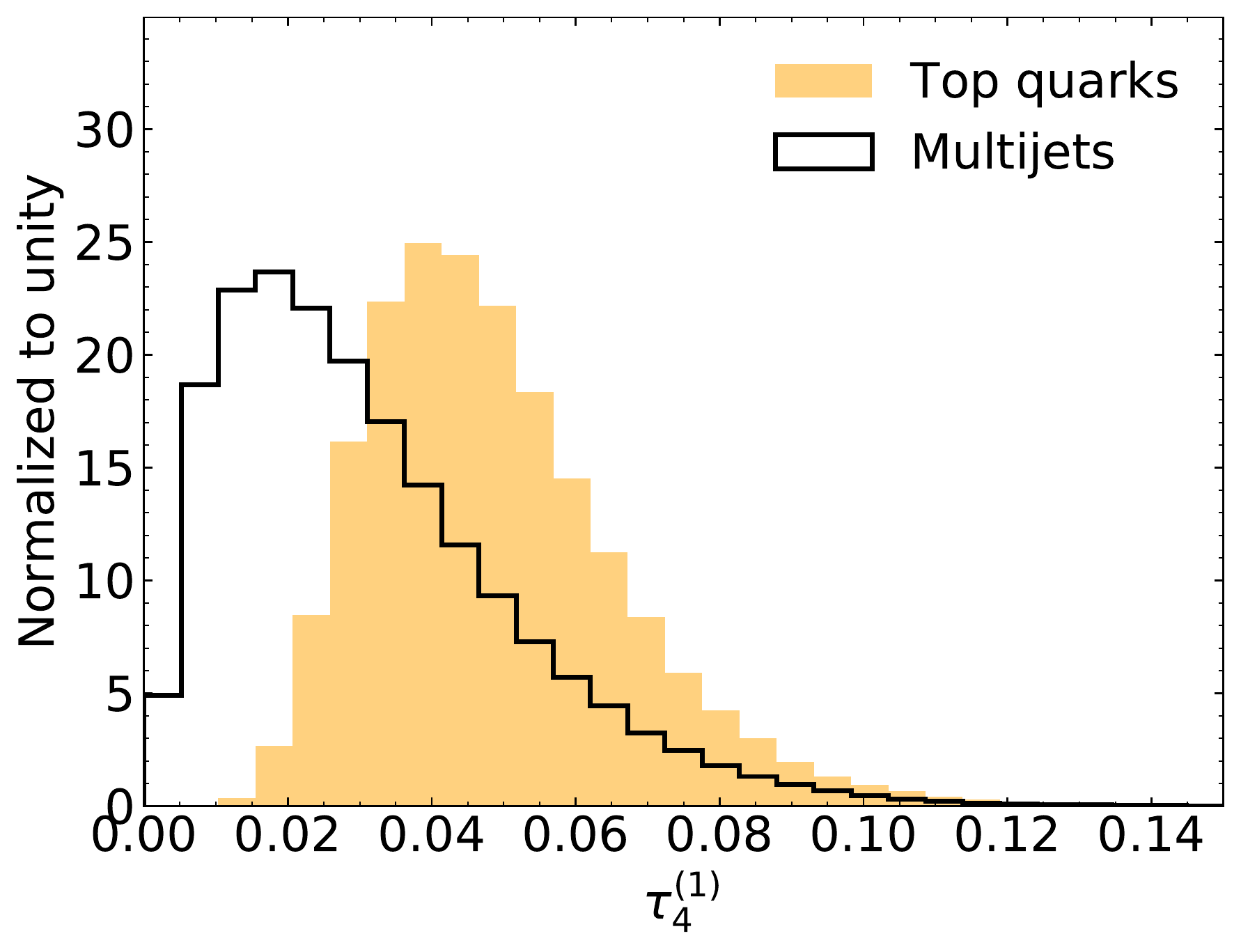}\\
\includegraphics[height=0.2\textwidth]{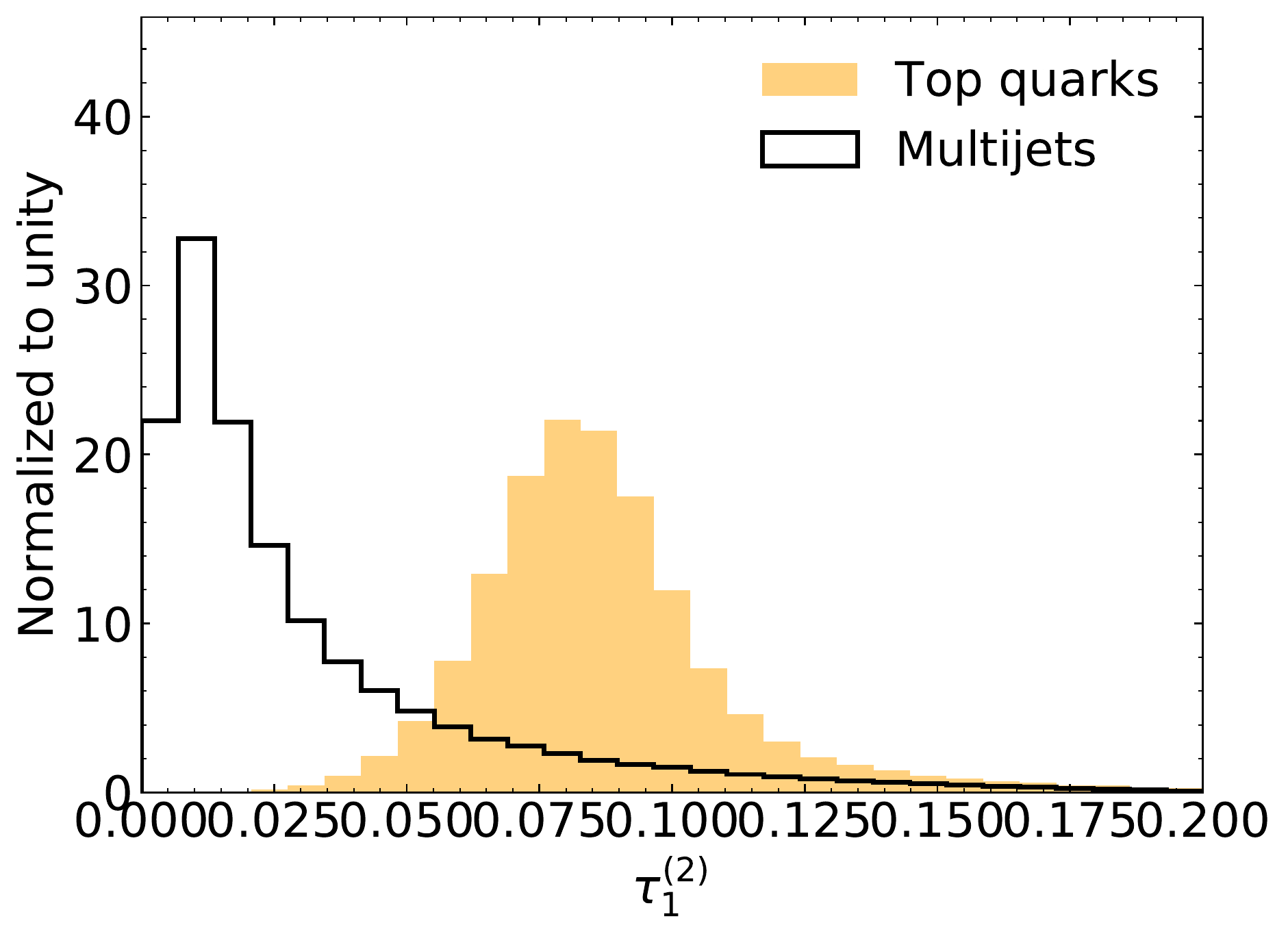}
\includegraphics[height=0.2\textwidth]{tau2_2}
\includegraphics[height=0.2\textwidth]{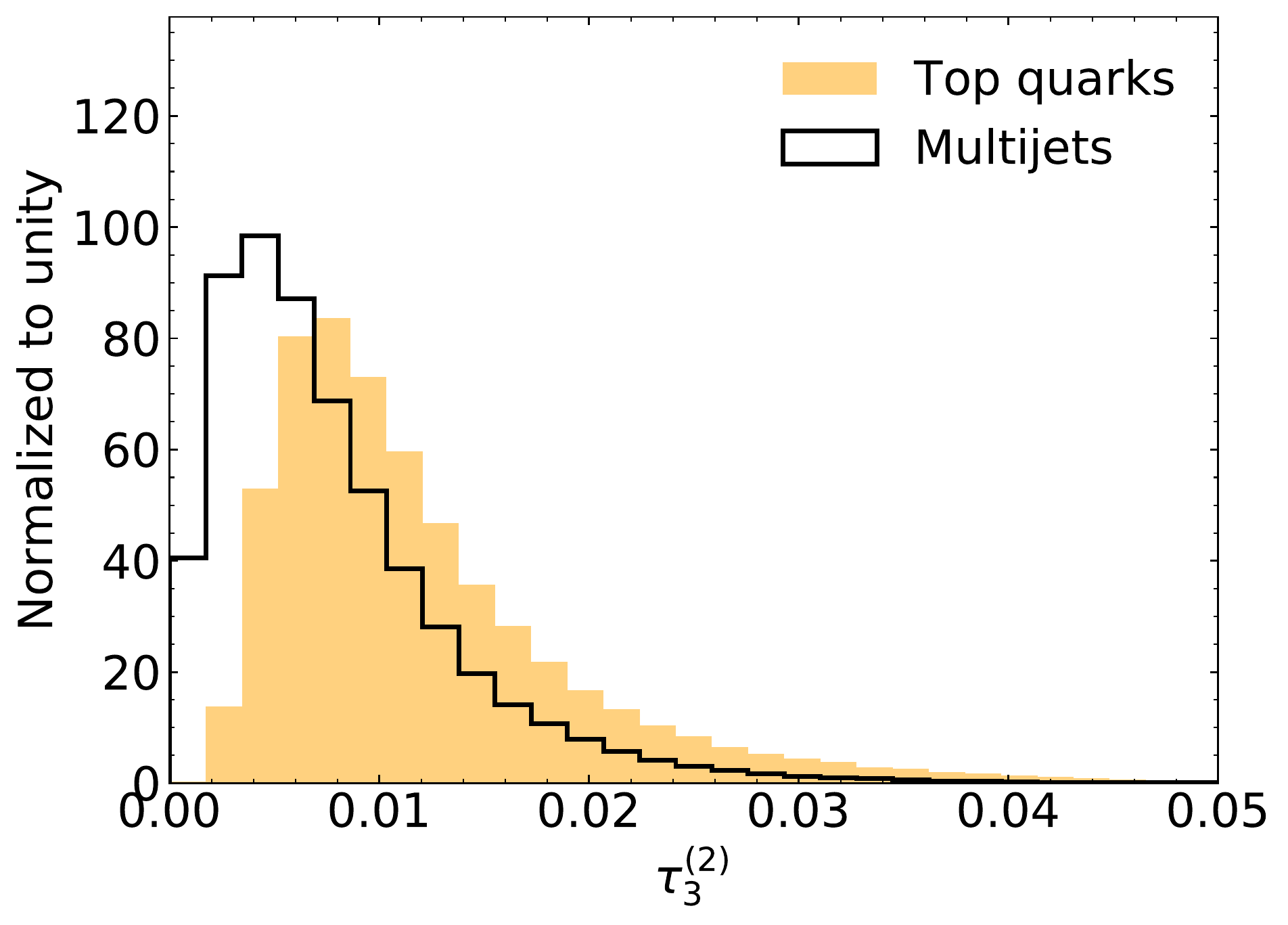}\\
\includegraphics[height=0.2\textwidth]{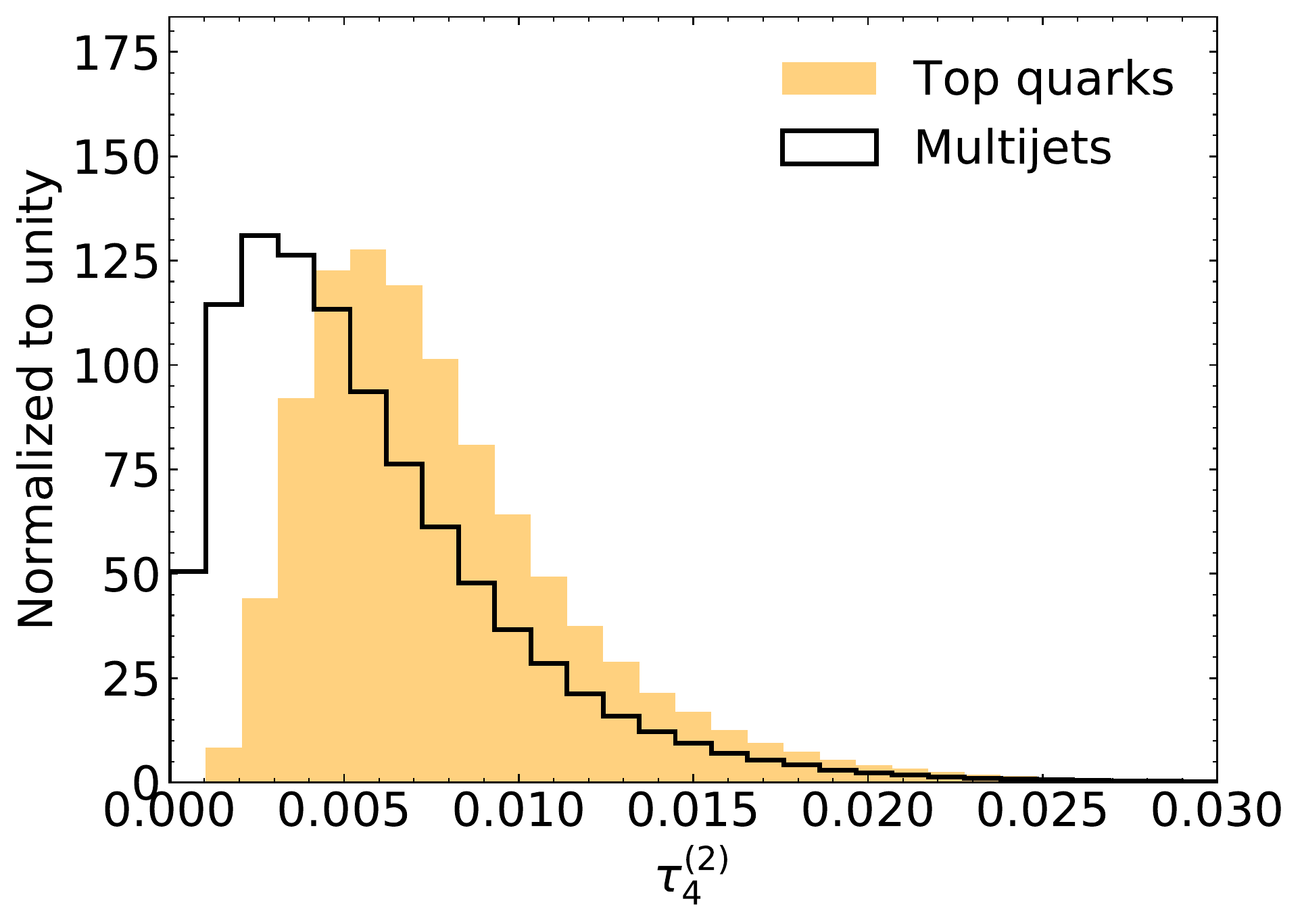}
\caption{The 13 features used for the boosted top analysis.
}
\label{fig:topplos}
\end{figure}

We compute the following set of high level features suggested by~\cite{Datta:2017rhs}
\beq
 \ym,\,\,\, p_T,\,\,\, \tau_1^{1/2},\,\,\, \tau_2^{1/2},\,\,\, \tau_3^{1/2},\,\,\, \tau_1^{1},\,\,\, \tau_2^{1},\,\,\, \tau_3^{1},\,\,\, \tau_4^{1},\,\,\, \tau_1^{2},\,\,\, \tau_2^{2},\,\,\, \tau_3^{2},\,\,\, \tau_4^{2}\,.
\eeq
Here, $\tau_N^a$ are the subjettiness variables introduced in~\cite{Thaler:2010tr,Thaler:2011gf} and are computed using {\sc fastjet}~\cite{Cacciari:2011ma}.  
This set of 13 variables is a complete basis for 5-body phase space and therefore it provides a complete description of the physics at the parton level~\cite{Datta:2017rhs,Datta:2017lxt,Moore:2018lsr,Datta:2019ndh}. It also offers a useful~\cite{Moore:2018lsr}
feature space for modelling the top quark jets and inclusive jets after hadronization.  Histograms of these features for signal and background are presented in Fig.~\ref{fig:topplos}. 

All the features are rescaled to be between 0 and 1. The neural network specification is 3 hidden layers of 64 nodes each, ReLU activations, and batch normalization after the first hidden layer. We train for 200 epochs with fixed learning rate of $10^{-3}$ and the default Adam optimizer. We use a large batch size of 10k to ensure an accurate DisCo sampling estimate.

\begin{figure}[t!]
\centering
\includegraphics[width=0.65\textwidth]{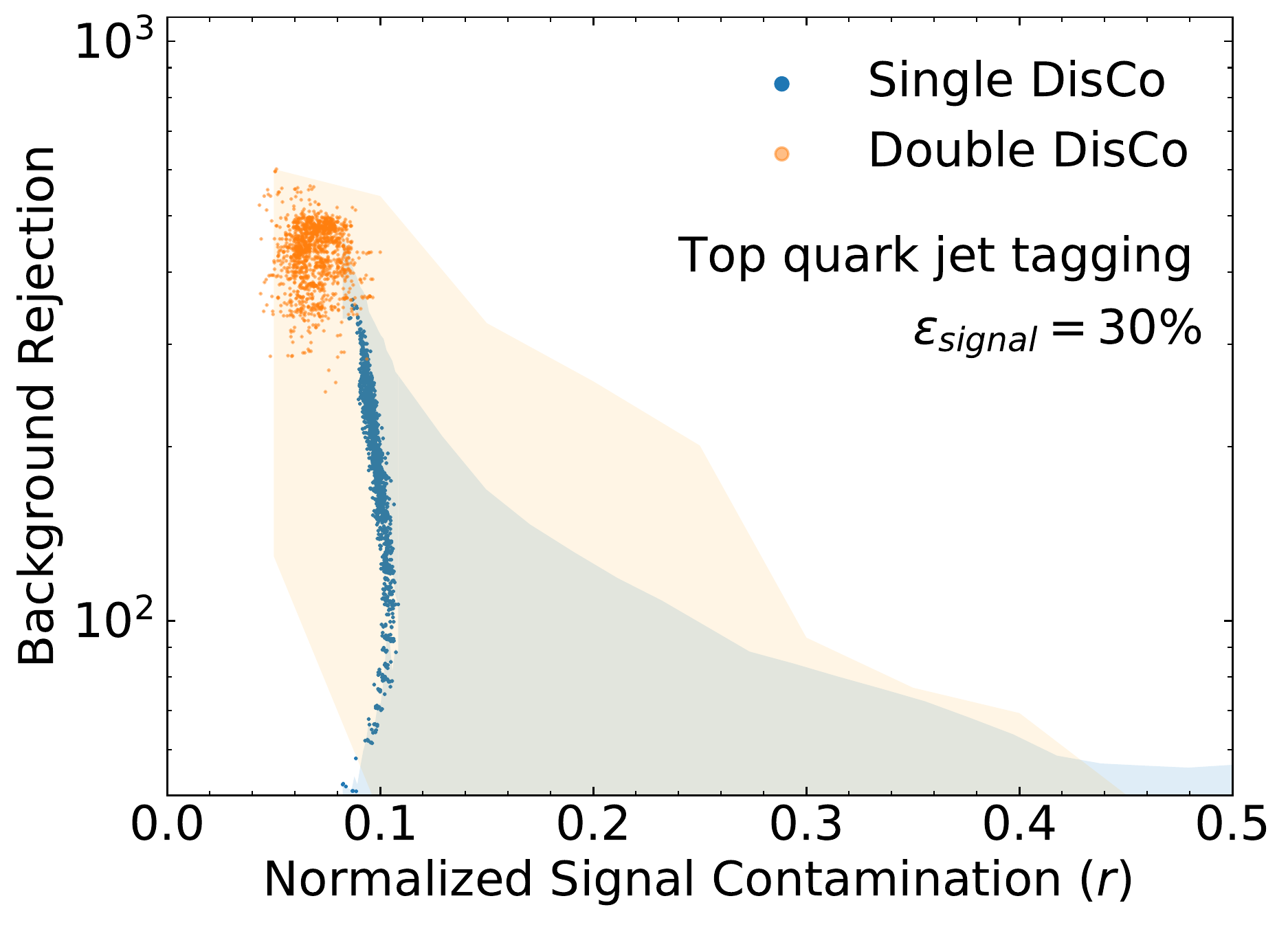}
\caption{A scatter plot of background rejection and normalized signal contamination ($r$) 
across DisCo parameters, epochs and thresholds on the two features, for $\epsilon_\text{signal}=30\%$ and background ABCD closure better than 10\%. 
High density regions are depicted with individual data points while low density regions are drawn as shaded regions.
\vspace{10mm}}
\label{fig:topR30vssc}
\end{figure}

\begin{figure}[t!]
\centering
\hspace{-5mm}
\includegraphics[width=0.49\textwidth]{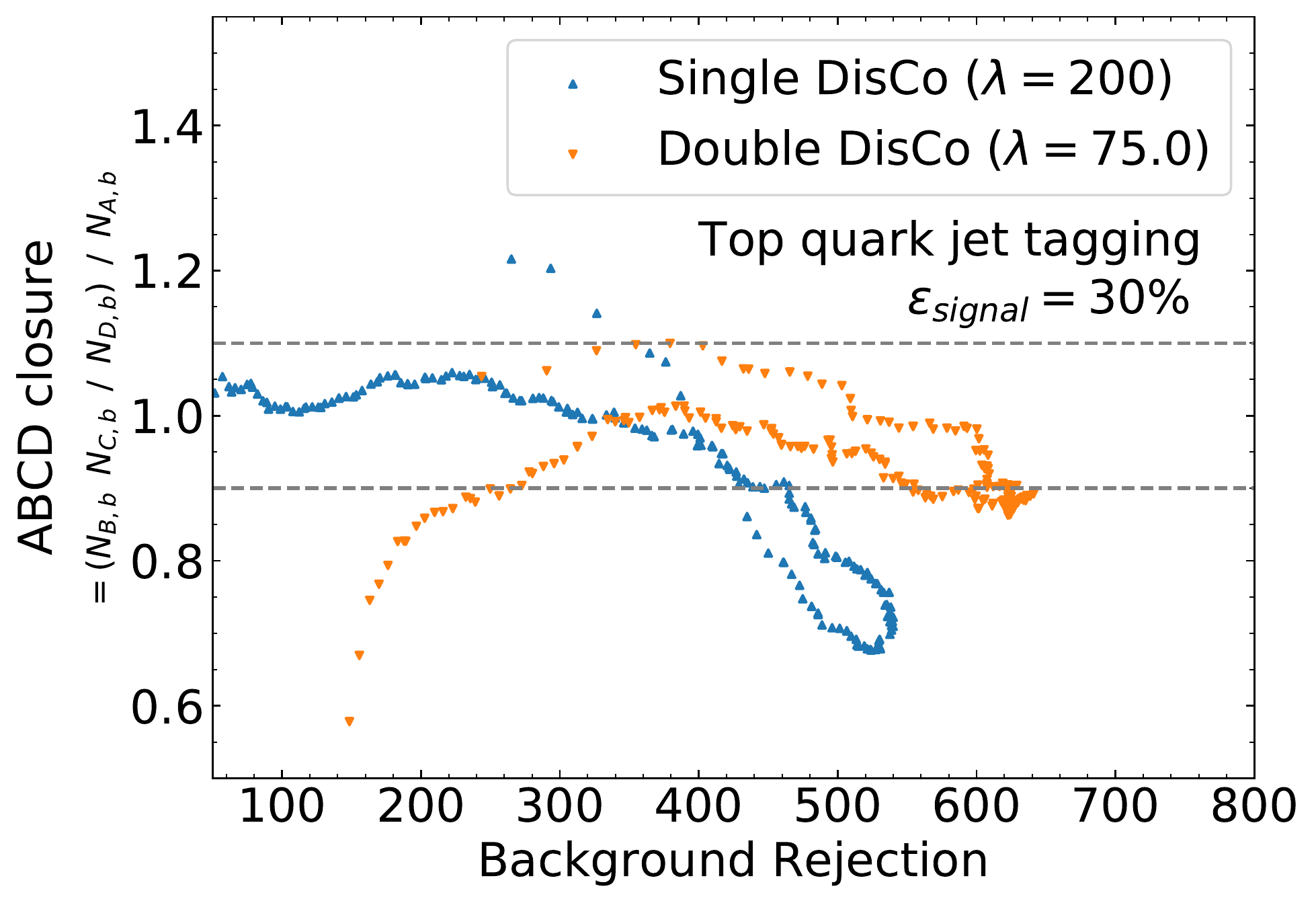}
\includegraphics[width=0.51\textwidth]{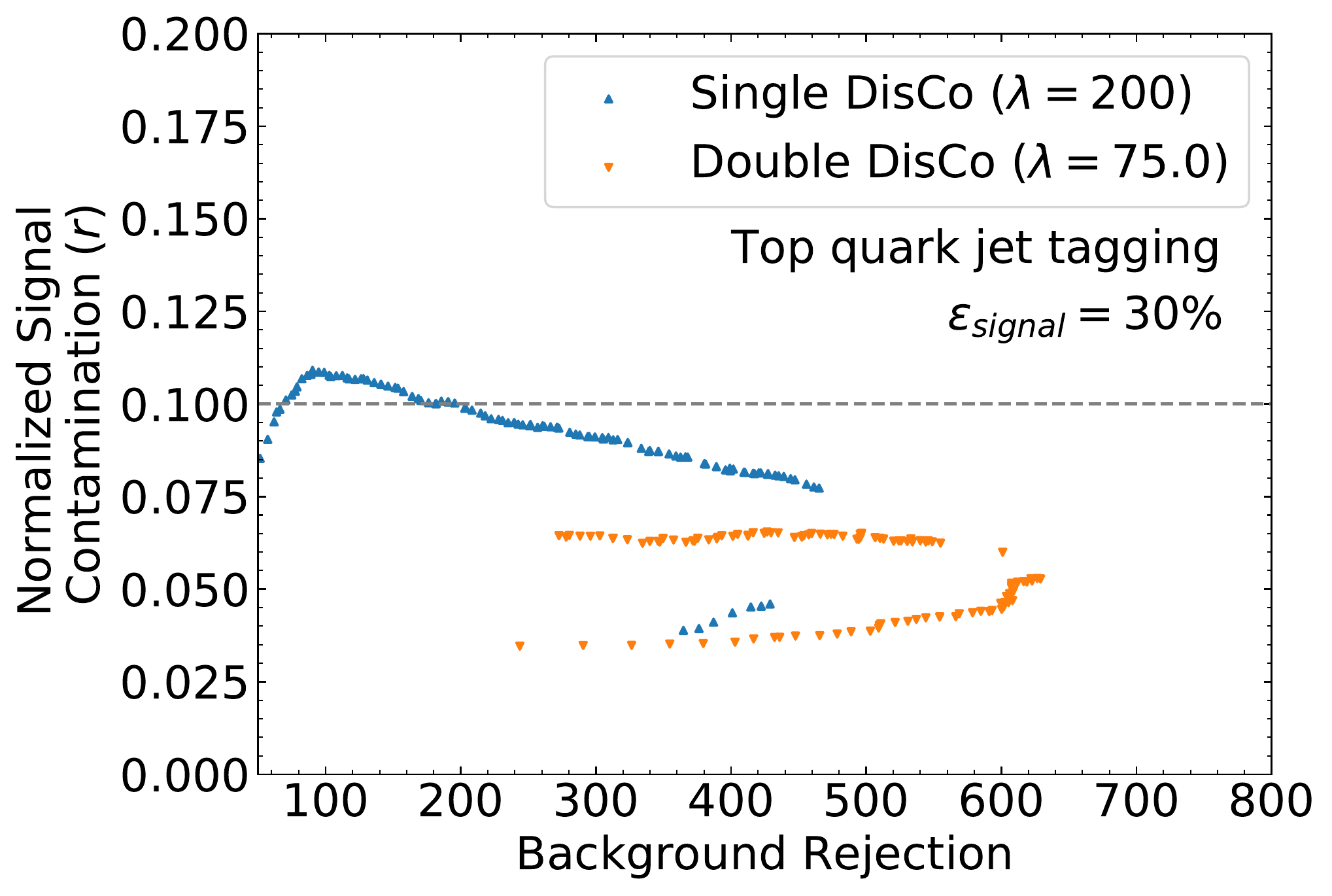}
\caption{Performance metrics for the boosted top analysis. Left: A scatter plot of the ABCD closure for the background versus the background rejection for $\epsilon_\text{signal}=30\%$ in the boosted top analysis.  Right: For the points in the top plot with ABCD closure within 10\% of unity, this is a scatter plot of the normalized signal contamination ($r$) versus the background rejection. 
\vspace{5mm}}
\label{fig:topbestepochs}
\end{figure}

\begin{figure}[t]
\centering
\includegraphics[height=0.5\textwidth]{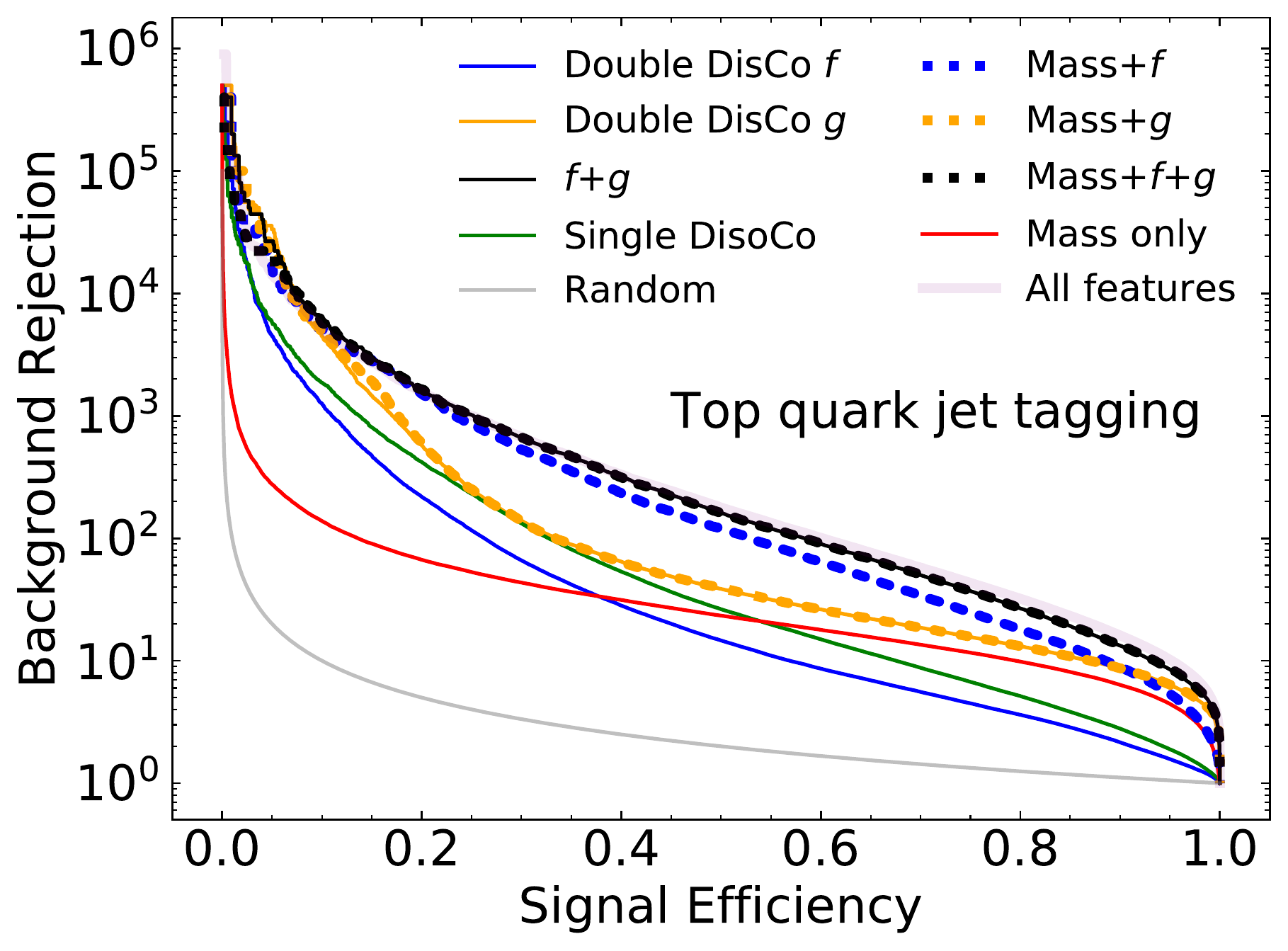}
\caption{ROC curve for the boosted top analysis. The background rejection is shown as a function of the signal efficiency for various combinations of Single and Double DisCo classifiers with or without mass. }
\label{fig:toprocmass}
\vspace{1cm}
\end{figure}

For Single DisCo, we train a single neural network on just the subjettiness variables (we could have included $\ym$ and $p_T$ too with little change).
For Double DisCo, we train two neural networks on all the features ($\ym$, $p_T$, and the subjettiness variables). The neural networks specifications, feature preprocessing, and training details are all the same for Single and Double DisCo. However, for Double DisCo, in addition to the usual DisCo loss term described in Eq. \eqref{eq:DDloss}, we include a {\it second} DisCo term which only takes the tail of the neural network outputs (again for background only) as inputs. This was found to help with the stability of the ABCD prediction for lower signal efficiencies, which can be sensitive to the extreme tails of the background. For the tail we required the simultaneous cuts of $y_1>(y_1)_{\text{bg},50}$ and $y_2>(y_2)_{\text{bg},50}$, where $y_{1,2}$ are the outputs of the two neural networks and ``bg,50" refers to the $50^\text{th}$ percentile cut on the background distributions.

\begin{figure}[t!]
\centering
\includegraphics[width=0.45\textwidth]{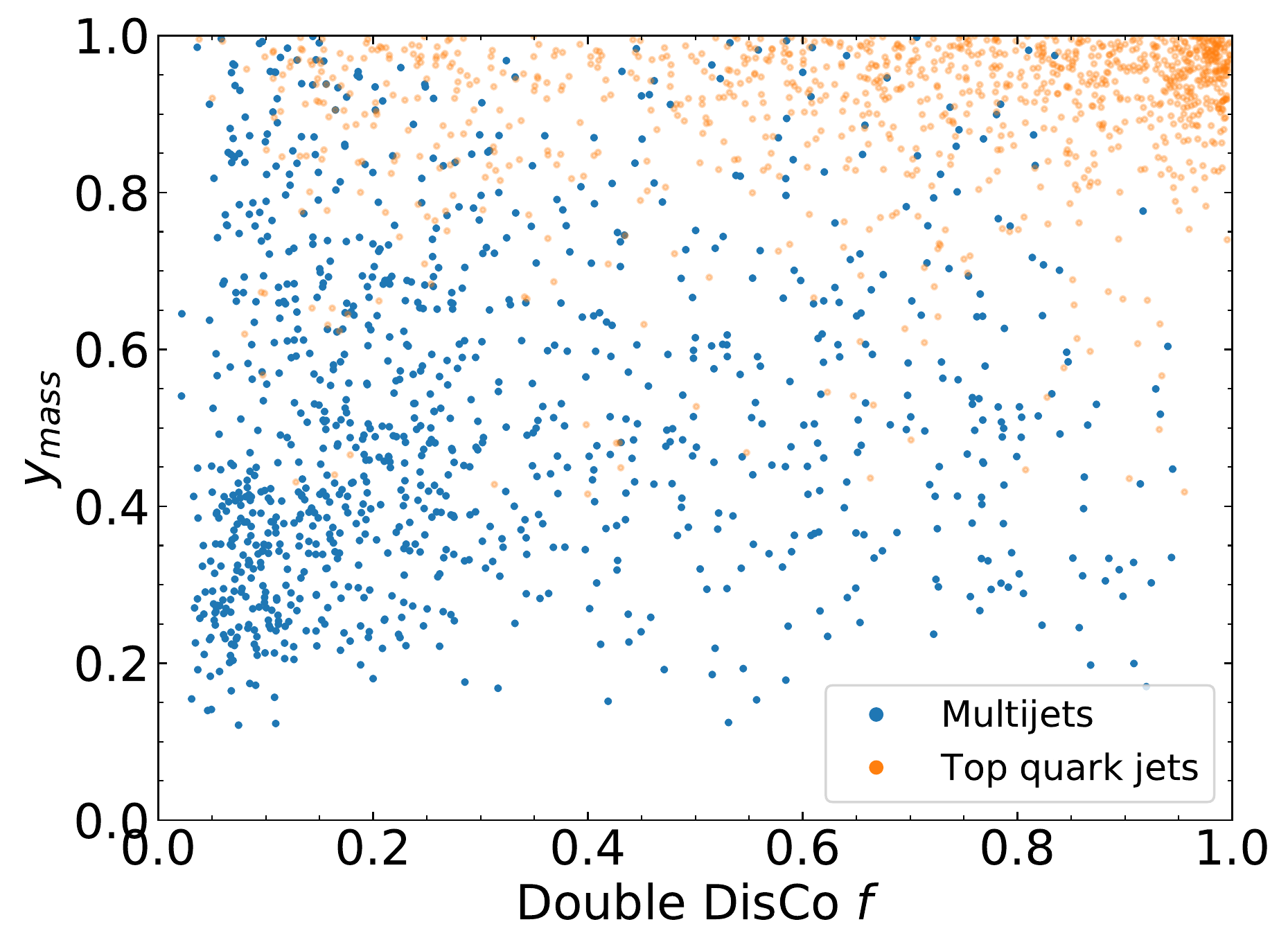}
\includegraphics[width=0.45\textwidth]{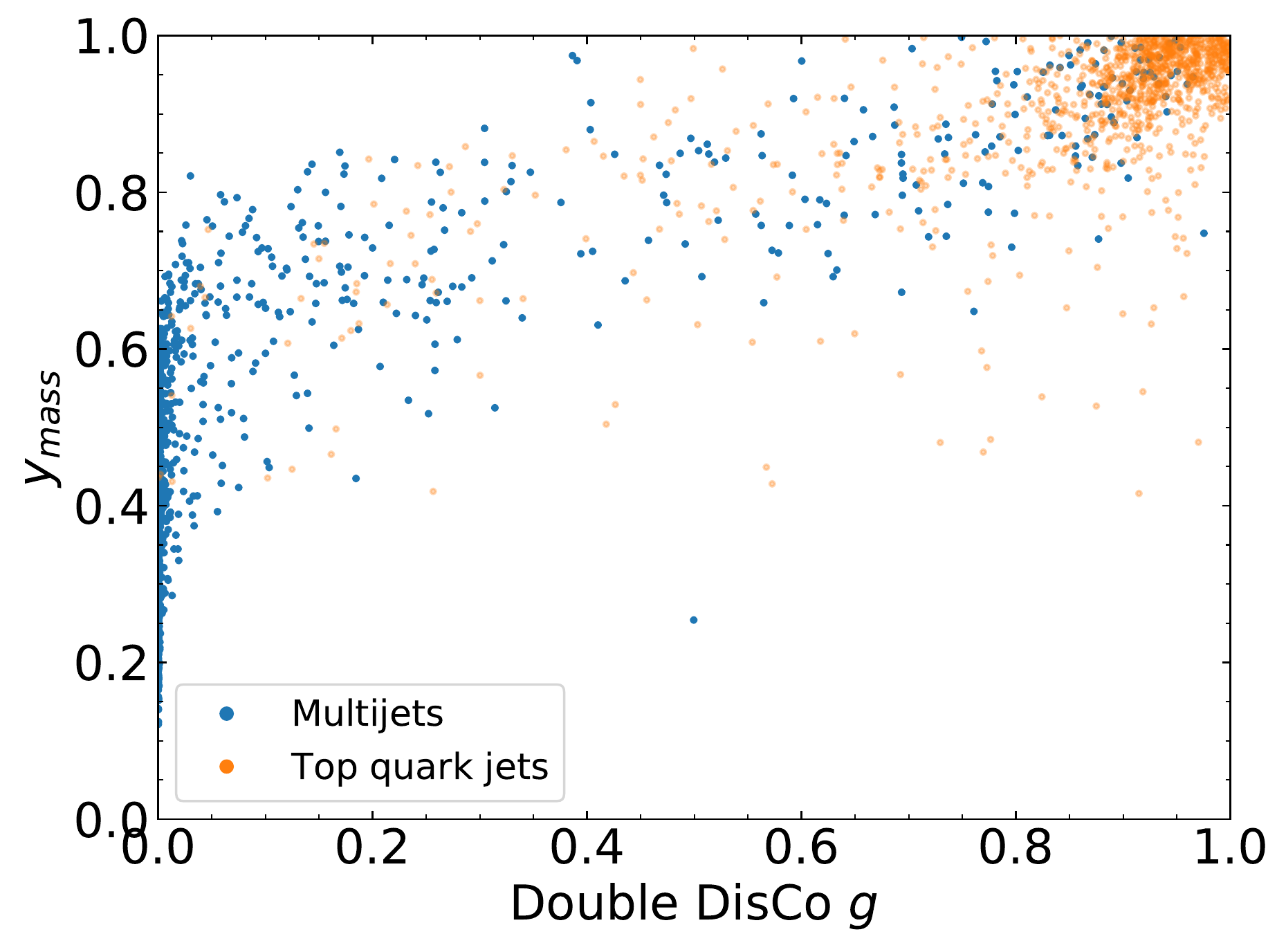}\\
\includegraphics[width=0.45\textwidth]{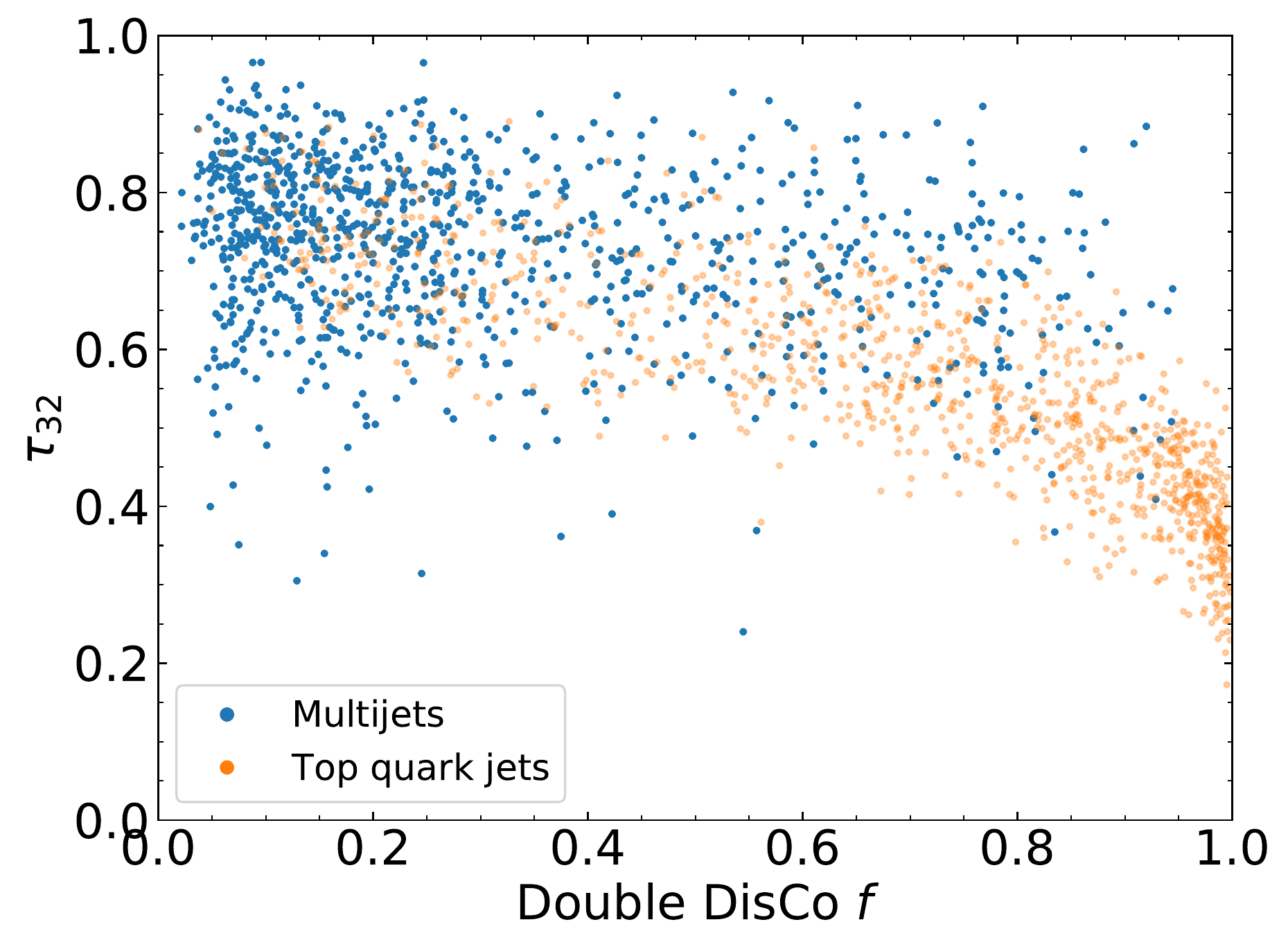}
\includegraphics[width=0.45\textwidth]{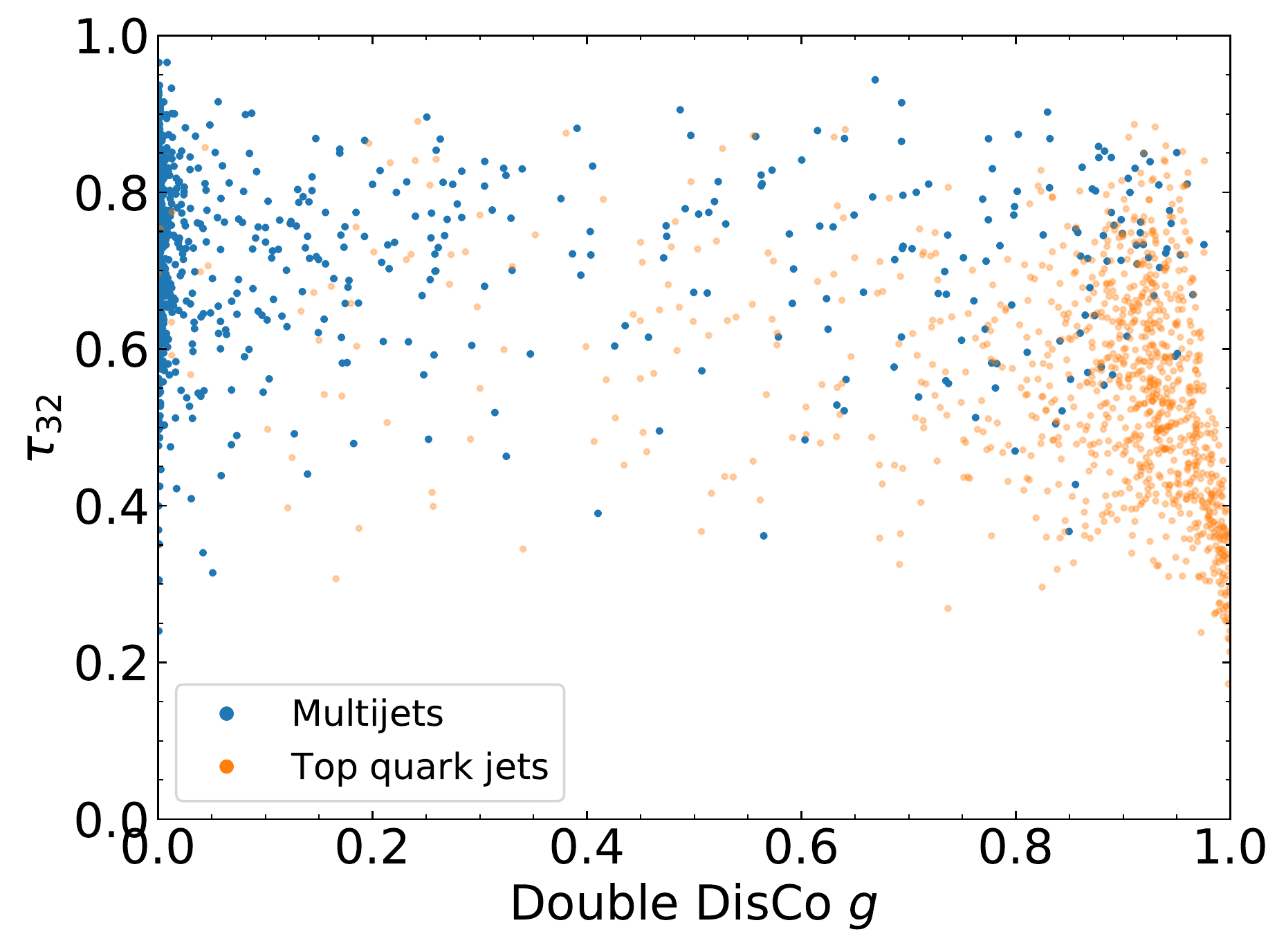}
\caption{Scatter plots of $y_\text{mass}$ (top) and $\tau_{32}$ (bottom) with the Double DisCo classifiers $f$ (left) $g$ (right) in the boosted top analysis.}
\label{fig:topscatter}
\end{figure}

For both Single and Double DisCo we have scanned over the following values of the DisCo parameter:
\beq
\lambda=25,\,\,50,\,\,75,\,\,100,\,\,150,\,\,200
\eeq
Values of $\lambda$ larger than 200 tended to destabilize the training. 
We show in Fig.~\ref{fig:topR30vssc} the background rejection at 30\% signal efficiency vs.\ the normalized signal contamination $r$ defined in Eq.~\eqref{eq:rdef},
for every epoch, DisCo parameter, and value of rectangular cuts on the two classifiers that achieves the required signal efficiency (same method as Sec.~\ref{sec:toy}), subject only to the requirement that the ABCD closure for the background is accurate to within 10\%: $|N_{A,b}-N_{A,b}^\text{predicted}|<0.1$ (see Eq.~\ref{eq:ABCDpred}).  We see that Double DisCo is able to achieve both higher background rejection and significantly lower signal contamination than Single DisCo.

Fig.~\ref{fig:topbestepochs} shows the ``best" models  for Single DisCo and Double DisCo, where ``best" corresponds to an epoch and $\lambda$ that robustly reaches the upper left corner of Fig.~\ref{fig:topR30vssc}. Here each point in the plot represents a choice of the rectangular cut that achieves 30\% signal efficiency. We see that both Single DisCo and Double DisCo are able to achieve accurate ABCD closure and low signal contamination across a wide range of rectangular cuts.

Next we turn to the question of what did Single and Double DisCo learn ---
specifically how the available information was used by the individual NNs.
Shown in Fig.~\ref{fig:toprocmass} are a number of ROC curves. This includes ROC curves for mass, the individual classifiers in Single and Double DisCo, as well as additional NN classifiers obtained from training simple DNNs on various combinations of mass and NN1, NN2 from Double DisCo.

A first observation is that one of the Double DisCo classifiers ($g$) outperforms all the other individual classifiers without explicitly added mass information for all values of the signal efficiency. The next best performance is achieved by
the Single DisCo Classifier, followed by the second one of the Double DisCo classifiers ($f$).\footnote{Both $f$ and $g$ started with equivalent initial conditions
and their symmetry was spontaneously broken during network training.}

Jet mass by itself is very effective for loose selections 
(corresponding to a high signal efficiency). This can be understood from
the good separation observed in Fig.~\ref{fig:topplos} (top left). However, for tighter selections additional substructure information is needed.

Combining mass with one of the Double DisCo classifiers ($g$) does not strongly alter
its performance. This implies that the information contained in mass is
learned by this NN. However, it clearly outperforms mass, 
meaning that $g$ contains more features than just mass. 
On the other hand, combining mass with the weaker Double DisCo classifier ($f$)
dramatically improves it --- it becomes almost, but not quite, optimal. 
This is to be expected as $f$ is forced to be independent from $g$ for background
examples. If $g$ contains mass completely, then $f$ should be mostly independent of mass, and adding it to $f$ should result in a major performance boost.

Finally, there is no real difference between a combination of the
two Double DisCo classifiers ($f+g$),
a further combination also including the mass (mass~$+f+g$),
and a direct training on all input features.
This further confirms that the mass information has been fully absorbed by $f+g$ ---
specifically $g$ via the argument above. The maximally inclusive mass~$+f+g$ classifier
of course should not be used as input to the ABCD method. However,
we can compare its performance to results on the same dataset in Ref.~\cite{Kasieczka:2019dbj}. A classifier based on multi-body N-subjettiness trained following the procedure suggested in Ref.~\cite{Moore:2018lsr} achieved a background rejection of up to around 1/900 for a signal efficiency of 30$\%$. We observe a 
slightly weaker 1/700 which is to be expected as a lower number of N-subjettiness
observables is used as inputs here. 

In the the scatter plots of the Double DisCo discriminators 
in Fig.~\ref{fig:topscatter}, we again observe the
larger discrimination power of $g$ compared to $f$. Looking 
at the top left distribution, we indeed see no dependence of
$f$ on the mass while in the top right a clear correlation is there for $g$.
On the other hand, in the bottom left, we see a trend between $f$ and $\tau_{32}$
which encodes to which amount the jet is compatible with a 3-prong substructure. 
This information is largely not learned by $g$.

We conclude that Double DisCo can do better than Single DisCo because it is partitioning the information differently than just mass versus everything else. 
 
\subsection{RPV SUSY}
\label{sec:RPV}

For our third example, we 
consider an actual ``real-life" application %
of the ABCD method on LHC data: the $\sqrt{s}=13$ TeV ATLAS search for paired dijet resonances~\cite{Aaboud:2017nmi}. Similar searches were conducted by CMS~\cite{Sirunyan:2018rlj} and by both experiments at $\sqrt{s}=8$ TeV~\cite{Aad:2016kww,Khachatryan:2014lpa}. These searches were motivated by pair production of identical squarks which each decay promptly to two jets via RPV couplings. For background estimation, these searches all used the standard ABCD method. In this section we will describe our recast of this search and the performance gains derived from training Single and Double DisCo on it.  

The ATLAS search consisted of the following steps:

\begin{itemize}

\item {\it Preselection}: Events are required to have at least four jets with $p_T>120$ GeV and $|\eta|<2.4$.   The leading four such jets are used to form two squark candidates based on nearest proximity in $\Delta R=\sqrt{(\Delta\phi)^2+(\Delta\eta)^2}$.  The minimum $\Delta R$ from the resulting pairings is defined as $\Delta R_\text{min}$ and the two dijet masses are used to form the average mass $m_\text{avg}= \frac{1}{2}(m_{\text{dijet 1}} + m_\text{dijet 2})$ and fractional mass asymmetry $\Am =\frac{1}{m_\text{avg}}
|m_\text{dijet 1}-m_\text{dijet 2}|$.  Events with $m_\text{avg}<255$ GeV must have $\Delta R_\text{min}<0.72-0.002(m_\text{avg}/\text{GeV}-255)$ and events with $m_\text{avg}\geq 255$ GeV must have $\Delta R_\text{min}<0.72-0.0013(m_\text{avg}/\text{GeV}-255)$.  

\item{\it Final selection}: For the final selection, the ATLAS search performs counting experiments in successive windows of $m_\text{avg}$, and for background estimation uses the ABCD method in $|\cos \theta^*|$ and $\Am$, where $\theta^*$ is the polar angle of one of the squarks in the squark-squark center-of-mass frame.  The signal region is defined as $\Am<0.05$ and $|\cos \theta^*|<0.3$.

\end{itemize}

\begin{figure}[h!]
\centering
\includegraphics[height=0.2\textwidth]{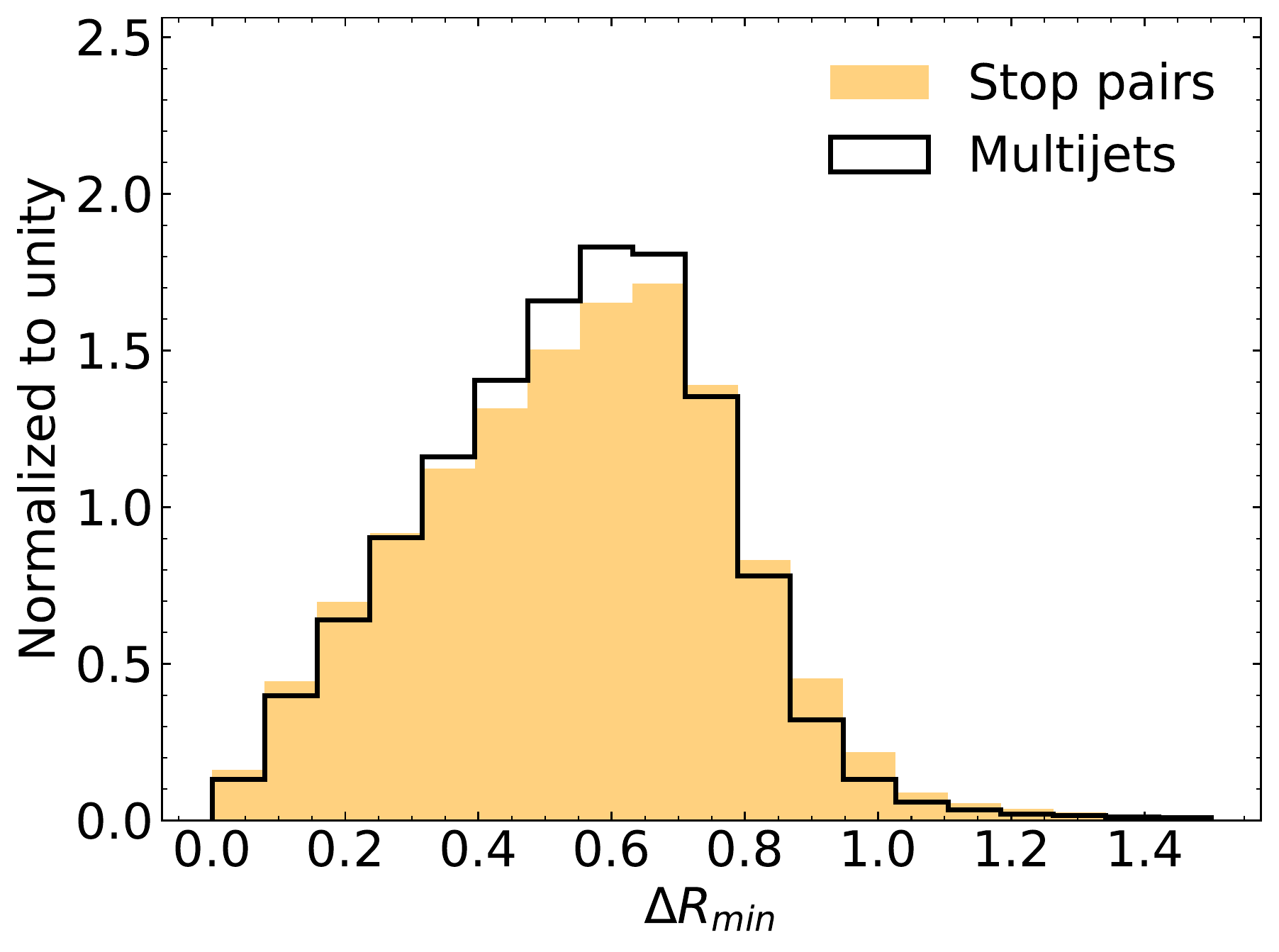}
\includegraphics[height=0.2\textwidth]{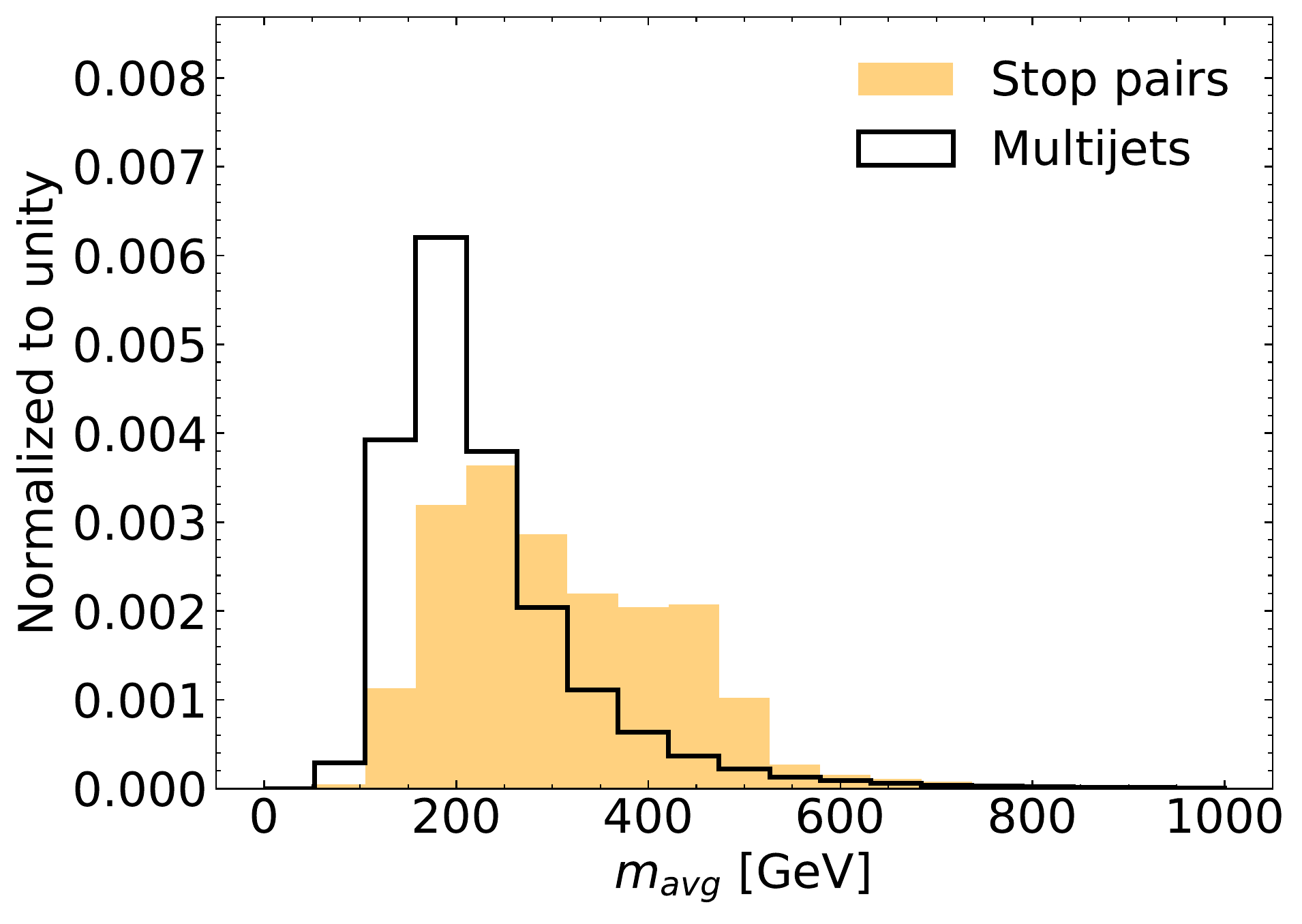}
\includegraphics[height=0.2\textwidth]{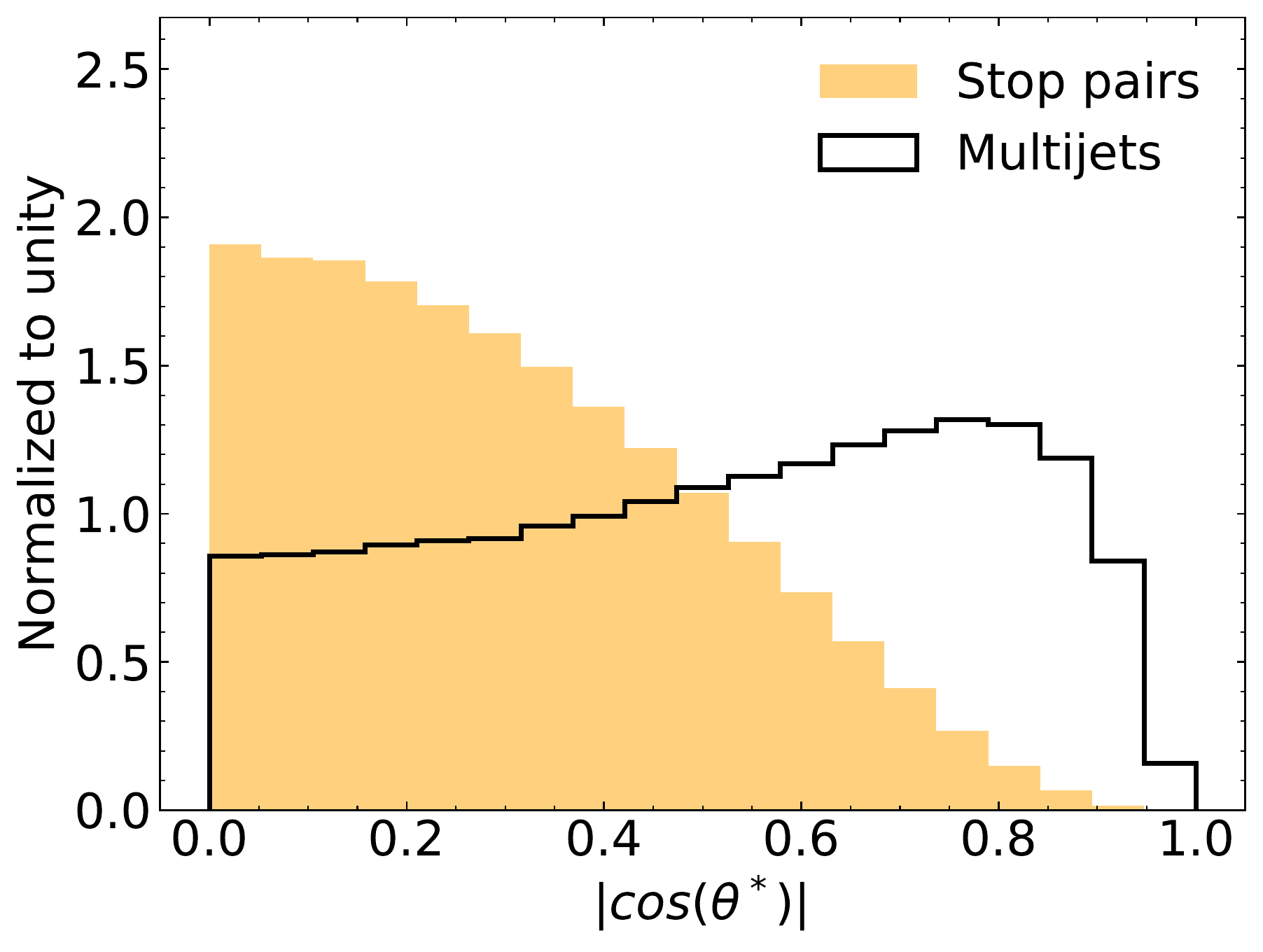}\\
\includegraphics[height=0.2\textwidth]{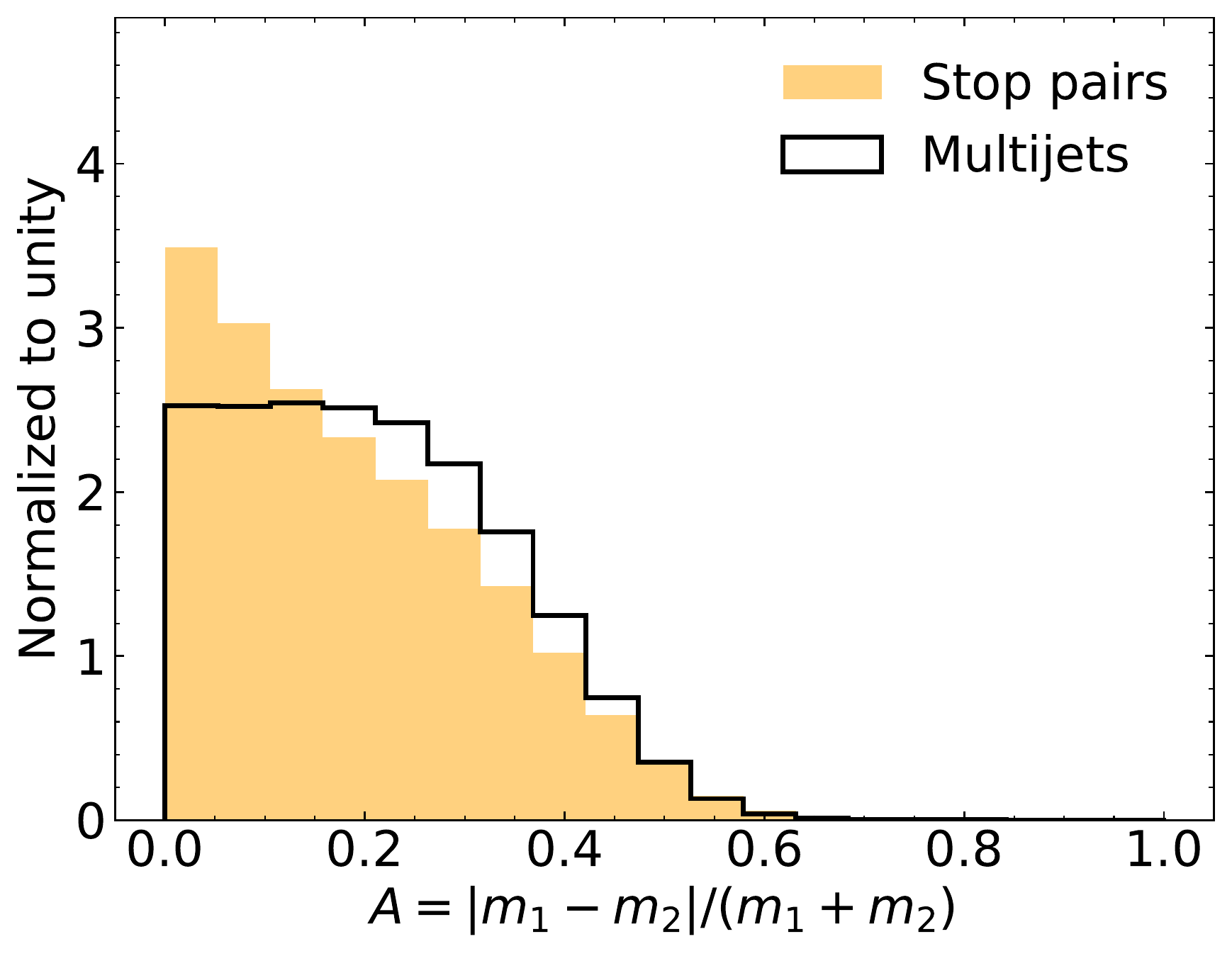}
\includegraphics[height=0.2\textwidth]{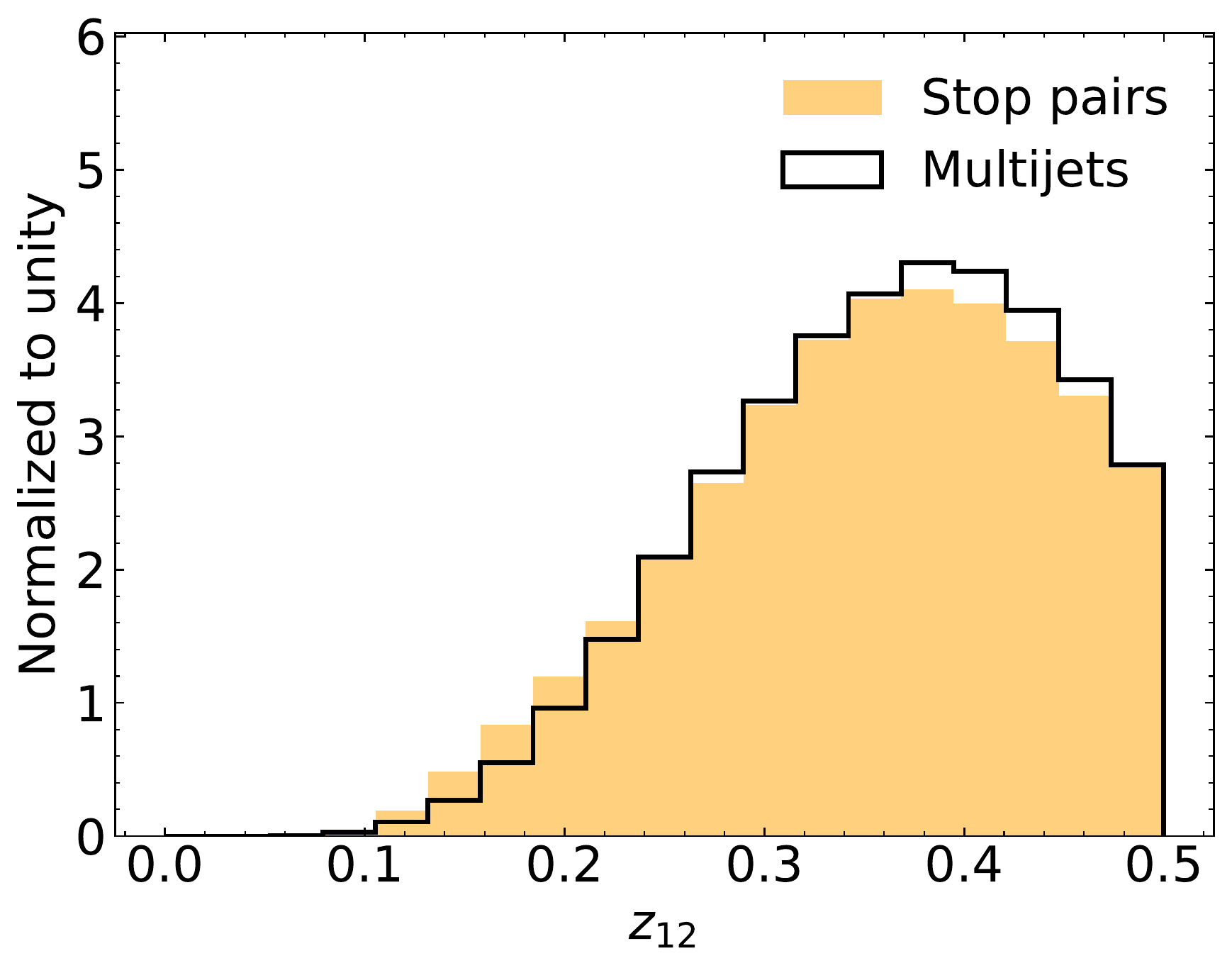}
\includegraphics[height=0.2\textwidth]{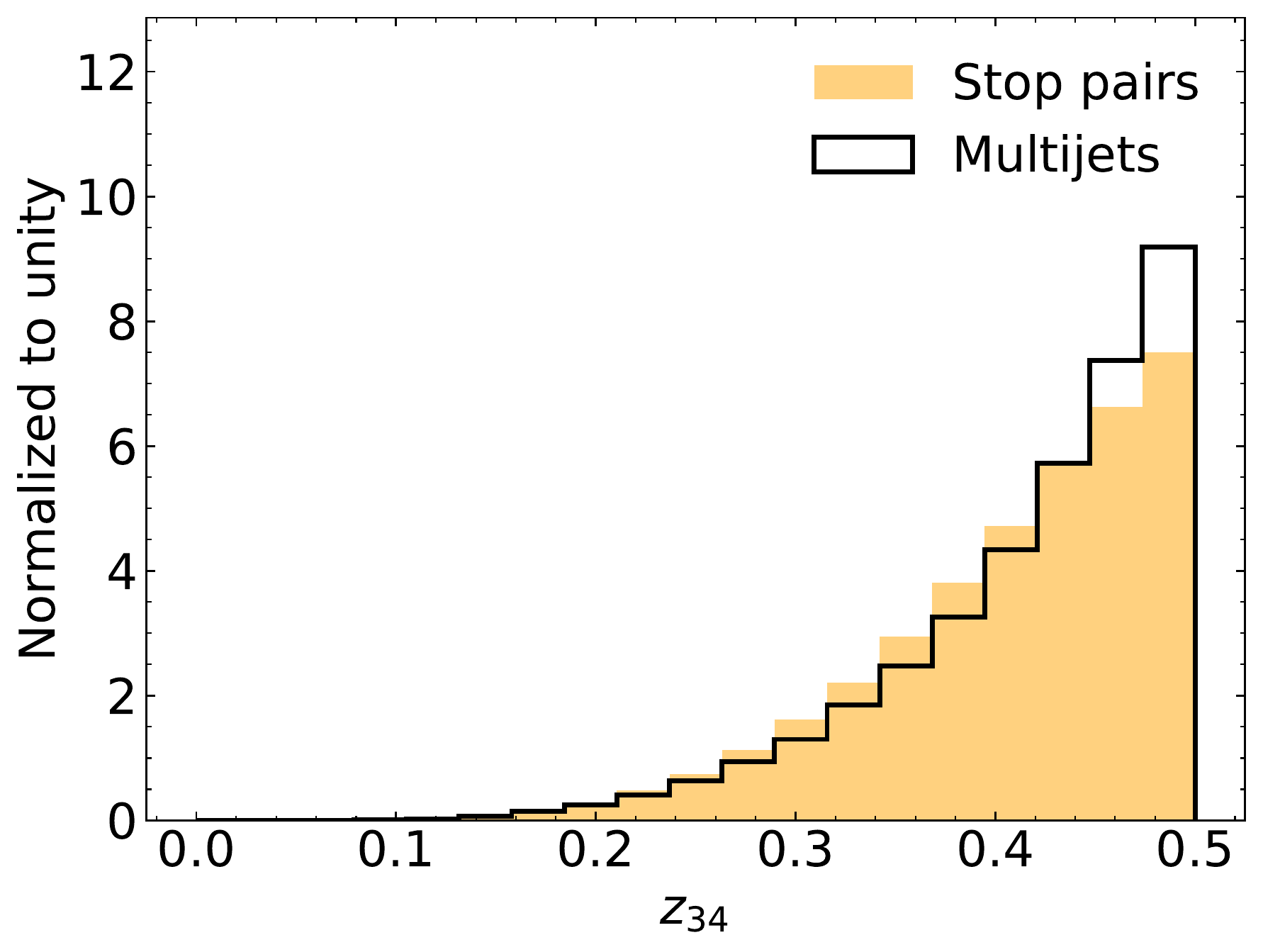}\\
\includegraphics[height=0.2\textwidth]{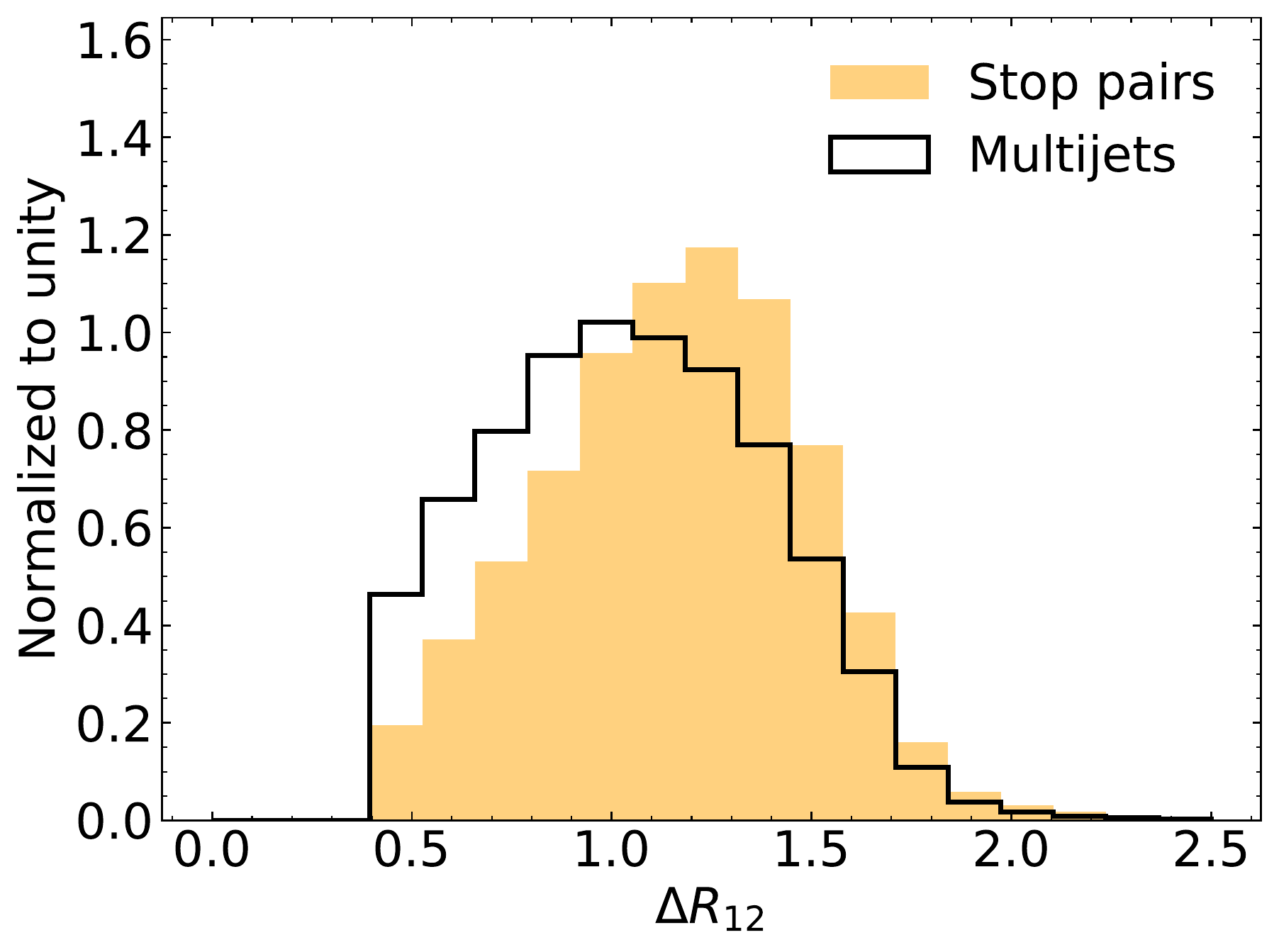}
\includegraphics[height=0.2\textwidth]{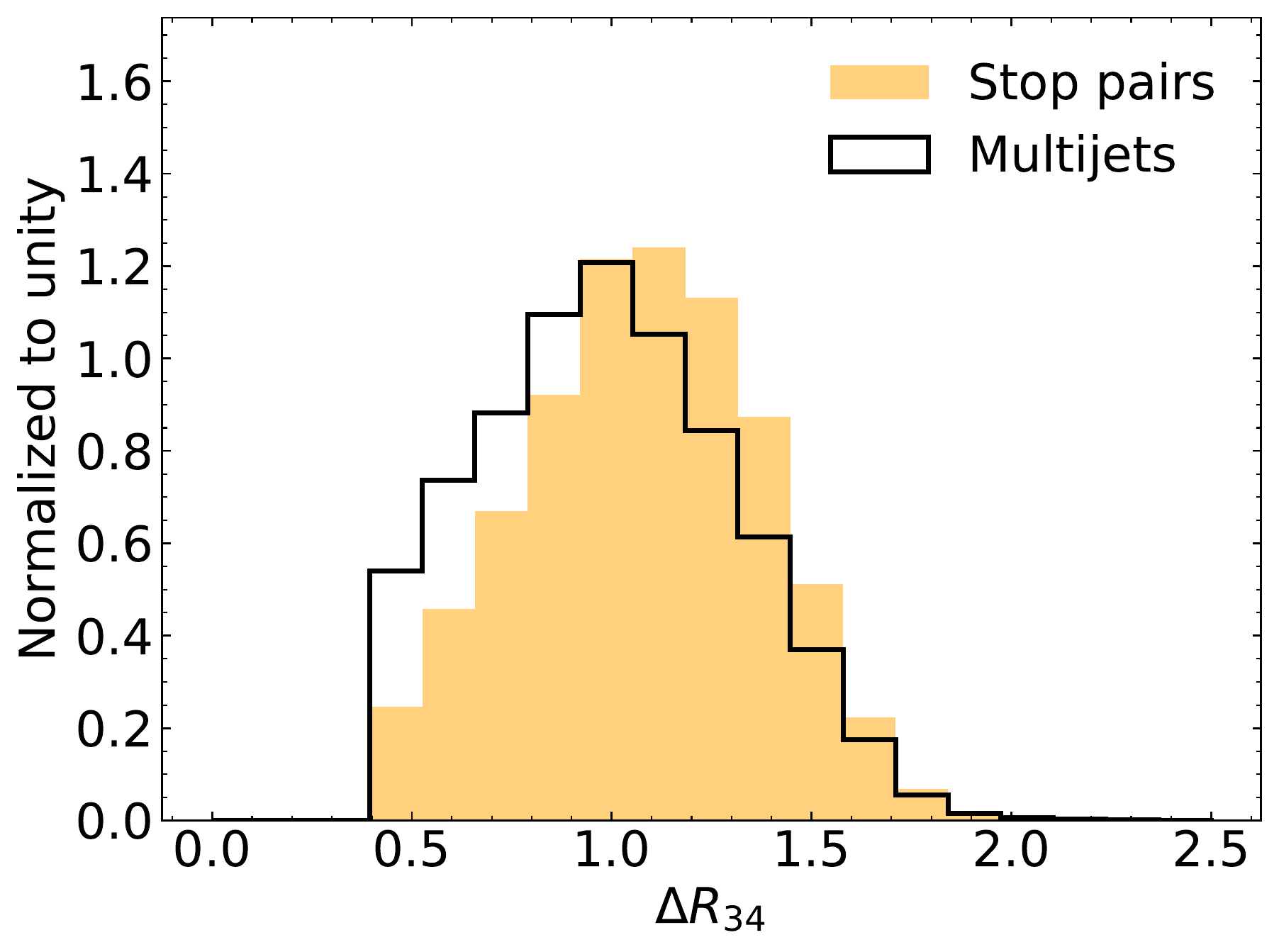}
\includegraphics[height=0.2\textwidth]{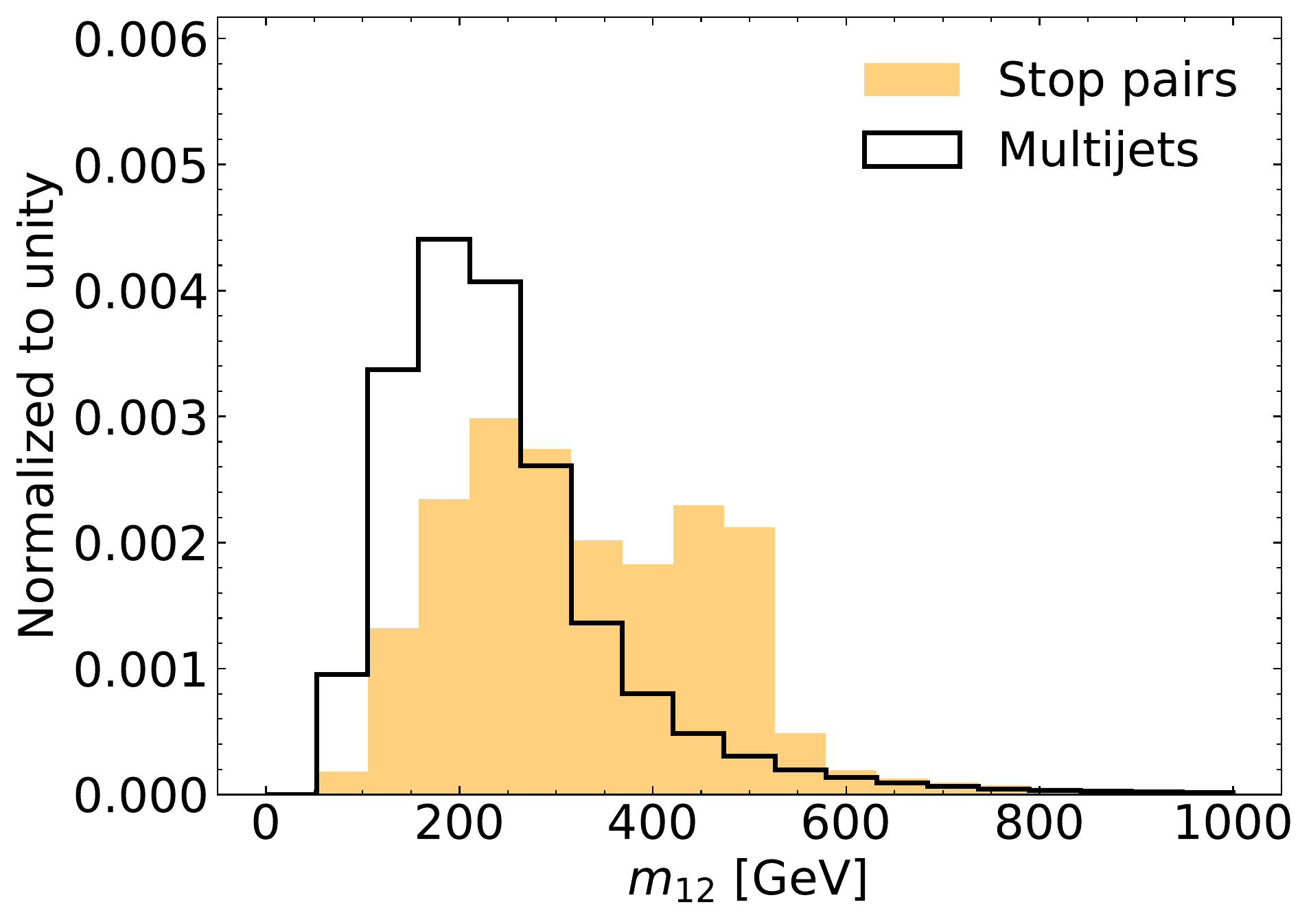}\\
\includegraphics[height=0.2\textwidth]{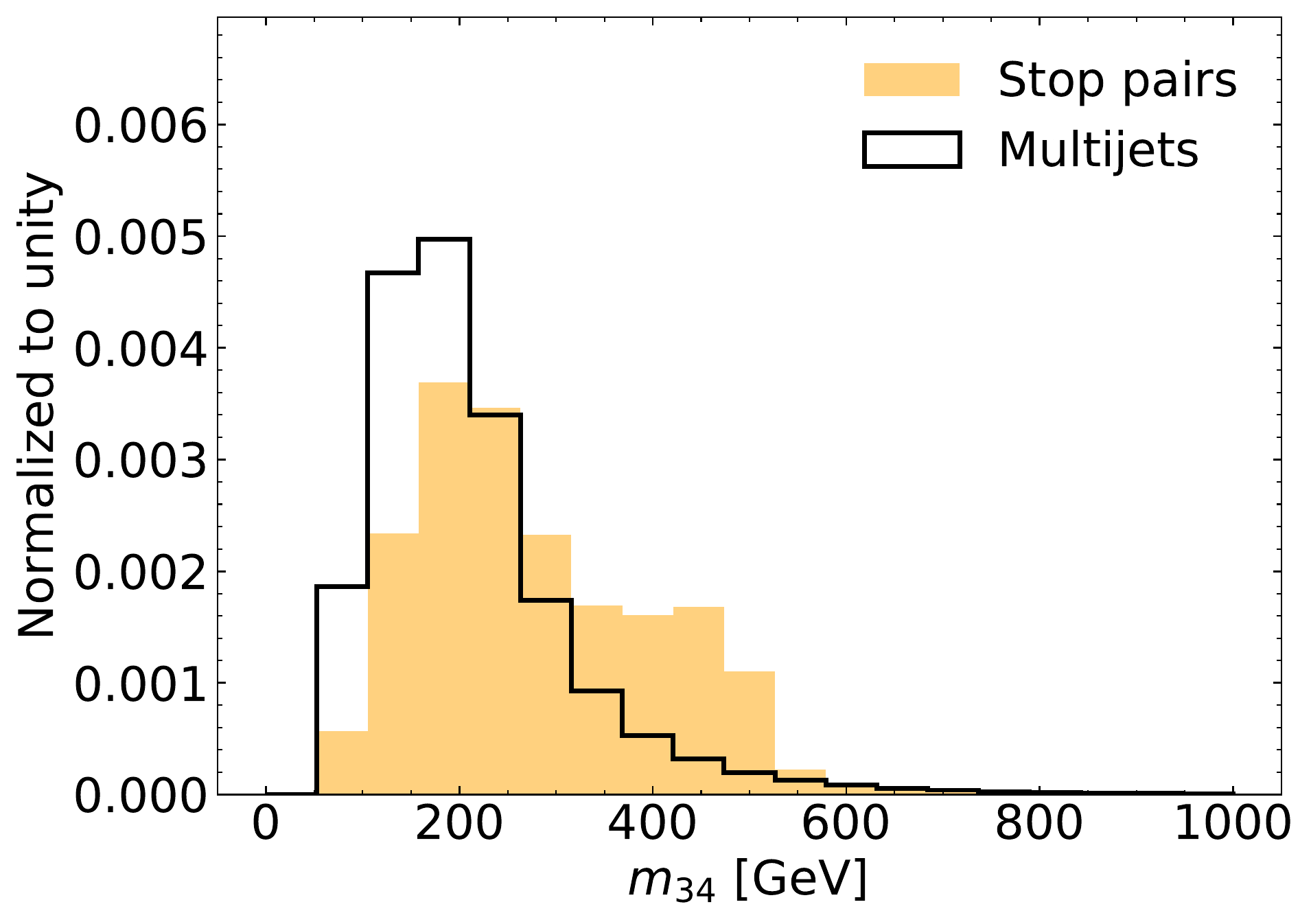}
\includegraphics[height=0.2\textwidth]{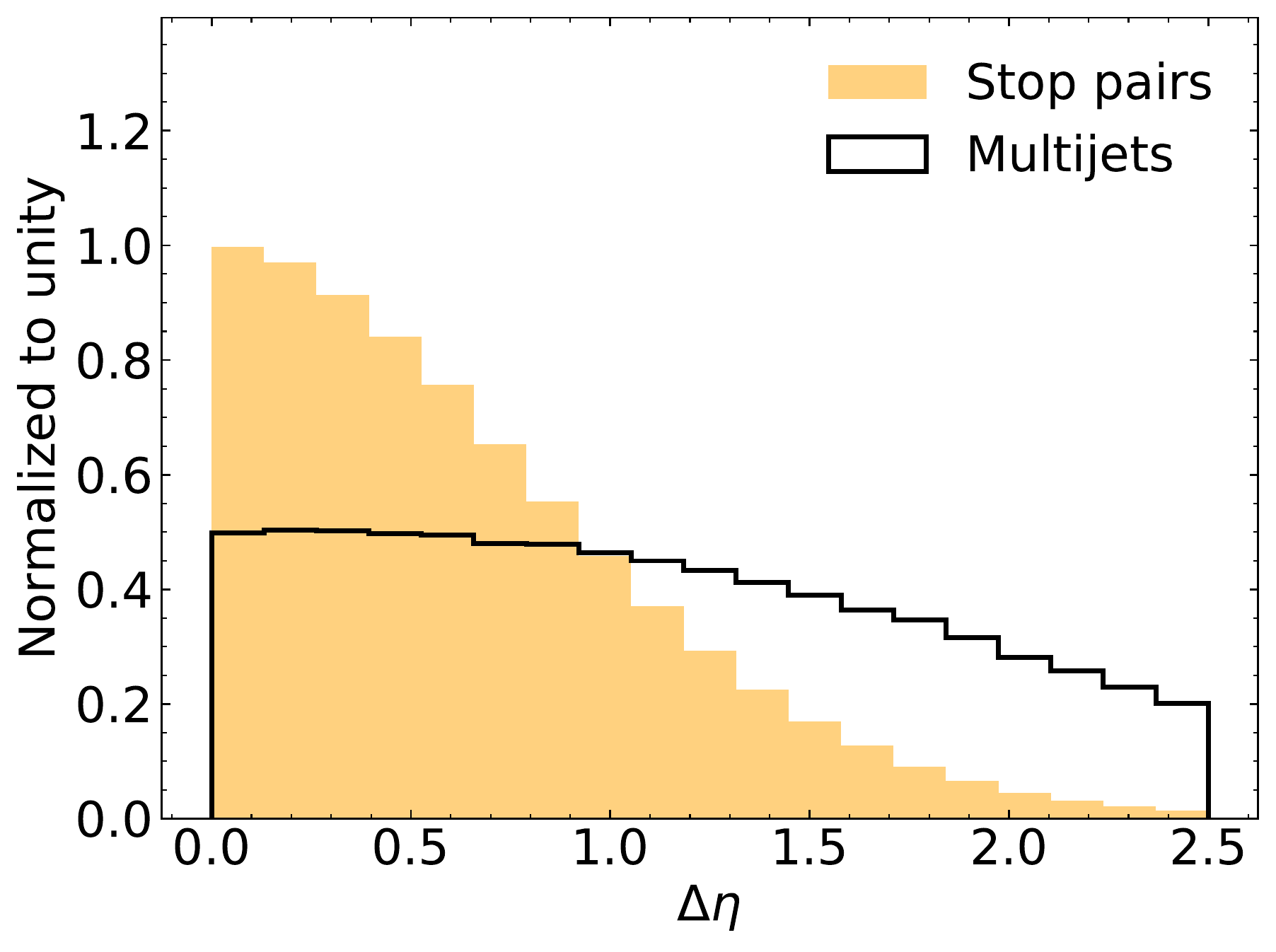}
\includegraphics[height=0.2\textwidth]{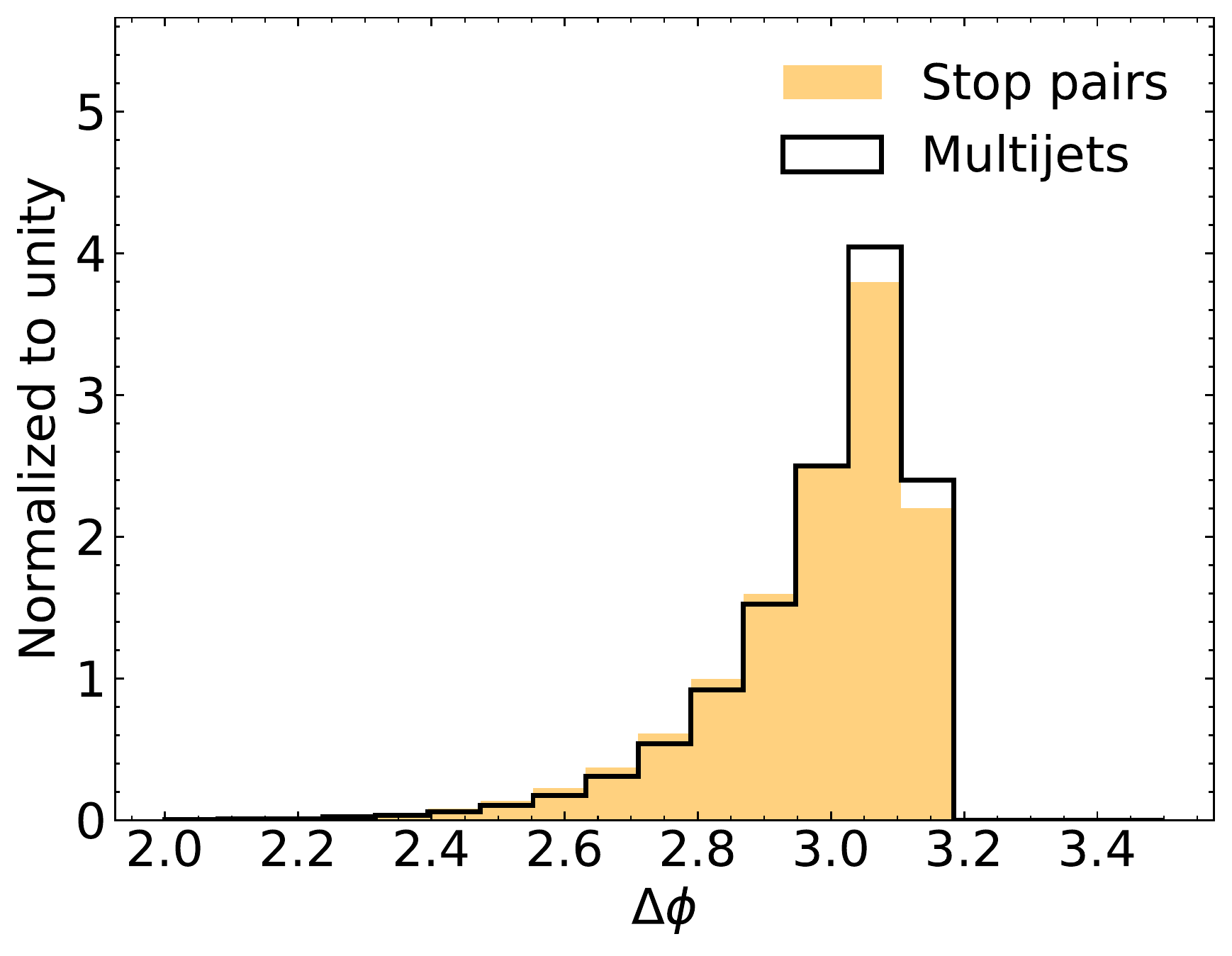}\\
\includegraphics[height=0.2\textwidth]{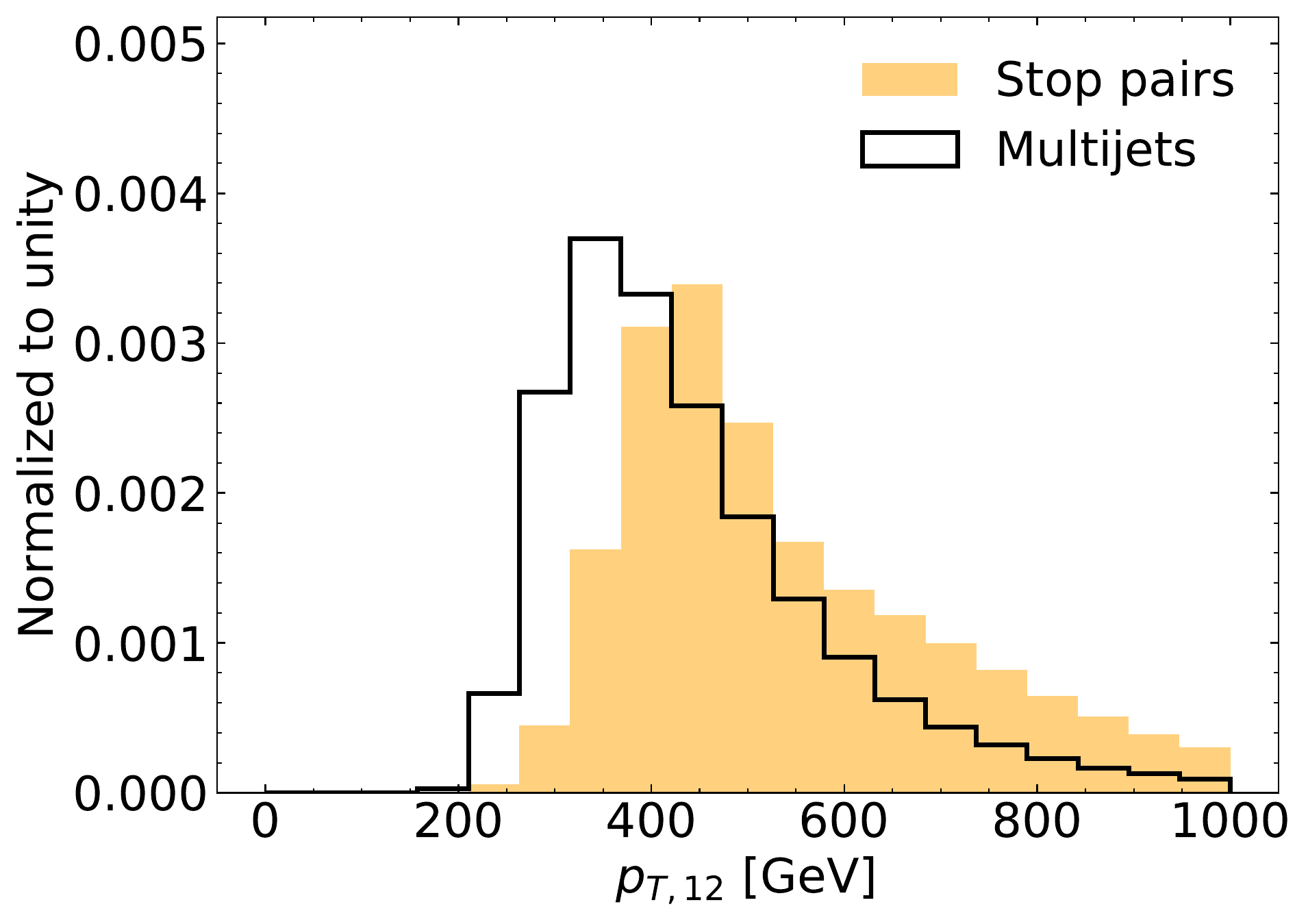}
\includegraphics[height=0.2\textwidth]{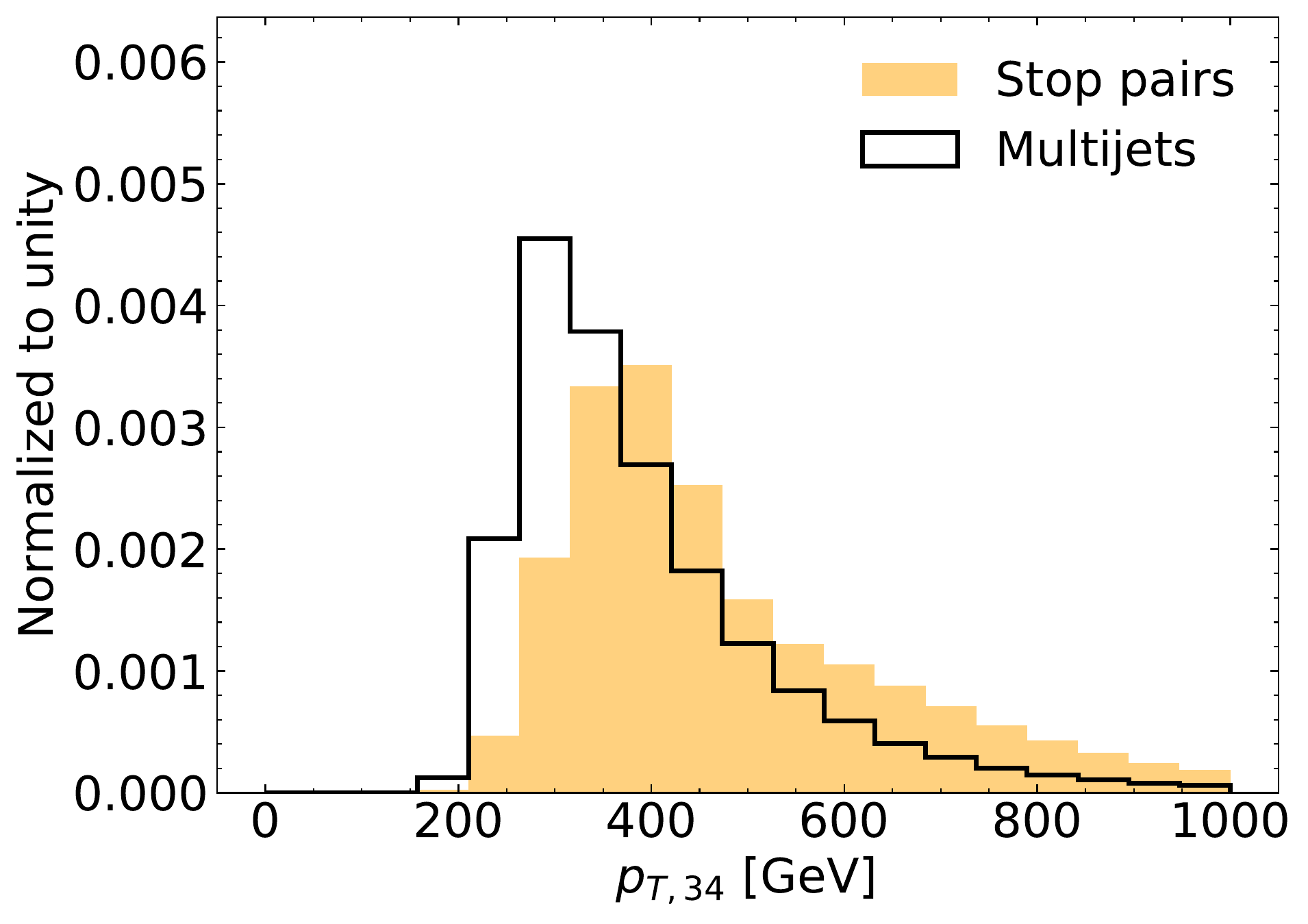}
\caption{The features used to train the RPV classification model.}
\label{fig:rpvplos}
\end{figure}

ATLAS ended up setting a limit at approximately $m_\text{squark}=500$~GeV, so we will also focus our analysis on this value of the squark mass. We repeat the preselection cuts but instead of the final selection on $m_\text{avg}$, $\Am$ and $\cos\theta^*$, we instead feed a list of inputs to Single and Double DisCo to learn the optimal features. The inputs are:
\beq
\Delta R_\text{min}, \,\, m_\text{avg}, \,\,  \cos\theta^*,\,\,  \Am,\,\,  z_{12},\,\,  z_{34},\,\,  \Delta R_{12},\,\,  \Delta R_{34},\,\,  m_{12},\,\,  m_{34},\,\,  \Delta\eta,\,\,  \Delta\phi,\,\,  p_{T,12},\,\,  p_{T,34}\,,
\eeq
where $z_{12}$ ($z_{34}$), $\Delta R_{12}$ ($\Delta R_{34}$), $m_{12}$ ($m_{34}$), $p_{T,12}$ ($p_{T,34}$) are the $p_T$ of the subleading jet divided by the sum of the transverse momenta of both jets, the opening angle between the two jets, the invariant mass of the two jets, and the $p_T$ of the two jets for the stop dijet pair with the leading small-radius jet (and the other stop dijet pair), respectively. Histograms of these features are shown in Fig.~\ref{fig:rpvplos}. All features are rescaled to the range $[0,1]$ before feeding to the NNs. For Single DisCo we use $\cos\theta^*$ rather than $\Am$ as the fixed variable $X_0$ ($\cos\theta^*$ is the stronger of these two features from the ATLAS RPV squark analysis) and feed everything else to the NN classifier. For Double DisCo we feed everything to the two NN classifiers.

Squark 
pair events and multijet events are generated with \textsc{Pythia} 8.230~\cite{Sjostrand:2014zea,Sjostrand:2006za} at a center-of-mass-energy of $\sqrt{s}=13$ TeV interfaced with \textsc{Delphes} 3.4.1~\cite{deFavereau:2013fsa} using the default CMS run card.  Jets are clustered using the anti-$k_t$ algorithm~\cite{Cacciari:2008gp} with radius parameter $R = 0.4$ implemented in \textsc{Fastjet} 3.2.1~\cite{Cacciari:2011ma,Cacciari:2005hq}. 1M signal events and 10M background events were generated, of which about 100k signal events and 60k background events pass the  preselection.  In order to ensure a high event selection efficiency for the background, events are generated using $2\rightarrow 3$ matrix elements with a minimum separation of $R = 0.8$ and minimum $\hat{p}_T$ of 100 GeV for the softest parton and 200 GeV for the hardest parton.  Signal events are produced using the SLHA~\cite{Skands:2003cj,Allanach:2008qq} card from the recent ATLAS search~\cite{Aaboud:2017nmi,1631641} in which the squark mass is 500 GeV and all other super partners are decoupled.

\begin{table}[h]
\centering
\begin{tabular}{|c|c|c|}
\hline
 cut & ATLAS & our recast\\
\hline
\hline
$\Delta R_{min}$ & 13.0\% & 11.9\% \\
\hline
inclusive SR & 10.2\% & 9.5\%\\
\hline
mass window & 25\% & 23.3\% \\
\hline
\end{tabular}
\caption{Relative efficiencies for each cut on the signal in the ATLAS RPV SUSY search and our recast.}
\label{tab:sigcuts}
\end{table} 

\begin{table}[h]
\centering
\begin{tabular}{|c|c|c|c|c|}
\hline
& \multicolumn{2}{|c|}{ATLAS} & \multicolumn{2}{|c|}{our recast} \\
\hline
 Region $i$ & $f_{i}$ & $\delta_i$ & $f_{i}$ & $\delta_i$ \\
\hline
D (SR) & 6.8\% & 6.3\% & 6.4\% &  6.3\% \\
\hline
A & 11.4\% & 3.1\% & 10.5\% & 3.4\% \\
\hline
F & 30.7\% & 0.2\% & 31.6\% & 0.3\% \\
\hline
C & 51.1\% & 0.07\% & 51.6\% & 0.2\% \\
\hline
\end{tabular}
\caption{Relative fractions $f_i$ of data in the regions $i=D$, $A$, $F$ and $C$ used in the ATLAS RPV SUSY analysis and our recast, and signal to background ratios $\delta_i$ in each region.}
\label{tab:bgcuts}
\end{table} 

The validation of the recasting  of the ATLAS analysis is shown in Table~\ref{tab:sigcuts} and Table~\ref{tab:bgcuts}. In the former we show the relative signal efficiencies after successive cuts. In the latter we show the relative fractions $f_i$ (since we do not attempt to get the overall normalizations of our simulations correct) of data in ATLAS regions $i=D$, $A$, $F$, $C$ (for ATLAS $D$ is the SR);
and the signal to background ratio $\delta_i$ in each region. Following ATLAS, for the data fractions, the counts are taken after the inclusive selection with no mass window cut, while for the signal to background ratios they are taken after the mass window cut. Overall, we see excellent agreement between the ATLAS numbers and our recasted numbers.

For training the NNs, we use 100k signal and 360k background events, while the validation sample consists of 25k signal and 250k background events. In the classifier loss, we rebalance the signal and background contributions as if they were 50/50.  

We used the same hyperparameters as the top tagging example. (We also explored using 128 nodes per hidden layer but found that it did not help.) 
For DisCo parameters we chose 
\beq
\lambda = 10,\,\,\,20,\,\,\,30,\,\,\,40,\,\,\,60,\,\,\, 100\,.
\eeq
Unlike the top tagging example we do not add the additional DisCo term sensitive to the tails of the background distributions when training Double DisCo; because the background rejections in this case were not as high as for top tagging, the additional term was found not to help. 

\begin{figure}[t]
\centering
\includegraphics[width=0.7\textwidth]{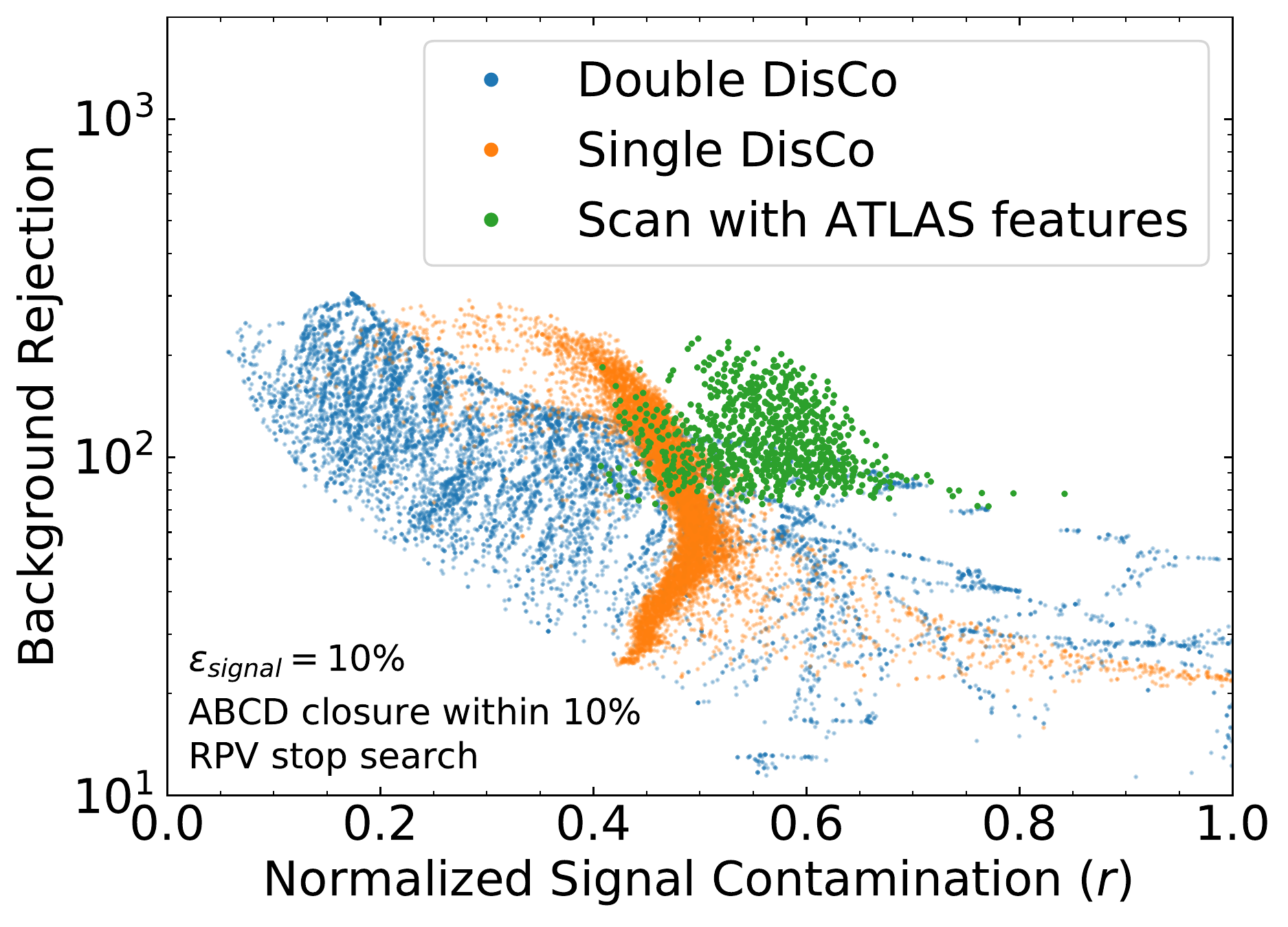}
\caption{A scatter plot of background rejection versus normalized signal contamination ($r$) in the RPV SUSY analysis for various epochs with Single and Double DisCo as well as a scan of three-dimensional thresholds on the features used by the ATLAS analysis.}
\label{fig:RPVstop_comparison_scatter_64}
\end{figure}

The comparison of Single and Double DisCo is shown in Fig.~\ref{fig:RPVstop_comparison_scatter_64}. As in the top tagging section, we have plotted every epoch and every rectangular cut and every value of the disco parameter satisfying the 10\% accuracy condition on the ABCD prediction. This shows the performance of the models in the plane of $R_{10}$ (background rejection factor at 10\% signal efficiency) vs total fractional signal contamination. We see that while Double DisCo cannot surpass Single DisCo in terms of raw performance (as measured by $R_{10}$), it can achieve dramatically lower signal contamination for roughly the same $R_{10}$. 

We have also included scans over the features used in the ABCD method as used in the ATLAS RPV search, these are the green points in Fig.~\ref{fig:RPVstop_comparison_scatter_64}.\footnote{The actual ATLAS analysis used a working point that corresponds to about 2.5\% signal efficiency.  We found this to be sub-optimal to a 10\% value, which is why it is used in Fig.~\ref{fig:RPVstop_comparison_scatter_64}.}  We note that ATLAS had significant normalized signal contamination with their selection (40-80\%). Both Single and Double DisCo offer a marked improvement in both signal contamination and background rejection compared to the standard ABCD method with manually-chosen high-level features.

\clearpage

\section{Conclusions} \label{sec:conclusions}

Estimating backgrounds is essential for every experimental analysis in particle physics. 
One of the most  well-established data-driven technique for background estimation is the ABCD method. In this paper we have re-examined the criteria for the ABCD method to be effective and proposed a way to find the variables used to establish the ABCD regions using machine learning.

A general observation we make in this paper is that 
the signal contamination in the background region \emph{normalized} to the signal 
fraction in the signal region drives the quality of the ABCD background estimate. This observation is independent of any machine-learning appraoches to determining the features. We argue that 
controlling this normalized signal contamination should become a default procedure in applying the ABCD method, since neglecting it can lead to incorrect, and typically overly conservative,
$p$-values.

Regardless of how one estimates contamination of the background samples, a necessary condition for the ABCD method to work is the availability of two independent classifiers.
These classifiers are usually found by guessing observables that, on physical grounds, seem like they would be independent, and then verifying their independence with simulations or validation regions. Such a procedure is by no means guaranteed to yield optimal results. Indeed, observables designed for classification, either by hand or learned by machine, easily have better discrimination power than observables chosen to be independent. However, optimal observables 
aim to make maximum use of available information and will in general exhibit complex dependencies with all 
other observables.  

In this paper, we proposed to use machine learning methodology to optimize the ABCD method. We considered two use cases: 1) Single DisCo, where a first variable (such as mass) is fixed and another is learned to be decorrelated with it and optimize discrimination and 2) Double DisCo, where both variables are learned. For both methods, our machine learning approach builds upon the DisCo loss term, a recently developed method for automated decorrelation.
This technique allows for the autonomous construction of a robust data-driven background estimation assuming
a specific signal model.

We considered three examples: 1) a simple model of correlated random variables that demonstrates how Single and Double DisCo work,  2) boosted top tagging and 3) an RPV squark search, based on an existing ATLAS analysis. 
We found that while  Single DisCo offers competitive 
performance in terms
of pure background rejection, Double DisCo achieves 
lower signal contamination levels in both of the physical examples considered.  We note that while DisCo was used to demonstrate decorrelation in this paper, the general idea can be combined with any decorrelation method~\cite{Louppe:2016ylz,Dolen:2016kst,Moult:2017okx,Stevens:2013dya,Shimmin:2017mfk,Bradshaw:2019ipy,ATL-PHYS-PUB-2018-014,DiscoFever,Xia:2018kgd,Englert:2018cfo,Wunsch:2019qbo,Rogozhnikov:2014zea,Sirunyan:2019nfw,clavijo2020adversarial,Aguilar-Saavedra:2017rzt,Sirunyan:2020lcu} and the best approach may be application-specific.

On the surface, one advantage of the traditional ABCD method has over the proposed automated approaches is that it is largely signal model independent. However, even there, it
is necessary to explicitly verify low signal contamination for all considered models using simulations. On the other hand, the training of Single DisCo or Double DisCo can be 
extended to a cocktail of signal models or parametrised as a function of the considered signal~\cite{Cranmer:2015bka,Baldi:2016fzo}.

While the Single and Double DisCo approaches achieve excellent performance,
even better sensitivity might be obtained by optimizing the necessary criteria
of low signal contamination and good ABCD closure more directly. We argued in earlier sections that the Single and Double DisCo loss qualitatively capture these requirements, but direct optimization of the conditions is challenging as they cannot be readily cast in a differentiable form. One might, for example, try an iterated learning approach or one based on reinforcement learning, where the final $p$-value for ABCD searches is used as a score.
Further studies in this direction are left to future work.

Finally, it is important to consider the task of background estimation in the broader context of analysis optimization.  A variety of methods have been proposed to directly optimize analysis sensitivity including uncertainty~\cite{Wunsch:2020iuh,deCastro:2018mgh,Elwood:2018qsr,Dorigo:2020ldg}.  Background estimation is a key part of analysis design and could be integrated into the ABCD method in order to further optimize the overall discovery potential. An orthogonal approach is to construct searches for new physics in a
model independent way~\cite{hepmllivingreview,DAgnolo:2018cun,Collins:2018epr,Collins:2019jip,DAgnolo:2019vbw,Farina:2018fyg,Heimel:2018mkt,Roy:2019jae,Cerri:2018anq,Blance:2019ibf,Hajer:2018kqm,DeSimone:2018efk,Mullin:2019mmh,1809.02977,Dillon:2019cqt,Andreassen:2020nkr,Nachman:2020lpy,Aguilar-Saavedra:2017rzt,Romao:2019dvs,Romao:2020ojy,knapp2020adversarially,collaboration2020dijet,1797846,1800445,Amram:2020ykb,Cheng:2020dal}.
Such searches will also require robust and automated data driven background predictions and --- at least partially --- can be trained with a Single or Double DisCo method.

In summary, we are able to increase the discovery potential of
physics analyses by enabling robust background estimates for more powerful classifiers.
This improvement is made possible by clearly defining the objectives and
then using automated tools to optimize a parametric function to achieve them.  
The present work shows that even time-tested and widely deployed analysis methods can benefit
from systematic optimization.

\section*{\label{sec::acknowledgments}Acknowledgments}

We would like to thank Alejandro Gomez Espinosa and Simone Pagan Griso for useful discussions and Simone for additionally providing feedback on the manuscript.  
We thank Olaf Behnke and Thomas Junk for helpful comments 
on the manuscript and especially for examples of early uses of the ABCD method.
BN, MS, DS were supported by the U.S.~Department of Energy, Office of Science under contract numbers DE-AC02-05CH11231, DE-SC0013607, and DOE-SC0010008, respectively.   BN would also like to thank NVIDIA for providing Volta GPUs for neural network training. GK acknowledges support by the Deutsche Forschungsgemeinschaft (DFG, German Re\-search Foundation) under Germany’s Excellence Strategy – EXC 2121 ``Quantum Universe'' – 390833306. DS is grateful to LBNL, BCTP and BCCP for their generous support and hospitality during his sabbatical year.

\appendix

\section{Distance Correlation}
\label{sec:discodef}

For two random variables $f$ and $g$, the distance covariance is defined as
\begin{multline}
    \text{dCov}^2[f,g]=
    \Big\langle |f-f'|\times|g-g'|\Big\rangle\\
    +\Big\langle |f-f'|\Big\rangle\times\Big\langle|g-g'|\Big\rangle
    -2\Big\langle |f-f'|\times|g-g''|\Big\rangle\,,
\end{multline}
where $(f,g)$, $(f',g')$, $(f'',g'')$ are all independent and identically distributed from the same joint distribution. In practice, we evaluate $\text{dCov}^2[f,g]$ by averaging $|f_i-f_j|\times |g_i-g_j|$, $|f_i-f_j|$ and $|g_i-g_j|$ over all pairs of events $i,j$ and  $|f_i-f_j|\times |g_i-g_k|$ over all triplets of events $i,j,k$.

The distance correlation is then defined analogously to the usual correlation:
\begin{align}
    \text{dCorr}^2[f,g]=\frac{\text{dCov}^2[f,g]}{\text{dCov}[f,f]\,\text{dCov}[g,g]}\,.
\end{align}

\section{Single DisCo in the Gaussian Case}
\label{sec:singleDisCoproof}

In Sec.~\ref{sec:toy}, we observed that for the simple Gaussian model with three Gaussian random variables $X_0$, $X_1$ and $X_2$, the Single DisCo classifier $f(X_1,X_2)$ trained to be independent of $X_0$ (which is correlated with $X_1$ but not $X_2$) is only a function of $X_2$ and does not depend on $X_1$.  The purpose of this appendix is to prove this. 

We start by rotating from $(X_0,X_1,X_2)$ into another set of three Gaussian random variables that are mutually independent: $X_0, W$, and $X_2$ with $X_1=\alpha X_0+\beta W$, where $\alpha,\beta$ depend on $\rho_b$ and $W$ is independent from $(X_0,X_2)$.  Then, we can also write $h(X_0,W,X_2)=f(\alpha X_0+\beta W,X_2)$.  Let $Q=(W,X_2)$. Suppose that $h(X_0,Q)$ and $X_0$ are independent. Then for all sets $A$ and $B$:
\begin{equation}
\label{eq:first}
\Pr\Big[g(X_0,Q)\in A\text{ and }X_0\in B \Big]=\Pr\Big[h(X_0,Q)\in A\Big]\times\Pr\Big[X_0\in B\Big]
\end{equation}
For any $B$, define $A_B=\{h(x_0,q):x_0\in B,\forall q\}$.  Then, the probability that $h(X_0,Q)\in A_B$ given $X_0\in B$ is unity:
\begin{align}
    \Pr\Big[h(X_0,Q)\in A_B\text{ and }X_0\in B\Big]&= \Pr\Big[h(X_0,Q)\in A_B|X_0\in B\Big]\times\Pr\Big[X_0\in B\Big] \nonumber \\
    &=\Pr\Big[X_0\in B\Big]\,,
\end{align}
and so Eq.~\eqref{eq:first} simply reduces to $\Pr[h(X_0,Q\in A_B]=1$. This means that $h(x_0,q)$ cannot depend on $x_0$.  Therefore, we conclude that if $h(X_0,Q)$ and $X_0$ are independent, then $h$ does not depend on $X_0$. The only way for $h$ to not depend on $X_0$ is for $f$ to not depend on $X_1$.\\

\bibliography{refs,HEPML}
\bibliographystyle{utphys}

\end{document}